\documentclass[onecolumn,useAMS,usenatbib,a4paper]{mn2e}
\usepackage{epsfig}
\usepackage{subfigure}
\usepackage{verbatim}
\usepackage{hyperref}
\usepackage{color}
\usepackage{multicol}

\newcommand{\beeq}{\begin{equation}}
\newcommand{\eneq}{\end{equation}}
\newcommand{\bear}{\begin{eqnarray}}
\newcommand{\enar}{\end{eqnarray}}
\newcommand{\bef}{\begin{figure*}}
\newcommand{\enf}{\end{figure*}}

\newcommand{\apjl}{ApJL}
\newcommand{\aj}{AJ}
\newcommand{\aap}{A\&A}
\newcommand{\mnras}{MNRAS}
\newcommand{\apjs}{ApJS}
\newcommand{\apj}{ApJ}
\newcommand{\apss}{Ap\&SS}
\newcommand{\physrep}{Phys. Rep.}
\newcommand{\prd}{Phys. Rev. D}

\newcommand{\jcap}{J. Cosmology and Astroparticle Phys.}

\newcommand{\kbf}{{\bf k}}
\newcommand{\xbf}{{\bf x}}

\newcommand{\iacons}{\frac{C_1\rho_{\rm crit}\Omega_M}{G(z)} }

\newcommand{\twia}{w_{j}^{{\rm IA}}}

\newcommand{\wipzLa}{\Pi^{(a)}_j}%{\int_{z_L-\Delta z}^{z_L+\Delta z}P_j(z) dz}
\newcommand{\wipzLb}{\Pi^{(b)}_j}%{\int_{z_L+\Delta z}^{z_T}P_j(z) dz}
\newcommand{\invls}{\Sigma_{c,j}^{-1}}

\title[Intrinsic alignments in groups and clusters]
  {Intrinsic alignments of group and cluster galaxies in photometric surveys}
\author[Chisari et al.]
  {Nora Elisa Chisari$^1$,\thanks{nchisari@astro.princeton.edu}
  Rachel Mandelbaum$^2$,
  Michael A. Strauss$^1$,
\newauthor
  Eric M. Huff$^3$,
  Neta A. Bahcall$^1$\\
  $^1$Department of Astrophysical Sciences, Princeton University, 4 Ivy Lane,
Princeton, NJ 08544, USA\\
  $^2$ McWilliams centre for Cosmology, Department of Physics, Carnegie Mellon University, 5000 Forbes Ave., 
Pittsburgh, PA 15213, USA\\
  $^3$The centre for Cosmology and Astro-Particle Physics, CCAPP,
The Ohio State University, \\
  Physics Research Building, 191 West Woodruff Avenue Columbus, Ohio 43210, USA}
\date{Accepted xxxxx Received xxxxx}

\begin{document}
\pagerange{\pageref{firstpage}--\pageref{lastpage}} \pubyear{2014}

\def\LaTeX{L\kern-.36em\raise.3ex\hbox{a}\kern-.15em
    T\kern-.1667em\lower.7ex\hbox{E}\kern-.125emX}

\newtheorem{theorem}{Theorem}[section]

\maketitle

\label{firstpage}

\begin{abstract}
Intrinsic alignments of galaxies have been shown to contaminate weak gravitational lensing observables on linear scales, $r>$ 10 $h^{-1}$Mpc, but studies of alignments in the non-linear regime have thus far been inconclusive. We present an estimator for extracting the intrinsic alignment signal of galaxies around stacked clusters of galaxies from multiband imaging data. Our estimator removes the contamination caused by galaxies that are gravitationally lensed by the clusters and scattered in redshift space due to photometric redshift uncertainties. It uses posterior probability distributions for the redshifts of the galaxies in the sample and it is easily extended to obtain the weak gravitational lensing signal while removing the intrinsic alignment contamination. We apply this algorithm to groups and clusters of galaxies identified in the Sloan Digital Sky Survey `Stripe 82' coadded imaging data over $\sim 150$ deg$^2$. We find that the intrinsic alignment signal around stacked clusters in the redshift range $0.1<z<0.4$ is consistent with zero. In terms of the tidal alignment model of Catelan et al. (2001), we set joint constraints on the strength of the alignment and the bias of the lensing groups and clusters on scales between $0.1$ and $10\,h^{-1}$ Mpc  , $b_LC_1\rho_{\rm crit} = -2_{-14}^{+14} \times 10^{-4}$. This constrains the contamination fraction of alignment to lensing signal to the range between $[-18,23]$ per cent below scales of $1$ $h^{-1}$ Mpc at 95 per cent confidence level, and this result depends on our photometric redshift quality and selection criteria used to identify background galaxies. Our results are robust to the choice of photometric band in which the shapes are measured ($i$ and $r$) and to centring on the Brightest Cluster Galaxy or on the geometrical centre of the clusters. 
\end{abstract}


\section{Introduction}

The intrinsic alignments of galaxies are an important contaminant in weak gravitational lensing measurements \citep{Hirata04}. The ellipticity of a galaxy can be subject to physical effects that stretch it and orient it in preferential directions with respect to the large-scale structure. These preferential alignments with large-scale structure can mimic the coherent galaxy alignments due to gravitational lensing. For a review on intrinsic alignments, see \citet{Troxel14}.

On large scales, the mechanisms responsible for alignment \citep{Catelan01} are the stretching by the tidal field of the large-scale structure, for galaxies that are supported by random motions (red ellipticals), and tidal torquing, for galaxies with significant angular momentum (blue spirals). Primeval magnetic fields \citep{Reinhardt71} can also align spiral galaxies. If disc galaxies acquire their angular momentum from the tidal field in which they form and this tidal field then exerts a torque on them, the alignment is a second order effect. Thus it is expected to be weaker than for elliptical galaxies.

Red elliptical galaxies separated by large distances ($>6$ Mpc$/h$) are observed to point radially towards each other \citep[][]{Hirata07,Okumura09}, and this alignment signal can be well described by the coherent effect of the tidal field of the large-scale distribution of matter in the Universe \citep[][]{Blazek11}. There is marginal evidence for the alignments between the direction of the angular momentum (`spin') of blue galaxies  (\citealt{Lee11} and references therein), which are the most relevant source for weak gravitational lensing surveys. However, their shapes do not display a significant correlation and only upper limits on their shape alignments are currently available \citep{Mandelbaum11,Heymans13}.

On small scales, other physical mechanisms are expected to influence the degree of alignment. \citet{Hawley75} suggested that the preferred orientation of the tidal field of a cluster can leave its imprint on the orientations of galaxies, if their orbits are radial. \citet{Ciotti94} quantified the strength of alignment using theory and simulations and showed that the effect is more significant outside the cluster core radius. \citet{Faltenbacher08} found that dark matter haloes of group galaxies in simulations indeed align towards the central substructure, but they found that alignment peaks at the centre of the cluster and extends up to six times the virial radius. Intriguingly, simulations suggest a much stronger alignment than is observed on these scales \citep{Pereira08}. \citet{Pereira10} suggested that this discrepancy could be reconciled if the tidal field of the cluster is also responsible for a significant misalignment between the dark matter component of a halo and its luminous component. They showed that this effect is particularly strong in the inner $50$ kpc of the cluster and they also noticed that, when the luminous component becomes misaligned, it also tends to become more spherical. More recently, \citet{Tenneti14} studied the relative alignment of the stellar and the dark matter components of galaxies in high-resolution hydrodynamic simulations. Their results indicate that the misalignment between the two components is larger at lower redshift and for lower halo masses.
The stellar component of satellites show larger misalignment than that of the central subgroup, which seems to be a consequence of the overall mass dependence of the effect. \citet{Codis14} studied the alignment of galaxy shapes with the tidal field in hydrodynamical cosmological simulations at $z=1.2$. Their results indicate that blue galaxies are subject to significant alignment as opposed to red galaxies, which show no alignment signal. However, their model for obtaining galaxy shapes relies on the spin of a galaxy, and it is hence designed to place upper limits on the alignments of the blue population only.

On these scales, the approximations of the linear alignment model of \citet{Catelan01} break down. A simple extension towards the non-linear regime was suggested by \citet{Hirata07}, and applied by \citet{Bridle07}. To go beyond these approximations, \citet{Schneider10} developed a halo model of alignments. In the non-linear regime, the impact of baryonic physics and non-linear clustering on intrinsic alignments is an open question. Understanding of intrinsic alignments on the smallest scales could allow this signal to be used as cosmological probe. Moreover, the strength of alignment around a cluster could possibly probe its dynamical state and the epoch of formation (alignment is indicative of coevality; \citealt{Djorgovski83}). Alignments could preserve the primordial direction of collapse \citep{Wesson84} and probe preferential directions of accretion towards overdensities \citep{Aubert04}. Poor modelling of intrinsic alignments on small scales can also bias estimates of $\sigma_8$ from weak lensing \citep{Heymans13}. 

Observational studies of alignments below separations of several Mpc have been inconclusive. \citet[]{Hawley75} found a weak tendency for alignment of galaxies in the Coma cluster, followed by \citet{Thompson76}, who reported a significant radial alignment in that cluster, which was confirmed by \citet{Djorgovski83}. \citet{Bernstein02} studied a sample of $\simeq1800$ satellites around isolated galaxies in the 2dF Galaxy Redshift Survey, and found their results to be consistent with no alignments at separations below $500 h^{-1}$ kpc. Studies with larger samples followed with less definite results. \citet{Pereira05} found a marginally significant signal (at $4\sigma$) in the alignment of galaxies in a spectroscopic sample of 85 X-ray selected clusters. However, \citet{Hao11} noticed that such alignment was present only when determined from isophotal fit position angles and that it was dependent on the apparent magnitude of the Brightest Cluster Galaxy (BCG), suggesting residual systematics (such as contamination from the extended envelope of the BCG) rather than a physical alignment.  

Recently, \citet{Schneider13} studied the alignments of galaxies in groups in the GAMA \citep{GAMA} survey. Their sample consisted of $3850$ galaxy groups and $13655$ sources, selected using spectroscopic redshifts. They observed a marginal trend for alignment ($\simeq 3\sigma$) when galaxy shapes were measured using two-dimensional profile fits; the significance decreases ($\simeq 2\sigma$) when using weak lensing shapes from the Sloan Digital Sky Survey (SDSS; \citealt{Ahn12}). They attributed this difference to the different radial weighting of the light in those methods. Because profile fits put more weight on the outskirts of a galaxy, the tidal effect could be stronger \citep{Ciotti94,Kuhlen07} when that method is used; but it would also be more subject to contamination from the light of nearby physically unassociated objects. \citet{Sifon14} measured the alignment of spectroscopically selected satellites around $91$ massive clusters between $0.05<z<0.55$. Their results show no alignment of satellites around massive clusters.

If present, alignments of cluster galaxies would be a contaminant of cluster-galaxy lensing measurements when we cannot separate cluster members from galaxies in the background. This situation arises when photometric redshifts (photo-$z$) are used as a proxy for the distance to a galaxy. When photometric redshift scatter is comparable to the redshift separation between lens and sources, galaxies intrinsically aligned at the cluster redshift are mistaken for galaxies behind the cluster that are lensed by its gravitational potential well. \citet[][]{Blazek12} developed a formalism to minimize the impact of intrinsic alignment contamination on galaxy-galaxy lensing and applied it to Luminous Red Galaxies in the SDSS. Their model-independent constraints indicate that intrinsic alignments contaminate the weak lensing signal around these galaxies at the $10$ per cent at most given their photometric redshift quality. 

In this work, we develop a method to measure the intrinsic alignment of galaxies around clusters while removing the weak lensing contamination to the intrinsic alignment signal. This work differs from that of \citet{Blazek12} in several ways. First, we choose the statistical weights of our estimator to optimize the recovery of the average shear, rather than the surface density profile of a cluster. Secondly, we incorporate the full photometric redshift posterior distribution for each galaxy, $P(z)$, rather than the redshift that maximizes the likelihood or the posterior. We take this approach based on the results of comparing photometric redshifts obtained using a publicly available Bayesian code \citep{Feldmann08} through a template fitting procedure to a set of galaxies with spectroscopic redshifts. Finally, we apply our method to set constraints on the intrinsic alignments of galaxies in and around clusters of galaxies. 

The layout of this manuscript is the following. In Section \ref{sec:clusters}, we describe the group and cluster catalog \citep[][]{Geach11} used in our study. We describe the sample of galaxies with shapes and their photometric redshift calibration in Section \ref{sec:shapesample}. In Section \ref{sec:formalism}, we present our formalism for estimating the intrinsic alignment of galaxies around clusters. Section \ref{sec:NLA} describes the `non-linear alignment model' (NLA, \citealt{Bridle07}), which approximately predicts the alignment signal in the non-linear regime (up to an overall normalization that we will fit to our results). 
We use two different coadditions of SDSS data in this work. One coaddition \citep{Annis11}, in five optical bands, is used to determine the photometric redshifts of the galaxies and it was used by \citet{Geach11} to contruct the lens catalog. The second coaddition \citep{Huff11} was used to determine the shapes of galaxies in two bands. This second coaddition, which includes corrections for the point spread function, is necessary to achieve sufficient precision in the determination of galaxy ellipticities for cosmic shear studies.
Our constraints on the intrinsic alignment of cluster galaxies from the photometric sample are presented in Section \ref{sec:results}. We discuss the implications of these constraints on contamination from intrinsic alignments on weak lensing observables in Section \ref{sec:contamination}. In Section \ref{sec:discuss}, we compare our results to those of previous studies. We conclude in Section \ref{sec:conclude}. Appendix A provides a computation of the error bars of our intrinsic alignment estimator when the statistics are dominated by the noise coming from the intrinsic scatter in the shapes of galaxies, `shape noise'. In Appendix B, we present an analysis of the $P(z)$ distributions obtained for the calibration set. By fitting the $P(z)$ with a combination of Gaussian functions, we reduce the number of parameters needed to describe each galaxy, which will be necessary for using this distributions in ongoing and upcoming weak lensing surveys.

Throughout this work, we use as our fiducial cosmology: $\Omega_M = 0.288$, $\Omega_\Lambda=0.712$, $H_0=100h$ km$/s$ Mpc$^{-1}$ and $h=0.6933$ \citep[][]{Hinshaw12}. Correlation functions are presented as a function of comoving separation, $r_p$, and line-of-sight comoving distance, $\Pi$.

%---------------------------------------------------------------------------------------------------
%NEW SECTION------------------------------------------------------------------------------
%---------------------------------------------------------------------------------------------------

\section{Group and cluster catalog}
\label{sec:clusters}

The SDSS has imaged a region of $\sim 275$ deg$^2$, `Stripe 82', repeatedly to a coadded depth $\sim 2$ mag below the typical depth of the full footprint in 5 filters. \citet[][]{Annis11} produced a catalog of sources in the Stripe based on this coaddition, effectively going to $m_r \simeq 23.5$ for galaxies. This area provides a unique opportunity for higher redshift studies than the full footprint of the SDSS survey, and it is the largest area covered at this depth by any imaging survey in $5$ photometric bands to date.

The group and cluster catalog used in this work was constructed on Stripe 82 by \citet[][]{Geach11} and is publicly available\footnote{\url{http://www.physics.mcgill.ca/~jimgeach/stripe82/}}. $4098$ clusters were identified with a median redshift of $z_{\rm med} \sim 0.32$. For the purpose of this work, we restrict our analysis to the $3099$ clusters in the redshift range between $z=0.1$ and $z=0.4$, since we lose accuracy in determining the weak lensing contamination for higher redshift clusters due to the paucity of source galaxies behind them. 

The clusters are found by applying Voronoi tessellation to a set of photometrically selected galaxies, with colors consistent with the cluster red sequence, calibrated from Abell 2631 to account for $90$ per cent of its members and assuming that its slope in the color-magnitude diagram does not evolve with redshift. The galaxies used for constructing the lens catalog are pre-selected to have $m_r<23.5$ to minimize the impact of photometric uncertainties in the identification of the red sequence. This cut is deeper than (and thus consistent with) our calibration set depth and it is the same cut as applied in the shape catalogue of \citet{Huff11}. The redshifts of the clusters are determined by a weighted mean of the available spectroscopic and photometric redshift information of the cluster members. The spectroscopic and photometric redshifts are obtained from SDSS DR7 \citep{Abazajian09} and with {\sc hyperz} \citep{Bolzonella00}, respectively. The richness of the clusters is defined as the number of red sequence galaxies in the connected Voronoi cells that constitute the cluster. A Voronoi cell is part of a cluster if the probability that it has been drawn from a random field given its area is less than a critical value. In Figure \ref{fig:geachz}, we show the distribution of redshifts of the clusters, which is not volume-limited. 

%-----------------------------------------
%cluster_props.pro
\bef
\includegraphics[width=0.45\textwidth]{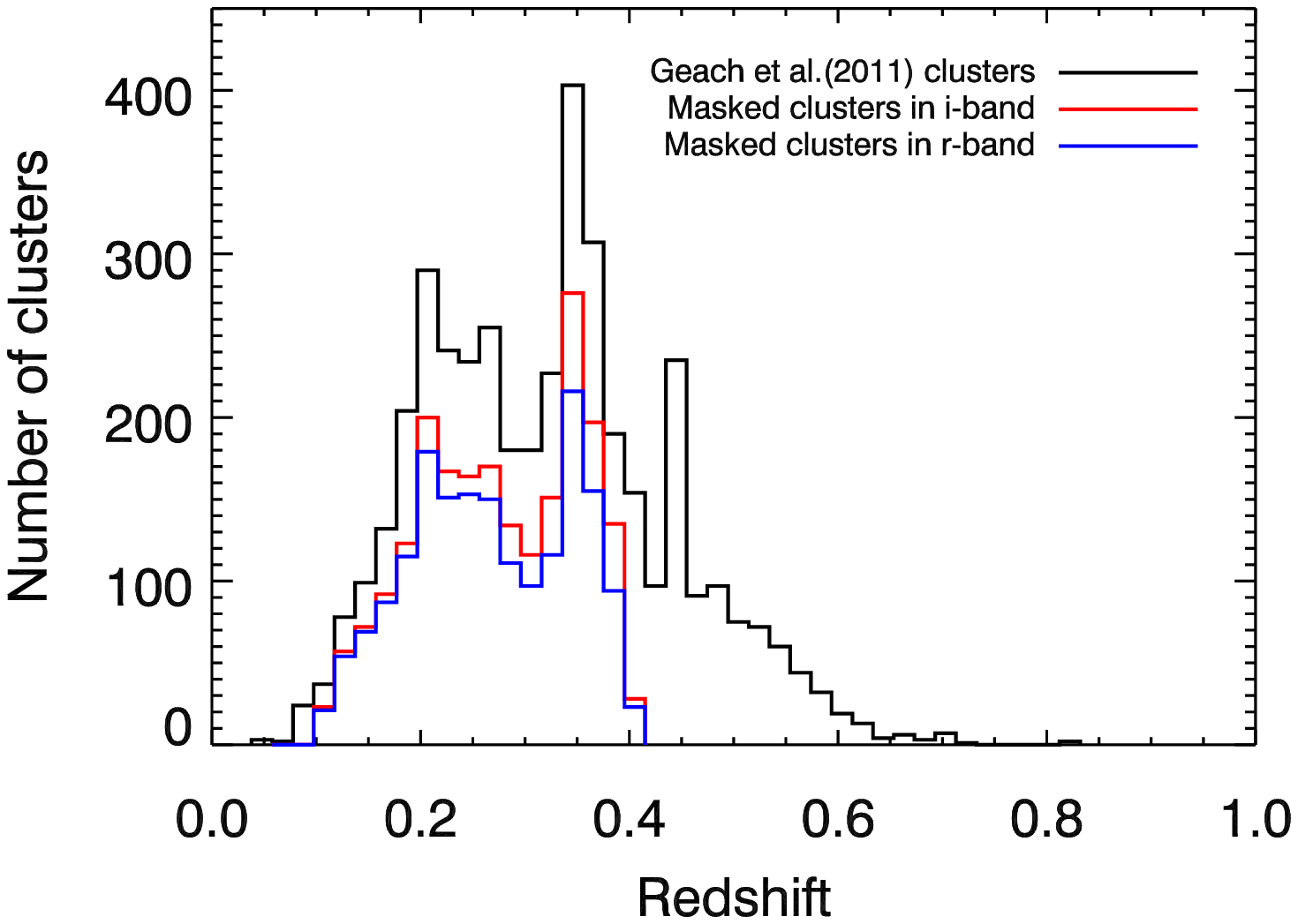}
\caption{Redshift distribution of the \citet{Geach11} group and cluster catalog. The red and blue lines indicate the effective redshift distributions of the clusters that are used for the stacks in the $i$-band and $r$-band, respectively, when masking is applied on the number of background galaxies used for lensing. }
\label{fig:geachz}
\enf
%-----------------------------------------

The BCG of each cluster is identified by \citet[][]{Geach11} and can be used as the centre of the cluster. This is an approximation, as there is a distribution of offsets to BCGs from the centre of their clusters \citep{Johnston07,George12,Zitrin12,Lauer14}. Alternatively, the centre of the cluster is determined by the mean of the positions of the member galaxies, which are identified by red sequence cuts. For this work, we choose the BCG as the cluster centre, but we discuss the effect of using the geometrical centre in Section \ref{subsec:centre}.  

A random catalog of clusters, which we will need to estimate the intrinsic alignment signal in Section \ref{sec:formalism}, is constructed matching the sky distribution of \citet[][]{Geach11} clusters and their joint richness and redshift distribution, with $10$ times as many points as clusters. The Stripe 82 area is subject to several systematics that need to be taken into account in our study. First, there is a variation of $\sim 0.4$ mag in depth across the Stripe due to the varying number of exposures used for the coadd as a function of RA (see Figure 1 of \citealt{Annis11}).  The depth variation produces an increase in the density of clusters with RA (a trend that is also seen in the density of galaxies in the source catalogue). We model the Stripe as segmented in three parts along RA, with boundaries $-50\degr<{\rm RA}<-15\degr$, $-15\degr<{\rm RA}<20\degr$ and $20\degr<{\rm RA}<40\degr$, and normalize the density of random points separately in each one. We do not consider clusters at ${\rm RA}<-50\degr$ or ${\rm RA}>40\degr$ due to high Galactic extinction in those ranges. Any systematics induced in the redshift distributions of the clusters and sources across the Stripe will be taken into account by our estimation of the error bars through the bootstrap method in Section \ref{sec:results}.

%---------------------------------------------------------------------------------------------------------------------------
%NEW SECTION----------------------------------------------------------------------------------------------------------------
%---------------------------------------------------------------------------------------------------------------------------
\section{Source catalog of galaxy shapes}
\label{sec:shapesample}

The coaddition of the Stripe 82 presented by \citet[][]{Huff11} is optimized (PSF-matched) for cosmic shear in the $r$- and $i$-bands in Stripe 82, and results based on this coaddition were presented in \citet[][]{Huff11b}. The moments and ellipticities of each galaxy are PSF-corrected by re-Gaussianization \citep[][]{Hirata03,Mandelbaum05} for both $r$-band and $i$-band photometry. There are $1,053,286$ and $1,234,521$ galaxies with measured shapes in Stripe 82 in $r$ and $i$-band to $m_r<23.5$ and $m_i<22.5$ over 168 and 140 deg$^2$, respectively. The ellipticity is defined as

\beeq
e = \frac{1-q^2}{1+q^2}
\eneq

\noindent where $q$ is the axis ratio of the galaxy. The ensemble ellipticity is related to the shear, $\gamma$, by a responsivity factor, $\langle e \rangle=\mathcal{R}\gamma$. The responsivity represents the response of a galaxy of given ellipticity to the shear and it depends on the population of galaxies and on the algorithm used to determine the shapes.

We use the coaddition by \citet{Annis11} to obtain photometric redshifts for the galaxies with shapes; unlike the \citet[][]{Huff11} coaddition,  all photometric bands are available in this case. We thus perform a separate query of Stripe~82 co-added data of \citet{Annis11} retrieving $u,g,r,i,z$ `modelMags' to compute photometric redshifts. modelMags are a Galactic extinction-corrected \citep{SFD} estimate of the magnitude of the galaxy given the best fitting profile in $r$-band \citep{Stoughton02}. The query requires that the $i$-band magnitude be in the range $13<m_i\leq25$, that the object is classified as a `GALAXY' (it is extended beyond the PSF) and detected as `BINNED1' (greater than $5\sigma$ peak after smoothing with the local point spread function) in both $i$ and $r$-bands. The effective depth of the source sample is determined by cross-matching to the shape catalogue, which has a limiting magnitude of $m_i<22.5$ and $m_r<23.5$. We remove objects flagged as `NOPROFILE', `PEAKCENTER', `NOTCHECKED', `SATURATED', `BLENDED', `CHILD', `BRIGHT' or `BAD$\_$COUNTS$\_$ERROR' \citep{Stoughton02} in either $i$ or $r$-bands. We reject objects where `DEBLEND$\_$NOPEAK' is set and the PSF magnitude error is large ($> 0.2$ mag) in those bands. With these cuts, we retrieve approximately one order of magnitude more galaxies than those with measured shapes in Stripe 82.

%=========================
%NEW SUBSECTION
%=========================

\subsection{Responsivity}
\label{ss:shearcalib}

In the \citet[][]{Huff11} shape catalog, the responsivity is obtained from simulated galaxy images that are processed by the same pipeline as the co-added galaxy images. The simulated galaxies are obtained by PSF-deconvolution of HST COSMOS images \citep{COSMOS}. A known shear is applied to these galaxies and they are then added using the SHERA processing pipeline \citep{Mandelbaum12} to the original coadded images, which were then re-analyzed. In this way, the SHERA galaxies inherit the properties of the coadd images, including shear measurement and selection biases, giving a very accurate measurement of the shear responsivity. The responsivity factor obtained in this way is $\mathcal{R} =1.776 \pm 0.043$ for the overall sample. This result is close to the analytic estimate for the shape measurement method used to generate the catalog  ($\mathcal{R}=2(1 -\langle e^2\rangle)$,  $\langle e^2 \rangle \sim 0.35^2$ per component, \citealt{Hirata03}), which gives $\mathcal{R} =1.755$.

Intrinsic alignments cause a small perturbation in the ellipticity of a galaxy. Weak lensing shapes can be used to estimate the alignment signal, and the shear calibration factor obtained by \citet{Huff11} is applicable. If alignments cause a small rotation of a galaxy to align it in a preferred direction, this would be equivalent to slightly changing its ellipticity, and a correspondence between intrinsic shear and position angle statistics can be obtained \citep{Hirata04b}. If tidal effects modify only the outer parts of the galaxy, alignments will be poorly estimated by weak lensing shapes because those are weighted by surface brightness. Specifically, the shapes used in this work are weighted by the best-fitting elliptical Gaussian to the light profile. \citet{Schneider13} have attributed the null detection of intrinsic alignments on group scales to this hypothesis. While this is an interesting question from the point of view of the physical impact of the tidal field on the shape of a galaxy in dense environments, the constraints derived in this work would still be valid. We would need to re-interpret them as the intrinsic alignment measurement derived from weak lensing shapes to determine their contamination on cluster mass estimates, rather than an optimal estimation of the degree of alignment of galaxies in groups and clusters.

%=========================
%NEW SUBSECTION
%=========================
\subsection{Photometric redshift calibration}
\label{ss:photoz}

The photometric redshifts for this work were obtained by applying the Zurich Extragalactic Bayesian Redshift Analyzer ({\sc zebra}\footnote{\url{http://www.exp-astro.phys.ethz.ch/ZEBRA}}, \citealt{Feldmann06}) on the $u,g,r,i,z$ photometric bands (SDSS modelMags) in our sample. {\sc zebra} finds the best fitting spectral energy distribution (SED) template for each galaxy in the photometric sample by interpolating between a fixed set of templates and convolving the templates with the SDSS $u,g,r,i,z$ filter transmission curves \citep{Gunn98}.

We use the templates from \citet[][]{Coleman80} for elliptical, Sbd, Sbc and irregular galaxies\footnote{\url{http://acs.pha.jhu.edu/∼txitxo/bayesian}} and two synthetic starburst templates from \citet{Kinney96}. We create $5$ interpolated templates between each pair of templates to yield a total of $31$ templates. We give each galaxy a template number; $0$ corresponds to an elliptical galaxy and $24$ to $30$ correspond to purely starburst galaxies (see Figure 9 of \citealt{Nakajima12}). \cite{Nakajima12} show that this template choice is adequate for the single-epoch SDSS photometric bands.

We determine several estimators of the photometric redshift and SED of a galaxy, which we will use in Section \ref{sec:formlism} to estimate the intrinsic alignment signal. We obtain the posterior of the photometric redshift of the galaxy and template, $P(z,t)$, and the posterior of the photometric redshift of the galaxy marginalized over template, $P(z)\equiv \sum_t P(z,t)$.  We span the redshift range $0<z<1.5$ with bins of width $0.015$. The likelihood, $\mathcal{L}(z,t)$, and the posterior are related through the photometric redshift prior, $p(z,t)$, by $P(z,t)=p(z,t)\mathcal{L}(z,t)$. We consider photometric redshifts obtained by the maximum of the likelihood marginalized over templates, $z^{\rm ML}_p$, and the maximum of the posterior marginalized over templates, $z_p$.We also consider the average photometric redshift weighted by the posterior, $\langle z \rangle \equiv \int z P(z) dz$. The prior marginalized over template is obtained by {\sc zebra} itself, with a redshift smoothing of $\Delta_S=0.05$, which smoothes over features in the redshift distribution due to the presence of large-scale structure in the small area probed by the calibration set. We also obtain the best fitting template number ($t^{\rm ML}_p$) at $z^{\rm ML}_p$ and the average template of a galaxy marginalized over redshift, $\langle t \rangle \equiv \sum_t^{N_t} t \int dz P(z,t)$. 
Ideally, we would like to incorporate the full $P(z,t)$ into our formalism, but this would be computationally very expensive. 

In the following subsections, we present our tests of the performance of {\sc zebra} on a subset of the photometric sample matched to a flux-limited spectroscopic sample.

%=========================
%NEW SUBSECTION
%=========================
\subsubsection{Calibration set}
\label{sss:tset}
We compile a redshift sample in the Stripe 82 area from a combination of the spectro-photometric PRIsm MUlti-object Survey\footnote{\url{http://primus.ucsd.edu/}} \citep[PRIMUS]{Coil11,Cool13} and the spectroscopic DEEP2 Galaxy Redshift Survey\footnote{\url{http://deep.ps.uci.edu/}} \citep[][]{Newman12}(PRIMUS+DEEP2). We use this sample to quantify the accuracy and precision of the derived photometric redshifts we obtain with {\sc zebra} and to construct the redshift prior that will be used to compute the photometric redshifts in the full Stripe 82 area. 

The DEEP2 survey is a spectroscopic survey with the DEIMOS\footnote{\url{http://www2.keck.hawaii.edu/inst/deimos/}} spectrograph on Keck II targeting galaxies at $0.75 < z < 1.4$, preselected using BRI photometry from \citet{Coil04}. PRIMUS targeted a complementary set of galaxies to DEEP2, covering the range $z<0.7$ \citep{Coil11} in two DEEP2 fields: $02$hr and $23$hr. In those fields, the combination of PRIMUS+DEEP2 yields an approximately magnitude-limited redshift survey to $R<23.3$ (AB magnitudes from CFH12k camera in the $3.6$-m Canada-France-Hawaii telescope, \citealt[][]{Coil11}, \citealt{Nakajima12}). The combined area of PRIMUS+DEEP2 is $1.25$ deg$^2$. Because of the higher quality of the DEEP2 redshifts, if an object has been observed by both surveys (within a matching radius of $2$ arcsec), we discard the PRIMUS observation and retain its DEEP2 redshift. In the case of PRIMUS, we restrict ourselves to objects classified as galaxies in the range $0<z<1.5$ and with ZQUALITY$\geq 3$. ZQUALITY of 4 corresponds to the highest-quality redshifts with $\sigma_z/(1+z)\sim0.0043$ and ZQUALITY of 3, to $\sigma_z/(1+z)\sim0.0051$ \citep{Coil11}, where $\sigma_z/(1+z)$ is $1.48\times{\rm median}(|z_s-z_p|/(1+z))$. For DEEP2, we only use observations with ZQUALITY$\geq 3$, which identify objects with secure redshifts. We match the combined catalog to SDSS data in Stripe 82 with a search radius of $1$ arcsec, after correcting the DEEP2 coordinates for a small offset in the astrometry.

The PRIMUS+DEEP2 combined catalog has $16,886$ galaxies matched to SDSS Stripe 82 coadds. This catalogue does not have the same photometric properties as the typical population of Stripe 82 coadd galaxies in the same area. Because we want the PRIMUS+DEEP2 photo-z prior to be representative of the galaxies in the coadds, we need to resample this set to match the Stripe 82 selection function.  We compute weights from comparing the number counts of galaxies in the matched PRIMUS+DEEP2 catalogue to the coadd galaxies in the same area in the $r$-band. The choice of band is driven by the fact that the PRIMUS+DEEP2 dataset is approximately magnitude limited in $r$, yielding reliable weights down to the faintest magnitudes ($m_r<23.3$). The weight function as a function of apparent magnitude is shown in Figure \ref{fig:rweight}. If defined from the $i$-band, weights are less reliable at the faint end, but our results are insensitive to the choice of band for the weighting.
We use these weights to resample, with repetition, the matched PRIMUS+DEEP2 dataset, yielding a catalogue of $41,300$ galaxies with spectroscopic redshifts from PRIMUS+DEEP2 and the selection function of SDSS Stripe 82 coadds in the range $18<m_r<23.3$. There are insufficient galaxies brighter than $m_r=18$ in the PRIMUS+DEEP2 set to define reliable weights, we thus do not consider galaxies in that range any further in our measurements. 

{\sc zebra} provides a photo-$z$ prior from this set, likelihoods and posteriors of redshift and template for each galaxy. We remove those objects where {\sc zebra} has produced a photo-$z$ likelihood peaked at the first or last photometric redshift bin, yielding $16,477$ galaxies. Of these, $\sim 73$ per cent are from the PRIMUS catalog. {\sc zebra} fails preferentially (64 per cent) for the DEEP2 objects, which are targeted to be at higher redshifts than the PRIMUS galaxies (36 per cent).
The resulting redshift distribution of the PRIMUS+DEEP2 calibration set is shown in Figure \ref{fig:below233}, along with the redshift distribution of the subset of those galaxies that have measured shapes in $i$-band or $r$-band. Figure \ref{fig:below233} shows that the DEEP2 and PRIMUS redshift distributions are almost disjoint. For galaxies with shapes, the resolution factor, $R_2$, compares the size of the galaxy image to that of the PSF. Our resolution cut was $R_2 > 0.333$ (corresponding, for the PSF of the coadds, to an effective radius of about 0.47 arcsec). The effect of PSF estimation errors is larger for poorly resolved galaxies, scaling as $R_2^{-1}$. We varied the resolution cuts, choosing the most inclusive threshold that was consistent with the systematic error requirements for cosmic shear \citep{Huff11,Huff11b}. In principle, more inclusive cuts could have been made for this measurement.

%%%%%%%%%%%%%%%%%%%%%%%%%%%%%%%%%%%%%%%%%%%%%%%%%%%%%%%%%%%%%%%%%%
\bef
\centering
\includegraphics[width=0.5\textwidth]{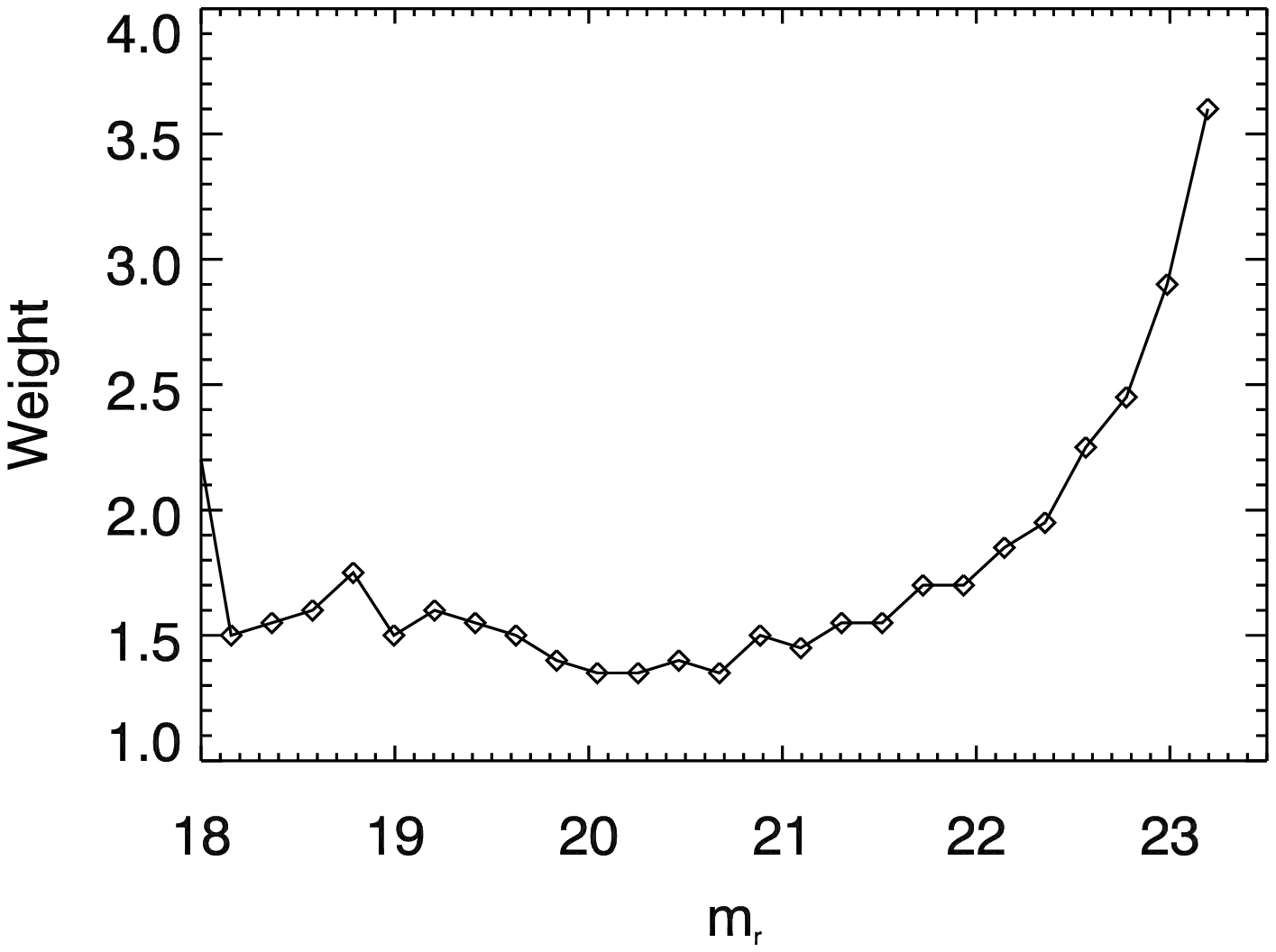}
\caption{Weights applied to resample the PRIMUS+DEEP2 dataset matched to Stripe 82 in order to reproduce the selection function of the Stripe 82 coadds. The weights are defined from the $r$-band photometry, where the PRIMUS+DEEP2 dataset is approximately magnitude-limited.}
\label{fig:rweight}
\enf
%%%%%%%%%%%%%%%%%%%%%%%%%%%%%%%%%%%%%%%%%%%%%%%%%%%%%%%%%%%%%%

%%%%%%%%%%%%%%%%%%%%%%%%%%%%%%%%%%%%%%%%%%%%%%%%%%%%%%%%%%%%%%%%%%
\bef
\centering
\includegraphics[width=0.5\textwidth]{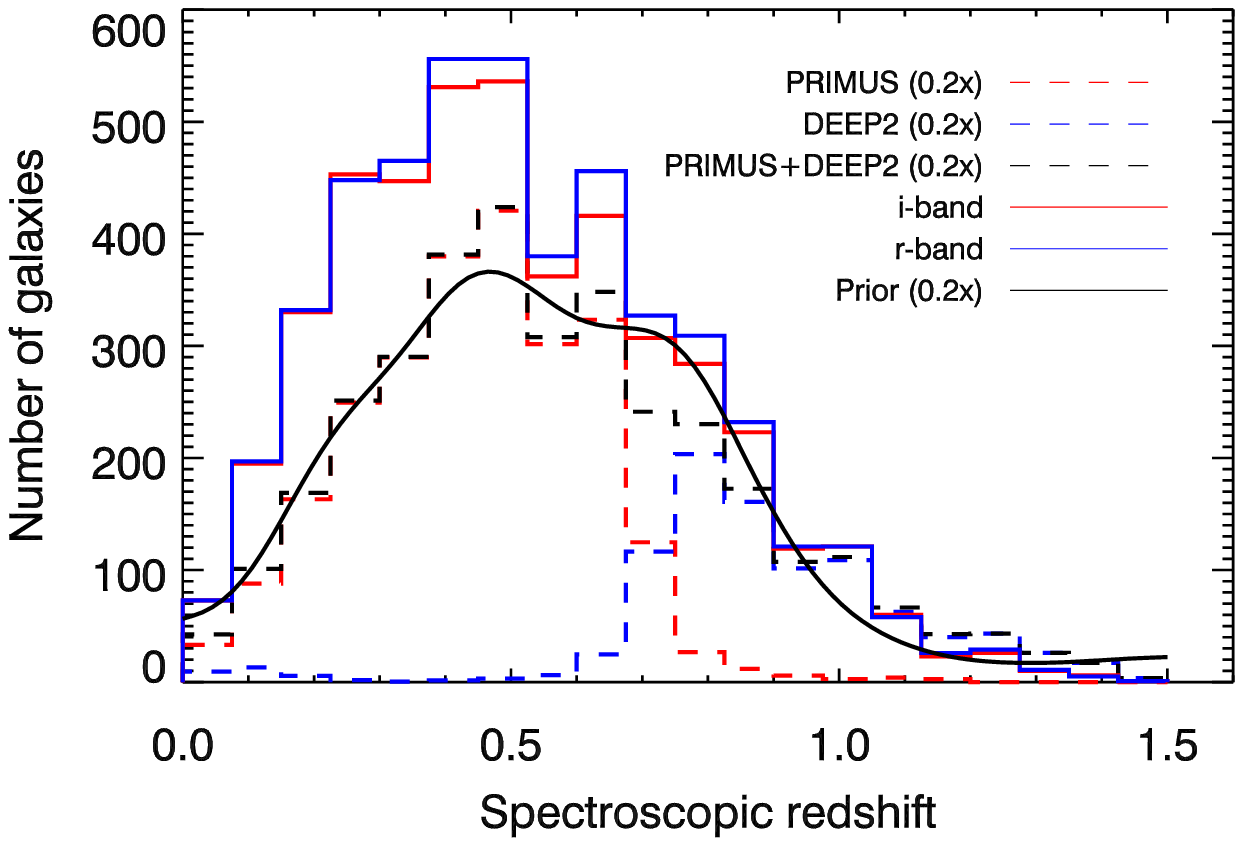}
\caption{Redshift distribution of the galaxies in the flux-limited calibration set with $m_R<23.3$ and ZQUALITY$\geq 3$. The percentage of galaxies from the DEEP2 survey in the training set is $27$ per cent, while $73$ per cent are objects from PRIMUS. The total number of galaxies in the set is $17,708$. The dashed lines show the distribution of galaxies in PRIMUS (red) and DEEP2 (blue) and the combined distribution (black), respectively, before applying quality cuts on the {\sc zebra} redshifts and arbitrarily normalized by a factor $0.2$ for visualization. The solid lines show the spectroscopic redshift distributions of galaxies with ZEBRA redshifts and shapes in $i$ (red) and $r$-bands (blue), respectively, among the PRIMUS+DEEP2 set. The black solid line shows the photo-$z$ prior of the calibration set.}
\label{fig:below233}
\enf
%%%%%%%%%%%%%%%%%%%%%%%%%%%%%%%%%%%%%%%%%%%%%%%%%%%%%%%%%%%%%%

%=========================
%NEW SUBSECTION
%=========================
\subsubsection{Photo-$z$ calibration results}
\label{subsec:speccalib}

For each galaxy in the calibration set, we compared three estimates of its photometric redshift: $z_p$, $z^{ML}_p$ and $\langle z \rangle$, to the spectroscopic redshift of the galaxies. For completeness, we present in this section a comparison of the results of the calibration for $z_p$, $z^{ML}_p$ and $\langle z \rangle$, but we use the full $P(z)$ for each galaxy in our alignment and lensing measurements in the following Sections of this paper. We have found $\langle z\rangle$ to give an unbiased estimate of the true redshift of the galaxy, $z_s$, in each $\langle z\rangle$ bin. 

In Figure \ref{fig:zMLvsz}, we show the result of the calibration based on $z^{ML}_p$. Between photometric redshifts of $0.25$ and $1$, the calibration is reliable. Below $z_s=0.25$, there is a large scatter in the results. The scatter in the calibration, represented by the $1\sigma$ dispersion around the mean, is $\simeq 0.1$ (no clipping). The bias starts positive at low redshift, it is zero between $z=0.5$ and $z=1$ and then becomes negative and steepens above $z=1$. Our results are in agreement with those of \citet[][]{Abdalla11}, who have tested the performance of {\sc zebra} on the SDSS Luminous Red Galaxy sample. For that sample, however, photometric redshifts are expected to perform particularly well due to the prominent $4000\AA$ break in passively evolving stellar populations. While our sample of galaxies is not selected by color, we have found that choosing galaxies with templates below $20$ reduces the bias in the $z^{\rm ML}_p$ calibration. In previous work, galaxies with $t_p^{\rm ML}>20$ \citep{Nakajima12} have been removed from the sample because they showed a large scatter around the median in the calibration (see their section 5.1). Approximately $15.7$ per cent of the unique $16,886$ galaxies in the calibration set have templates $t_p^{\rm ML}>20$. 
Figure \ref{fig:zIvsz} is analogous to Figure \ref{fig:zMLvsz} for the $\langle z \rangle$ photometric redshift estimate, which shows less bias and smaller scatter. The $\langle z \rangle$ calibration is also superior to that with the maximum of the posterior (not shown), suggesting that the $P(z)$ encodes additional information that is not captured in the peak photo-$z$. This is the motivation for incorporating the $P(z)$ to the formalism developed in Section \ref{sec:formalism}. These results are in agreement with a recent analysis by CFHTLenS \citep[][]{Benjamin13}. 

%%%%%%%%%%%%%%%%%%%%%%%%%%%%%%%%%%%%%%%%%%%%%%%%%%%%%%%%%%%%%%%%%%
\bef
\centering
\subfigure[Photometric redshifts correspond to the maximum of the likelihood for each galaxy, $z^{\rm ML}_p$.]{
\includegraphics[width=0.32\textwidth]{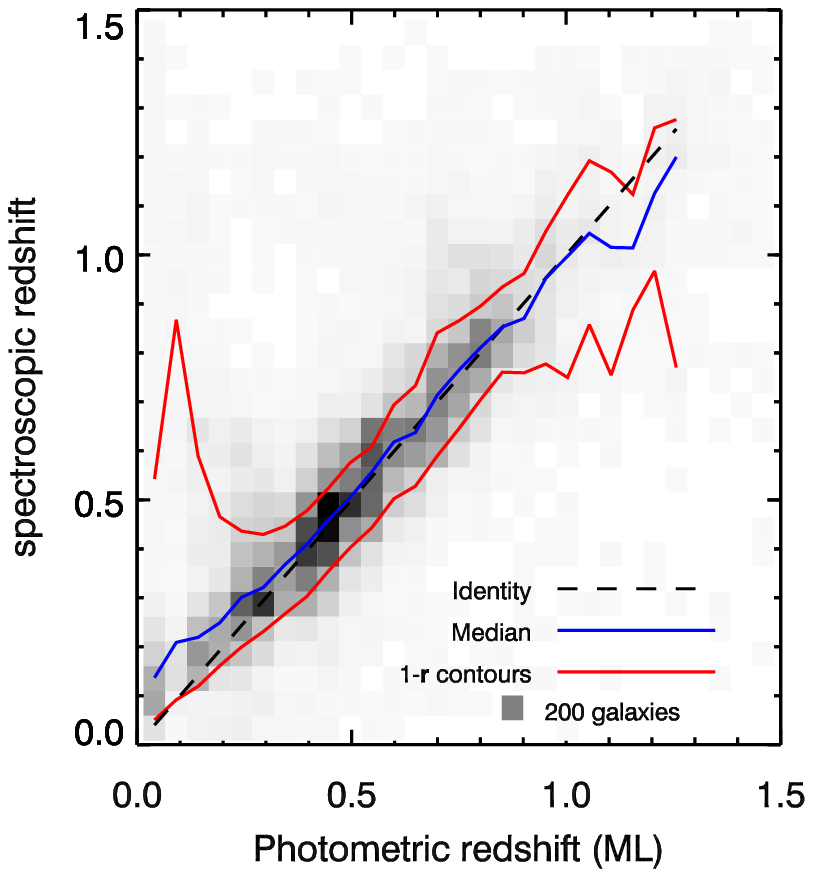}
\label{fig:zMLvsz}
}\hfill
\subfigure[Photometric redshifts corresponding to the redshift weighted over the posterior, $\langle z \rangle$.]{
\includegraphics[width=0.32\textwidth]{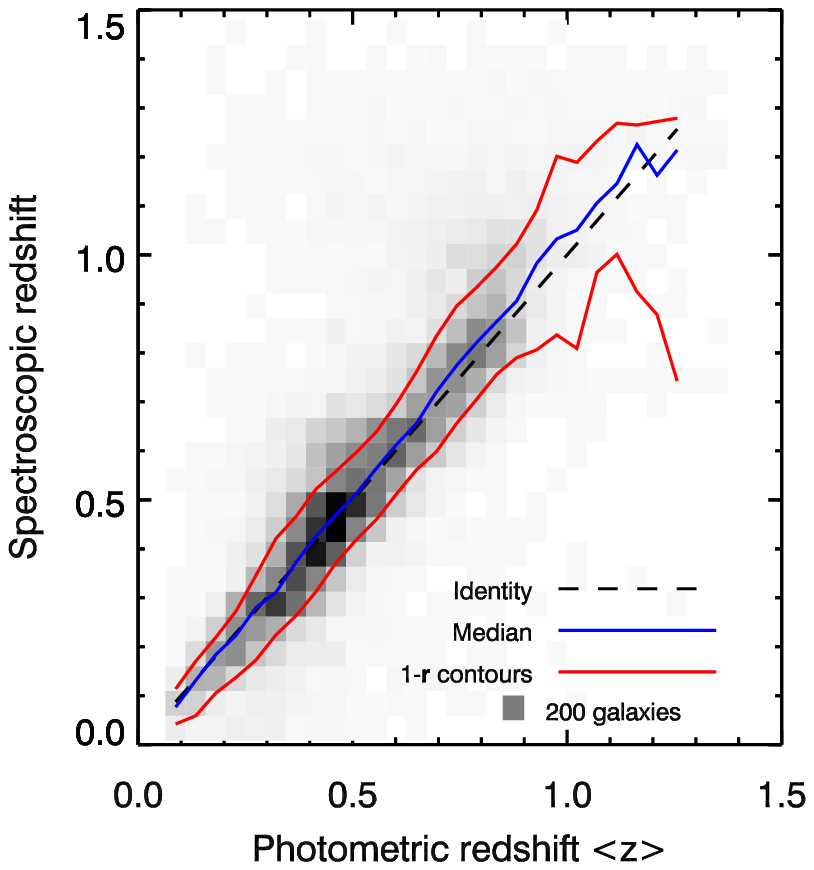}
\label{fig:zIvsz}
}\hfill
\subfigure[$\langle z \rangle$ for galaxies with shapes in $i$-band.]{
\includegraphics[width=0.32\textwidth]{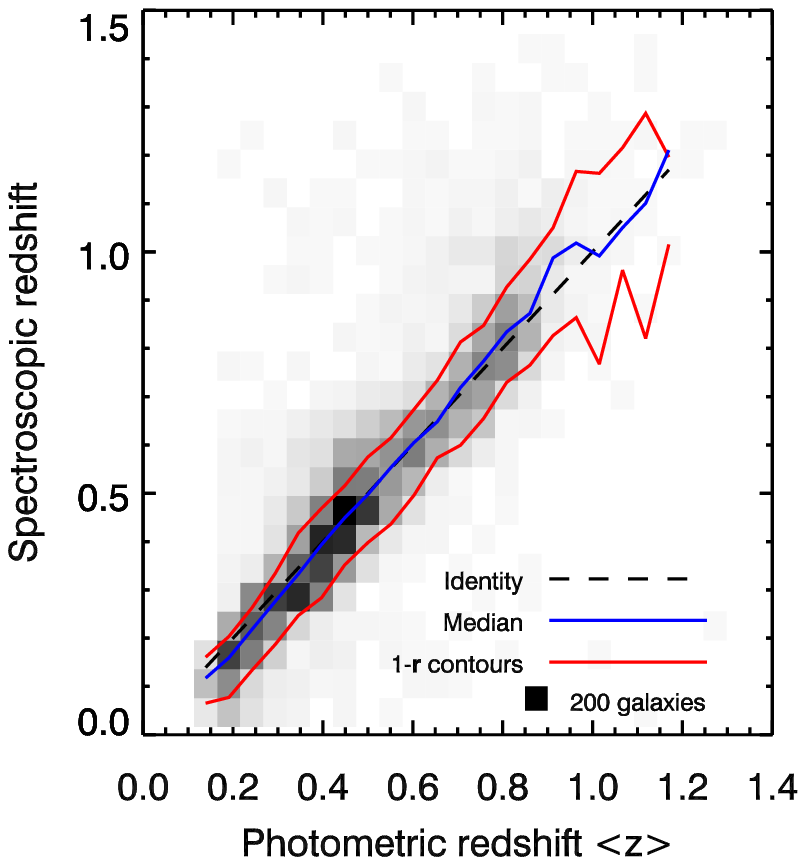}
\label{fig:zIvsz_i}
}
\caption{Results from the photometric redshift calibration of the SDSS Stripe 82 compared to spectroscopic redshifts measured by the PRIMUS and DEEP2 surveys using all galaxies with photometric redshifts (panels (a) and (b)) and galaxies with shapes in the $i$-band (panel (c)). The shading indicates the number of galaxies in each bin; the grayscale is linear. The comparison between panels (a) and (b) reveals that $\langle z \rangle$ gives a better agreement with spectroscopic redshift of a galaxy than does $z^{\rm ML}_p$. It is also superior to the redshift at the maximum of the posterior (not shown). The comparison of panel (c) with panel (b) demonstrates that the photometric redshifts for galaxies with measured $i$-band shapes are as good as in the general case of panel (b). We find similar results for galaxies with measured $r$-band shapes.}
\label{fig:zvsz}
\enf
%%%%%%%%%%%%%%%%%%%%%%%%%%%%%%%%%%%%%%%%%%%%%%%%%%%%%%%%%%%%%%%%%%

We also find that when $\langle z \rangle$ is used as an estimator for the photometric redshift, there is no improvement on the calibration when removing $t^{\rm ML}_p>20$ galaxies and we thus keep them in our sample. Consistent with this fact, we find a large bias between $t^{\rm ML}_p$ and $\langle t \rangle$ for the Kinney starburst templates, as shown in Figure \ref{fig:ttscatter}. Galaxies that have $t^{\rm ML}_p>20$ at their $z_p^{\rm ML}$  actually have a very broad $P(z,t)$. As a consequence, when averaged over all redshifts and templates, $\langle t \rangle$ yields a very different value from $t^{\rm ML}_p$. In other words, there are no galaxies in the sample that can be fit solely by the Kinney starburst templates. Rather, all galaxies at the blue end prefer a combination of logarithmically interpolated templates\footnote{See Section 3.7 of the {\sc zebra} user manual \citep{Feldmann08}.}.

%%%%%%%%%%%%%%%%%%%%%%%%%%%%%%%%%%%%%%%%%%%%%%%%%%%%%%%%%%%%%%%%%%
\bef
\centering
\includegraphics[width=0.5\textwidth]{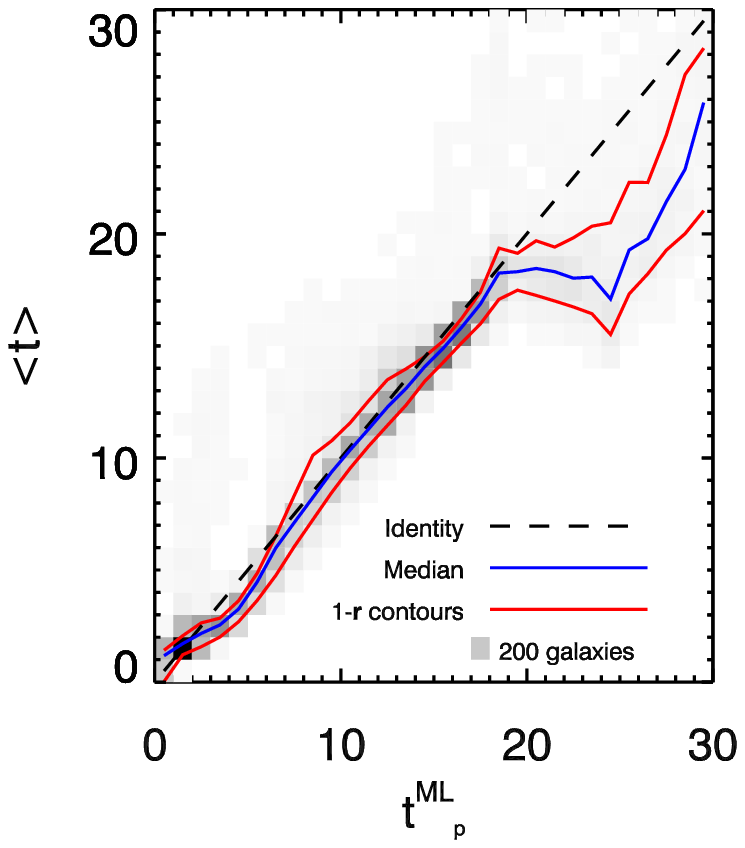}
\caption{Comparison of the best fitting template corresponding to the maximum of the likelihood marginalized over template, $t^{\rm ML}_p$ (i.e., best fitting template at $z_{\rm ML}$), and the template averaged over $P(z,t)$, $\langle t \rangle$ for galaxies in the calibration set that have a successful photo-$z$ computation. The large bias for high $t^{\rm ML}_p$ suggests that blue galaxies tend to prefer a combination of templates.}
\label{fig:ttscatter}
\enf
%%%%%%%%%%%%%%%%%%%%%%%%%%%%%%%%%%%%%%%%%%%%%%%%%%%%%%%%%%%%%%

\subsection{Photometric redshifts in Stripe 82}
\label{subsec:finalpz}

The calibration presented in Section \ref{subsec:speccalib} is valid for galaxies with $m_R<23.3$, where $m_R$ was the apparent magnitude from PRIMUS+DEEP2. There is a typical $\sim0.3$ dex $1\sigma$ scatter between $m_R$ as measured by DEEP2 and PRIMUS and $m_r$ from SDSS, with a slight bias for bright objects that is within the $1\sigma$ quoted. To be conservative in our photometric redshift assignment, we only trust the photometric redshifts and $P(z)$ for galaxies in Stripe 82 with $m_r<23.3$. As discussed previously, we further discard the photometric redshift information for galaxies where the peak of the likelihood computed by {\sc zebra} and marginalized over template occurred in the first or the last photometric redshift bin. 

We match the \citet[][]{Huff11} shape catalog to galaxies with photo-$z$s within $1$ arcsec. $86$ per cent of the galaxies with shapes in both bands have $P(z)$. The main source of mismatch are galaxies outside of the apparent magnitude range we consider for estimating the intrinsic alignment and lensing signals ($m_r<18$ or $m_r>23.3$).

Now that we have defined the group and cluster catalogue and the sample of galaxies with shapes, we will describe the methodology with which we will explore intrinsic alignments.

%---------------------------------------------------------------------------------------------------------------------------
%NEW SECTION----------------------------------------------------------------------------------------------------------------
%---------------------------------------------------------------------------------------------------------------------------

\section{Intrinsic alignment estimator}
\label{sec:formalism}

In the presence of photometric redshift uncertainties, a given sample of galaxies will contain a mixture of some that are intrinsically aligned with a particular foreground object, and others that are gravitationally lensed by it. In the next generation of imaging surveys, separating these two effects will be crucial to determine cluster masses. Intrinsic alignments probe the tidal field of a cluster, an astrophysical effect that has not been explored in depth. \citet{Blazek12} constructed an estimator for intrinsic alignments using a point estimate of the photo-$z$, with the goal of determining the impact of intrinsic alignments on weak gravitational lensing estimates of the mass. In this work, we focus on extracting the intrinsic alignment signal around a set of lenses (in our case, groups and clusters). We use the posterior probability of a photo-$z$ for each galaxy, $P(z)$, to model the redshift information contained in the photometry. Because of the intrinsic scatter of the $P(z)$, it is impossible to unambiguously separate the galaxies in the physical environment of the cluster from galaxies in the background. We do not attempt to isolate cluster members, but construct an statistical estimator to separate the weak lensing and the intrinsic alignment signal. 

When a galaxy physically associated\footnote{We refer to galaxies at the same redshift as the cluster and embedded in the same large-scale structure as being `physically associated' with the cluster. We emphasize that we are not referring only to satellite galaxies within the cluster potential, but we also include those at the same redshift and in the more extended large-scale structure. The motivation for including these galaxies is the fact that alignments of Luminous Red Galaxies are observed to extend to very large scale \citep{Hirata07,Okumura09,Blazek11}.} with the clusters cannot be distinguished from a galaxy in the background, the average tangential shear of galaxies around an overdensity of matter has two contributions,

\beeq
\langle \gamma_+ \rangle = \gamma^{\rm IA} + \gamma^{\rm G}.
\label{eq:shape}
\eneq

\noindent Images of galaxies behind the cluster are tangentially sheared by the gravitational potential, $\gamma^{\rm G}$, and galaxies physically associated with the cluster can be aligned with it due to the tidal field in the region or to infall occurring in preferential directions, giving rise to a contribution to the shear, $\gamma^{\rm IA}$. If alignments of satellite galaxies are radial, $\gamma^{\rm IA}$ and $\gamma^{\rm G}$ have opposite signs. We build an estimator for the average measured $+$ component of the ellipticity of galaxies, $\tilde{\gamma}$, in annuli of projected radius, $r_p$, around the cluster centre. The measurement of  the average $\times $ component of the ellipticity constitutes a null test, since a curl component is not expected by symmetry \citep{Schneider10}. 

%-----------------------------------------
%CORRELATION FUNCTION
%------------------------------------------
Analogously, we can define the correlation function of lenses and galaxy ellipticities, $g\gamma$, which also has contributions from lensing and alignments, $g\gamma = gI + gG$. Although the average shear $\langle \gamma_+ \rangle$ and the correlation $g\gamma$ are representations of the same physical mechanisms, there is a practical advantage to working with correlation functions in the context of this work. We are interested in measuring the $g\gamma$ correlation up to large separations, where the excess of clustered galaxies around the lens becomes very small and noisy. Thus, we prefer to work with correlations. Throughout this work, we will construct the correlation functions using a catalog of $N_R$ points distributed  randomly on the survey area and following the same richness and redshift distribution of the real $N_L$ lensing clusters.

\subsection{Separating weak lensing and intrinsic alignments}
\label{subsec:separate}

Consider two redshift bins, one centred on the cluster, $(a)$, and one centred behind it, $(b)$. In the bin centred on the cluster, the $g\gamma$ correlation for galaxies will have a contribution from intrinsic alignments, $gI^a$, and a contribution from the lensing contamination from galaxies behind the lens scattered into this bin due to photo-$z$ uncertainties, $gG^{b\rightarrow a}$. The latter also accounts for lensing of galaxies within $(a)$ when $(a)$ is very broad. In the bin behind the cluster, the lensing correlation, $gG^b$, will be the dominant signal, while the intrinsic alignments create a contamination, $gI^{a\rightarrow b}$. Instead of assigning galaxies to the two bins based on their photometric redshift, we consider the probability distribution of photometric redshifts for all galaxies and we weight them by the integral of that probability over the redshift range of the bin \citep{Sheth10}. For bin $(a)$, we integrate the $P(z)$ in the interval $[z_L-\Delta z,z_L+\Delta z]$ and for bin $(b)$, in the interval $[z_L+\Delta z,z_T]$, where $z_L$ is the redshift of the lensing cluster and $z_T$ is the redshift up to which we trust the photo-$z$ calibration. We choose $\Delta z$ to minimize the cluster-source position correlation function in bin $(b)$ (this method is similar to the `pair ratio test' of \citealt{Hirata07}). For convenience, we define $\Pi_j\equiv\int_{z_{\rm min}}^{z_{\rm max}}P_j(z) dz$ and specifically, the following two integrals of the $P(z)$ of galaxy $j$ in bins $(a)$ and $(b)$:

\bear
\Pi^{(a)}_j&\equiv&\int_{z_L-\Delta z}^{z_L+\Delta z}P_j(z) dz,\nonumber\\
\Pi^{(b)}_j&\equiv&\int_{z_L+\Delta z}^{z_T}P_j(z) dz
\enar

We can then construct the observed correlation of lens positions and galaxy shapes in each bin and as a function of projected radius from the centre of the stacked lenses, 

\beeq
gI^a (r_p)+ gG^{b\rightarrow a}(r_p) \equiv
\frac{N_R}{N_L}\frac{\sum_{j,r_p}^{\rm lens} \twia\wipzLa \tilde\gamma_j}{\sum_{j,r_p}^{\rm random} \twia\wipzLa}  
\label{eq:corra}
\eneq
\beeq
gG^b (r_p)+ gI^{a\rightarrow b}(r_p) \equiv
\frac{N_R}{N_L}\frac{\sum_{j,r_p}^{\rm lens} \twia\wipzLb \tilde\gamma_j}{\sum_{j,r_p}^{\rm random} \twia\wipzLb} .
\label{eq:corrb}
\eneq

\noindent where the sums are over galaxies in a given $r_p$ bin and each galaxy is weighted by the inverse variance of the shear estimate: the sum in quadrature of the shape measurement noise, $\sigma_{\gamma,j}^2$, and the intrinsic shape noise, $\gamma^2_{\rm rms}$, i.e.,
 
\beeq
\twia=\frac{1}{\sigma_{\gamma,j}^2+\gamma^2_{\rm rms}}.
\label{eq:wia}
\eneq

\noindent This weight is optimal for retrieving the average shear of a sample of galaxies when the galaxy redshifts are perfectly known. Optimal weighting for the recovery of surface mass density requires the use of the lensing kernel \citep[i.e.][]{Blazek12}. In the case of alignments, if the redshifts were known perfectly, the optimal weighting to obtain the average shear around a point is just the inverse shape noise. 

Throughout this paper, we will focus on estimating and modelling $gI^a(r_p)$ for a sample of stacked clusters. With our conventions, when intrinsic alignments are radial, $gI^a(r_p)$ is negative. The amplitude of $gI^a(r_p)$ gives the strength of the intrinsic alignment, and the radial dependence probes the tidal field of the cluster and of the surrounding large scale structure \citep{Ciotti94,Catelan01,Schneider10}. Because the sums in Eqs. \ref{eq:corra} and \ref{eq:corrb} include a dilution due to photometric redshift contamination, we will need to model this effect when relating $gI^a(r_p)$ to the matter power spectrum of the Universe in Section \ref{sec:NLA}.

Obtaining $gI^a(r_p)$ from Eq. \ref{eq:corra} requires an estimation of the lensing contamination to bin $(a)$, $gG^{b\rightarrow a}$. The shear caused on a single galaxy by a lens with surface mass density profile $\Delta\Sigma$ is $\gamma^G = \Delta\Sigma\Sigma_c^{-1}$, where 

\beeq
\Sigma_c = \frac{c^2}{4\pi G} \frac{D_S}{(1+z_L)^2 D_L D_{LS}},
\eneq

is the comoving lensing efficiency, $c$ is the speed of light, $G$ is Newton's gravitational constant, $D_S$ is the angular diameter distance to the source galaxy, $D_L$ is the angular diameter distance to the lens, and $D_{LS}$ is the angular diameter distance between the lens and the source. 
For an ensemble of stacked lenses with average surface mass density profile $\Delta\Sigma^{\rm stack}$, the shear at some projected radius $r_p$ is given by 

\beeq
\gamma^G = \Delta\Sigma^{\rm stack}\,\langle \Sigma_c^{-1}\rangle,
\eneq

\noindent where $\langle \Sigma_c^{-1}\rangle$ is the average of the inverse comoving lensing efficiency over the population of lenses and source galaxies. We have assumed here that the redshift evolution of the surface mass density profiles of the lenses, $\Delta\Sigma$, is small compared to the evolution of $\Sigma_c^{-1}$. We emphasize that there is a difference between estimating the lensing efficiency using photo-$z$s ($\tilde\Sigma_c$) or using the true redshifts of galaxies ($\Sigma_c$). A priori, it seems that we do not have the information of the true efficiency of each source galaxy. We will show in Section \ref{ss:lenseff} that it is possible to calculate the effective lensing efficiency of the population of sources using a representative subset of galaxies with known redshifts. 

Since we choose $\Delta z$ such that there is little contamination from cluster galaxies in bin $(b)$, we can safely neglect $gI^{a \rightarrow b}$ in Eq. \ref{eq:corrb}  and obtain the mean projected mass density profile of the lenses directly from that equation,

\beeq
\Delta\Sigma^{\rm stack}(r_p) = \frac{N_R}{N_L}\frac{\sum_j^{\rm lens} \twia\wipzLb \tilde\gamma_j}{\sum_j^{\rm random} \twia\wipzLb\invls}.
\label{eq:DS}
\eneq

\noindent This allows us to find an expression for $gG^{b\rightarrow a}$,

\beeq
gG^{b \rightarrow a}(r_p) = \Delta\Sigma^{\rm stack}(r_p)\frac{\sum_j^{\rm lens} \twia\wipzLa\invls}{\sum_j^{\rm random} \twia\wipzLa}.
\label{eq:lensa}
\eneq

\noindent Substituting in Eq. \ref{eq:corra}, we obtain the final expression for $gI^a$,

\beeq
gI^a(r_p) = \frac{N_R}{N_L} \left[ \frac{\sum_j^{\rm lens} \twia\wipzLa \tilde\gamma_j}{\sum_j^{\rm random} \twia\wipzLa} - \Delta\Sigma^{\rm stack}(r_p)\frac{\sum_j^{\rm lens} \twia\wipzLa\invls}{\sum_j^{\rm random} \twia\wipzLa}\right].
\label{eq:wfinal2}
\eneq

\noindent Note that if intrinsic alignments and lensing act with different radial dependence on the ellipticity of a galaxy, one would need to make a more careful subtraction of the lensing signal in Eq. \ref{eq:wfinal2}. 

We have built an estimator for the intrinsic alignment signal using photometric redshift posteriors assuming that there are no systematic errors in the average shear around the lenses and random points. In practice, systematics could be present, for example, from anisotropies in the PSF or from masking bias, the selection bias for galaxies with ellipticities parallel to the boundaries of masked regions \citep{Huff11b}. We remove this type of shear systematic errors from shear cross-correlation functions by subtracting the shear cross-correlation function around random points. This procedure is valid under the assumption that systematics arising from PSF anisotropy are uncorrelated with the positions of the lenses. This assumption holds when the main source of systematics is masking bias, since this is produced by a coherent PSF along the scan direction, which is a feature of how the observations were taken.

We could have chosen to leave a gap between bin $(a)$ and bin $(b)$ instead of making the bins contiguous in redshift. The sample of galaxies used in bin $(a)$ and bin $(b)$ is the same, only the range of integration of the $P(z)$ changes from one bin to the other. Throughout this work, we use a typical bin half-width for $(a)$ of $\Delta z=0.2$, which we will justify in Section \ref{sec:results}. The typical photo-$z$ scatter is $\simeq 0.1$ ($1\sigma$), and this is partly driven by the redshift smoothing of the prior, with $\Delta_S=0.05$ (see Section \ref{ss:photoz}). We expect that unless the gap chosen is much greater than the width of the $P(z)$, there will be no significant impact on the results. In the redshift range of our sample, it is impossible to consider a gap between bins $(a)$ and $(b)$ greater than $0.25$. For example, for clusters with $z_L=0.4$, the maximum considered in our cluster sample, a redshift gap of $\Delta_G=0.25$ implies that bin $(b)$ would have a lower redshift bound of $z=0.85$, which is very close to the limit up to which we trust our calibration (only $\sim15$ per cent of the galaxies in the calibration set lie above that redshift). Hence, the incorporation of a gap between bin $(a)$ and $(b)$ is only advantageous when the width of the $P(z)$ is much smaller than the gap, which is not feasible in our case.

We can easily re-write our formalism in the case where point estimates of the photo-$z$ are used, as in previous work \citep[i.e.][]{Blazek12}. This procedure simply requires identifying galaxies in bin $(a)$ as those with $z_L-\Delta z<z_p<z_L+\Delta z$ and galaxies in bin $(b)$, as those with $z_p>z_L+\Delta z$, where $z_p$ is a point estimate of the photo-$z$. This is equivalent to assuming that $P(z)$ is a Dirac-delta function centred on $z_p$. We compare the results of the estimator using $P(z)$ with this version in Section \ref{ss:point}.

\subsection{Lensing efficiency from spectroscopic redshifts}
\label{ss:lenseff}

The remaining question is how to estimate the sums over the lensing efficiency of the galaxies in Eqs. \ref{eq:DS} and \ref{eq:lensa}. When calculated in bin $(b)$, this factor represents the lensing efficiency of galaxies that, regardless of their true redshift, have a non-negligible probability $\wipzLb$ that contributes to the sum in that bin. Effectively, because galaxies at $z\leq z_L$ will not be lensed, this will be the lensing efficiency of galaxies behind the lens, which is also equivalent to the lensing efficiency of galaxies behind the points in the random catalog. Calculated for bin $(a)$, $\langle \Sigma_c^{-1} \rangle$ represents the lensing efficiency of galaxies that truly belong in bin $(b)$ but have been scattered to bin $(a)$ due to photo-$z$ uncertainty. The ratio between $\langle \Sigma_c^{-1} \rangle$ in $(a)$ and $(b)$ represents the percentage of the shear due to gravitational lensing that contaminates the alignment signal in $(a)$. To estimate this contamination we need to know the true redshift of the scattered galaxy. This suggests that we can use a calibration set of galaxies with spectroscopic redshifts, such as the one we use to determine the overall photo$-z$ quality, to estimate this sum statistically.

The $P(z)$ are calibrated using a spectroscopic sample of galaxies complete to a specific limiting magnitude (the `calibration set', see Section \ref{sss:tset}). The spectroscopic sample allows us to construct a redshift prior and to compute the typical lensing contamination to the intrinsic alignment signal. The formalism derived in this section relies on the true underlying redshift distribution of the galaxies with measured shapes, i.e., the redshift prior of the `source catalog' (Section \ref{sec:shapesample}), and thus it can only be applied to galaxies brighter than the limiting magnitude of the calibration set.

Since the galaxies that are clustered with the lens do not contribute to the lensing signal, this factor can be computed over all the area of the calibration sample as in \citet{Nakajima12}, as long as we correct for the relative area factor between the annulus we are considering and the total area of the calibration sample.  We rewrite the numerator of Eq.  \ref{eq:lensa} as

\bear
\mathcal{C}(r_p) &\equiv& \sum_j^{\rm lens} \twia\wipzLa\invls =  \frac{N_L}{N_R} \sum_j^{\rm random} \twia\wipzLa\invls \nonumber\\
&=& \frac{\tilde{A}_{\rm annulus}(r_p)}{A_{\rm calib}} \int dz_L \frac{dN}{dz_L} D_c^{-2}(z_L) \sum_i^{\rm calib} \mathcal{W}(m_i) w_{i}^{{\rm IA}} \Pi_i^{(a)} \Sigma_{c,i}^{-1}(z_L,z_{i,s}),
\label{eq:calib}
\enar

\noindent where in the first line, we have used the fact that excess galaxies are not lensed, and thus have $\Sigma_c^{-1}=0$. In the second line of Eq. \ref{eq:calib}, $\tilde{A}_{\rm annulus}(r_p)$ is the area of the annulus centred on comoving radius $r_p$, corrected for the area coverage of the survey, $D_c$ is the comoving distance to the lens, $j$ are galaxies in the source catalog around each lens and $i$ are galaxies in the calibration set that also have shapes and $P(z)$ information. 

However, galaxies in the source catalog will inevitably have different properties from the spectroscopic calibration sample we are using to construct $\mathcal{C}(r_p)$. We correct for this mismatch by defining a weight function, $\mathcal{W}(m_i)$, that depends on the apparent magnitude of the galaxy and the redshift of the lens. It is given by the ratio between the number of galaxies in the calibration set area with $P(z)$ and valid shapes as a function of magnitude, and the number of galaxies that pass the same criteria and have spectroscopic redshifts,

\beeq
\mathcal{W}(m_i) = \frac{\#\,{\rm Galaxies\,with\,shapes\,and\,photo-}z}{\#\,{\rm Galaxies\,with\,spec-}z{,\rm\,shapes\,and\,photo-}z}.
\label{eq:weight}
\eneq

\noindent Our source catalogue has shapes in two bands: $r$ and $i$. Consequently, we define a weight function separately in each band. We have found there is no dependence of $\mathcal{W}$ on color and we thus do not include color in the definition of $\mathcal{W}$.
%

%----------------------------------------------------------
%NEW SECTION
%----------------------------------------------------------

\section{Non-linear alignment model}
\label{sec:NLA}

The linear alignment model of \citet{Catelan01} relates the $\gamma^{\rm IA}$ component of the ellipticity of a galaxy to the tidal field of the gravitational potential, $\phi_p$, at some primordial redshift, $z_P$, as

\beeq
\gamma^I(\xbf,z) \equiv (\gamma_+^I,\gamma_\times^I) = \frac{C_1}{4\pi G}(\nabla_x^2-\nabla_y^2,2\nabla_x\nabla_y)\phi_p(\xbf)
\label{eq:hirata_gammaI}
\eneq

\noindent where $C_1$ is an unknown constant that quantifies the response of a galactic halo to the tidal field. The $+$ ellipticity is negative if galaxies point towards each other. This model has been successful in describing alignments across $10-100$ Mpc$/h$ \citep{Hirata07,Okumura09,Blazek11} and has been used to determine the level of contamination that can be expected from intrinsic alignments to weak gravitational lensing correlations in future surveys \citep{Hirata04}, and for forecasting future extraction of cosmological information from the intrinsic alignment signal \citep{Chisari13}. A non-linear extension of this model (hereafter the `non-linear alignment model' or NLA model) to smaller scales was suggested by \citet{Hirata07} and \citet{Bridle07}. The NLA model provides a better description of the ellipticity correlations of Luminous Red Galaxies on scales below 10 $h^{-1}$ Mpc than the linear alignment model. 

In the tidal alignment model, the position-intrinsic ellipticity power spectrum is given by \citep{Catelan01,Blazek11}

\beeq
P_{g+}(\kbf,z) = b_L \iacons \frac{k_x^2-k_y^2}{k^2} P_{\delta}(\kbf,z)\mathcal{S}(k),
\label{eq:pgplus}
\eneq

\noindent where $b_L$ is the effective linear bias of the lens sample, the $k_z$-axis is along the line of sight, $G$ is the growth function of matter perturbations, $\rho_{\rm crit}$ is the critical density of the Universe today, $\Omega_{M}$ is the fractional density of matter today and $\mathcal{S}$ is a smoothing filter, which defines the typical scale of a galactic halo. In the NLA model, $P_\delta$ is the non-linear matter power spectrum at redshift $z$. The combination of $b_LC_1$ is unknown a priori and this is what we will constrain from our results. We will usually express these constraints as constraints on $b_LC_1\rho_{\rm crit}$, rather than $b_LC_1$, since the former is a dimensionless quantity.

The NLA model is not meant to capture all the physics that is relevant for determining intrinsic alignments on the scales that we use here; however, we use it primarily because it is likely relevant on the larger scales we probe, and because it conveniently allows us to represent our constraints in terms of a single number. The relevant limitations of the NLA model in the context of this work are as follows. First, on scales beyond the cluster virial radius, the linear alignment model is missing several higher-order terms, and the NLA model only includes one of them while ignoring others at the same order (Blazek et al. 2014, in preparation). Second, our measurements go below the halo virial radius, where the linear alignment model is not at all valid. Although this lacks strict physical motivation, this approximation provides a reasonable description of intrinsic alignments of Luminous Red Galaxies in the non-linear regime \citep{Okumura09}.

Because we are exploring correlations in the non-linear regime, we set the smoothing scale to $10$ kpc$/h$, an order of magnitude smaller than the minimum scale we probe observationally. Defining this filter in Fourier space avoids spurious features due to ringing in the Fourier transform. For the purpose of this work, we do not include the effect of large scale linear peculiar velocities \citep{Kaiser87}, since it is only relevant scales larger than those probed in this work \citep{Blazek11,Chisari13}. Fingers-of-god \citep{Jackson72,Kaiser87} can have an impact on the small scales probed in this work, but we can safely ignore them due to the smoothing induced by the typical photo-$z$ scatter.

We construct the three dimensional correlation between position and ellipticity in real space as a function of projected separation and line-of-sight distance by transforming the power spectrum in Eq. \ref{eq:pgplus},

\beeq
\xi_{g+}(r_p,\Pi,z) = - b_LC_1 \frac{\rho_{\rm crit}\Omega_{M}}{2\pi^2 G(z)}
\int_0^\infty dk_z
\int_0^\infty d\kappa
\frac{\kappa^3}{\kappa^2+k_z^2} P_\delta({\kbf},z) \mathcal{S}(\kbf) \cos(k_z\Pi) J_2(\kappa r_p)
\label{eq:correl}
\eneq

\noindent where $\kappa^2=k_x^2+k_y^2$  and $J_2$ is the second spherical Bessel function of the first kind. We relate $\xi_{g+}(r_p,\Pi,z)$ to the estimator of the correlation function $gI^a$ constructed in Section \ref{sec:formalism}.

\subsection{Correlation function estimator}

We build an estimator for the annular average of the correlation function that combines the positions of lenses, $g$, with the ellipticities of the sources, $I$, as

\beeq
gI(r_p) = \frac{\sum^{\rm lens}_{r_p} \gamma_+}{\sum^{\rm random}_{r_p} 1}
\label{eq:wgp0}
\eneq

\noindent in bins of projected radius, $r_p$. We are assuming the distances to galaxies are known, and we will incorporate the $P(z)$ in the next section. 
Since sources are not clustered with the random points, the probability of finding a random lens in a volume $dV_L$ separated by $(r_p',\Pi)$ at $z$ from a source which is in volume $dV_I$ is

\beeq
\sum^{\rm random}_{dV_I,dV_L} 1 = n_L n_I dV_L dV_I 
\label{eq:rd}
\eneq

\noindent where $n_L$ and $n_I$ are the comoving number density of lenses and sources, respectively. The excess probability of finding a lens-source pair, $gI$, separated by $(r_p',\Pi)$ at $z$, where the source galaxy has ellipticity $I$, is

\beeq
\sum^{\rm lens}_{dV_I,dV_L} \gamma_+ = n_L n_I  \xi_{gI}(r_p',\Pi,z) dV_L dV_I
\label{eq:dd}
\eneq

\noindent where $r_p'$ is projected comoving radius and $\Pi$ is the comoving distance along the line of sight. Both are integration variables for $dV_I$.

The predicted correlation function, Eq. \ref{eq:wgp0}, can then be estimated as

\beeq
gI(r_p) = \frac{\int dV_L \int dV_I n_L n_I \xi_{g+}(r_p',\Pi,z_L) }{\int dV_L \int dV_I n_L n_I}
\label{eq:gIaxi}
\eneq

\noindent where $\xi_{g+}$ is given by Eq. \ref{eq:correl}. The integral over $V_L$ runs over the volume of the survey and the integral over $V_I$ runs over a hollow cylinder around the lens of inner radius $r_p-\Delta r$ and outer radius $r_p+\Delta r$ and length $2L_{\rm coh}$, the coherence length of $\xi_{g+}$. The coherence length represents the scale over which the correlation is significant. Typically, a coherence length of $L_{\rm coh}= 80$ Mpc$/h$ is used for intrinsic alignment studies using spectroscopic galaxies \citep{Blazek11}. If the length of the cylinder is much smaller than the redshift separation of the lenses \citep{Limber53}, $\Pi$ and $z_L$ are independent variables, resulting in the following expression for the estimator,

\beeq
gI^a(r_p) = \frac{\int dz \frac{dV_c}{dz} n_L(z) n_I(z)  \int_{r_p-\Delta r}^{r_{p}+\Delta r} dr_p' r_p' 
\int_{-L_{\rm coh}}^{L_{\rm coh}} d\Pi \xi_{g+}(r_p',\Pi,z) }
{\int dz \frac{dV_c}{dz} n_L(z) n_I(z)  [(r_p+\Delta r)^2-(r_p-\Delta r)^2] L_{\rm coh}}
\label{eq:whatspec}
\eneq

Following \citet{Bartelmann01}, we assume that position-alignment correlations have significant power only over the coherence scale. We define $\Pi_{I}$ and $\Pi_L$ as the comoving distance along the line-of-sight at $z_I$ and $z_L$, respectively. The correlation function of position-alignment is only relevant when $|\Pi_I-\Pi_L|<L_{\rm coh}$. Because $L_{\rm coh}$ represents a distance that is small compared to the change in redshift needed to see a cosmological evolution of $\delta$, $\xi_{g+}$ is evaluated at $z_L$ in Eq. \ref{eq:whatspec}. The final assumption of the Limber approximation is that the weight functions, in this case, $n_L(z_L)$ and $n_I(z_I)$, do not vary appreciably on scales $|\Pi_I-\Pi_L|<L_{\rm coh}$. \footnote{If the $(r_p,\Pi)$ bin is small, the resulting redshift weighting of Eq. \ref{eq:whatspec} can be written as

\beeq
W_z(z) = \frac{p_L(z)p_I(z)}{D_c^2(z)\frac{dD_c}{dz}(z)}\left[ \int dz \frac{p_L(z)p_I(z)}{D_c^2(z)\frac{dD_c}{dz}(z)} \right]^{-1},
\label{eq:wweight}
\eneq

\noindent where $p_L(z)\equiv N_L^{-1} dN_L/dz$ is the redshift distribution of the lenses and $p_I(z)$ is the redshift distribution of the sources, both normalized to unity. This constitutes a generalization of the weighting derived in Eq. (A7) of \citet{Mandelbaum11} from the case where lenses and sources are the same population to the case where they are different populations.}

%----------------------------------------------------------------------------------------------------
%NEW SUBSECTION--------------------------------------------------------------------------------------
%----------------------------------------------------------------------------------------------------

\subsection{Modeling of photometric redshifts}

In practice, we know the redshifts of the lenses to good accuracy, but there is significant scatter in the photometric redshifts of the sources. To take this scatter into account, we define $P(z_p|z_I)dz_p$ as the probability that an object with spectroscopic redshift $z_I$ in the calibration set has a photometric redshift within $dz_p$ of $z_p$. This function is a two dimensional matrix of $z_I$ and $z_p$ that is constructed by adding the photometric redshift posteriors of galaxies in a given spectroscopic redshift interval centred on $z_I$. The matrix can be computed from the calibration set assuming that it is representative of the full sample of source galaxies. 

We can extend Eq. \ref{eq:whatspec} to this case by considering the probability that a galaxy $I$ with true redshift $z_I$ is placed outside the `coherence volume' due to its photometric redshift. We construct a second function, $\tilde{P}(z_I,z_L)$, that yields the probability that an object with $z_I$ lies within a given $z_p$ interval, weighted by $\twia$,

\beeq
\tilde{P}(z_I,z_L)=\int_{z_L-\Delta z}^{z_L + \Delta z} dz_p P(z_p|z_I) \twia .
\label{eq:ptilde}
\eneq

\noindent While \citet{Blazek12} define $P(z_p|z_I)$ by counting galaxies in bins of $z_I$ and $z_p$, we add the photometric redshift posteriors in each bin:

\beeq
\tilde{P}(z_I,z_L)=\sum_{z_I-\Delta z_I<z_{I,j}<z_I+\Delta z_I}\mathcal{W}(m_j) \int_{z_L-\Delta z}^{z_L + \Delta z} dz_p P_j(z_p) \twia,
\eneq

\noindent where we have incorporated the weight function defined for the calibration factor in Section \ref{sec:formalism} to match the calibration set to the set of galaxies with shapes and photo-$z$.
The probabilities computed in Eq. \ref{eq:rd} and Eq. \ref{eq:dd} have to be accordingly modified to take into account the uncertainty in the source redshift, hence,

\bear
n_L n_I dV_L dV_I&\rightarrow &n_L(z_L) n_I(z_I) \tilde{P}(z_I,z_L) dV_L dV_I\nonumber\\
n_L n_I \xi_{g+}(r_p,\Pi,z) dV_L dV_I& \rightarrow & n_L(z_L) n_I(z_I) \xi_{g+}(r_p,\Pi_I,z_L) \tilde{P}(z_I,z_L) dV_L dV_I
\enar

When the photo-$z$ uncertainty described by $\tilde{P}$ is taken into account, Eq. \ref{eq:gIaxi} is modified to

\beeq
gI^a(r_p) = \frac{\int dV_L \int dV_I n_L(z_L) n_I(z_I) \tilde{P}(z_I,z_L) \xi_{g+}(r_p',\Pi_I,z) }{\int dV_L \int dV_I n_L n_I \tilde{P}(z_I,z_L)}.
\label{eq:whatphot}
\eneq

Can we still use the Limber approximation in the presence of photo-$z$ scatter represented by $\tilde{P}(z_p|z_I)$? We need to consider two cases. First, galaxies with $|\Pi_I-\Pi_L|<L_{\rm coh}$ are within the assumptions of the Limber approximation. Their contribution to the projected correlation is modulated by $\tilde{P}(z_I,z_L)\simeq \tilde{P}(z_L,z_L)$. In the second case, galaxies with $|\Pi_I-\Pi_L|>L_{\rm coh}$ are outside of the interval where we can apply the Limber approximation. Another interpretation of this result is that the coherence scale has to be recalculated due to the photo-$z$ scatter. When photometric redshifts are used, the effective coherence length, $L_{\rm coh}'$, is based on $\xi_{g+}(r_p,\Pi(z_I),z_L)\tilde{P}(z_I,z_L)$.  However, we know that $\xi_{g+}\simeq 0$ for these galaxies, so they will not contribute to the numerator of Eq. \ref{eq:whatphot} even if $\tilde{P}(z_I,z_L)$ is large. In other words, once we have integrated over the $\Pi$ range that adds all the information in the correlation function, there is no need to integrate further, even if the photo-$z$ scatter is large. This statement holds assuming that the photometric redshifts are not significantly biased with respect to the spectroscopic redshifts of the calibration set. We conclude that we can apply the Limber approximation also in the photo-$z$ case, provided that $L_{\rm coh}$ and $L_{\rm coh}'$ do not differ significantly.

The Limber approximation applied to Eq. \ref{eq:whatphot} is 

\beeq
gI^a(r_p) = \frac{2\pi \int dz \frac{dz}{dV_c} \frac{dN_L}{dz}(z) \frac{dN_I}{dz}(z) \tilde{P}(z,z) \int_{r_p-\Delta r}^{r_{p}+\Delta r} dr_p' r_p' 
\int_{-L_{\rm coh}}^{L_{\rm coh}} d\Pi\,\xi_{g+}(r_p',\Pi,z) }
{\int dV_L \int dV_I n_L n_I \tilde{P}(z_I,z_L)}.
\label{eq:modelgp_limber}
\eneq

In Eq. \ref{eq:modelgp_limber}, the integral over $\Pi$ can be computed analytically. If the Limber approximation were applied, we would be miscounting the galaxies used to compute the intrinsic alignment signal and we would be underestimating the denominator. The denominator can be obtained by summing over the galaxies in the calibration set and the lens sample,

\beeq
\int dV_L \int dV_I n_L n_I \tilde{P}(z_I,z_L)  = \pi [(r_p+\Delta r)^2-(r_p-\Delta r)^2] \int dV_L n_L(z_L) D_c(z_L)  \int dz_I n_I(z_I) \tilde{P}(z_I,z_L) 
\eneq

The NLA model template predicted by Eq. \ref{eq:modelgp_limber} is computed entirely from the calibration set, similarly to the calibration factor of Eq.\ref{eq:calib}. To assess the uncertainty due to the finite sample of galaxies in the calibration set, we bootstrap over those galaxies in Eq. \ref{eq:modelgp_limber}.

%---------------------------------------------------------------------------------------------------------------------------
%NEW SECTION--------------------------------------------------------------------------------------------
%---------------------------------------------------------------------------------------------------------------------------
\section{Results}
\label{sec:results}

In this section, we present the results of applying the intrinsic alignment estimator of Section \ref{sec:formalism} to the group and cluster catalog in Stripe 82 described in Section \ref{sec:clusters} and the shape catalogue of \citet{Huff11}.

We choose the value of $\Delta z$, defined in Section \ref{sec:formalism} as the width of bin $(a)$, such that the intrinsic alignment contamination in bin $(b)$ can be safely neglected. We compute the position correlation of cluster centres and galaxies with shapes in bin $(b)$ for three possible choices of $\Delta z$: $0.1$ (the typical $1\sigma$ scatter in the photo-$z$ calibration), $0.15$ and $0.2$. When $\Delta z=0.2$ is used, the clustering is consistent with $0$ within the $1\sigma$ uncertainty (from Poisson error in number counts in radial bins). We thus adopt this value for the rest of this section.

%=========================
%NEW SUBSECTION
%=========================
\subsection{Calibration factor}
\label{subsec:boostcalib}

%Calibration factor
To compute the calibration factor of Eq. \ref{eq:calib}, $\mathcal{C}(r_p)$, we identify the galaxies in the PRIMUS+DEEP2 set of spectroscopic redshifts that also have shape information in our source catalog. There are $4,558$ and $4,738$ galaxies with spectroscopic redshifts and shapes in the $i$-band and $r$-band shape catalogs, respectively. To construct this set, we do not restrict ourselves to $m_r<23.3$ because the PRIMUS+DEEP2 set already has the $m_R<23.3$ cut in place; the fraction of galaxies with $m_r>23.3$ and $m_R<23.3$ is negligible.

The distributions of these subsets of galaxies are, however, biased in redshift and magnitude with respect to the full catalogues of galaxies with shapes that are used to reconstruct the intrinsic alignment signal. To correct for this bias, we need to assign a weighting, $\mathcal{W}$ in Eq. \ref{eq:weight}, to galaxies with shapes and spectroscopic redshifts relative to the galaxies with shapes in the PRIMUS+DEEP2 area (restricting to the same area minimizes the impact of cosmic variance on the determination of the weights). We have verified that the weighting of Eq. \ref{eq:weight} is a function of apparent magnitude but almost independent of the color of the galaxies. This function is very similar to that of Figure \ref{fig:rweight}. It takes values from $1.5$ to $5$ for increasing $m_r$. In $r$-band, the function is flatter by definition, since the PRIMUS+DEEP2 matched set was resampled using $r$-band number counts in Section \ref{sss:tset}. 

To estimate the uncertainty in the calibration factor, we bootstrap over the galaxies in the calibration set, drawing $200$ samples with replacement from the galaxies of the spectroscopic calibration set in $i$-band. We do the same for galaxies with shapes in $r$-band. We use the bootstrap to estimate $\mathcal{C}/\tilde{A}_{\rm annulus}(r_p)$ and its uncertainty, which is independent of $r_p$. We obtain
$\mathcal{C}^{(a)}/\tilde{A}_{\rm annulus}^{(a)}=22.85\pm0.38$,
$\mathcal{C}^{(b)}/\tilde{A}_{\rm annulus}^{(b)}=76.62\pm0.87$ and 
$\mathcal{C}^{(a)}/\tilde{A}_{\rm annulus}^{(a)}/(\mathcal{C}^{(b)}/\tilde{A}_{\rm annulus}^{(b)})=0.2982\pm0.0073$ for $i$-band, and
$\mathcal{C}^{(a)}/\tilde{A}_{\rm annulus}^{(a)}=22.92\pm0.37$,
$\mathcal{C}^{(b)}/\tilde{A}_{\rm annulus}^{(b)}=79.24\pm0.79$ and 
$\mathcal{C}^{(a)}/\tilde{A}_{\rm annulus}^{(a)}/(\mathcal{C}^{(b)}/\tilde{A}_{\rm annulus}^{(b)})=0.2892\pm0.0068$ in $r$-band.
The ratio of the calibration factor at the lens redshift and behind it represents the percentage of the total shear behind the lens that leaks into the intrinsic aligment total shear due to photometric redshift scatter. The estimated uncertainties do not account for cosmic variance due to the small area of the calibration set. In other words, we assume that the galaxies with shapes in the calibration area are a representative subset of galaxies with shapes across all of Stripe 82.

%=========================
%NEW SUBSECTION
%=========================
\subsection{Intrinsic alignments}
\label{subsec:intrinsic}

In this section, we present the results of applying our formalism described in Section \ref{sec:formalism} to obtain constraints on the intrinsic alignment of galaxies in and around clusters given by the correlation function defined in Eq. \ref{eq:wfinal2}. 

In what follows, we fit a power-law to the measured correlation of the form

\beeq
gI^a = A_{g+}\left(\frac{r_p}{1\,h^{-1}{\rm Mpc}}\right)^{\alpha_{g+}}
\eneq

\noindent Here, $A_{g+}$ is a dimensionless constant. Note that, as opposed to Eq. \ref{eq:modelgp_limber}, the power-law fit includes the effect of photometric redshift dilution in $gI^a$. Therefore, the power-law fit is not intended to provide an alternative to the NLA model, but it is only used in this work to assess the significance of the detection. 

We compute the $\Delta\chi^2=\chi^2(A_{g+},\alpha_{g+})-\chi^2({\rm best\,fit})$ on a grid of $(A_{g+},\alpha_{g+})$ spanning $-4<\alpha<2$ and using the bootstrap covariance matrix. This gives us an assessment of the significance of the difference in goodness of fit for each pair of parameters, compared to the best fit value. A detailed description of this method can be found in \citet{Mandelbaum06}. Notice that because the covariance matrix is noisy, the standard $\chi^2$ distribution is not applicable here. We simulate the $\chi^2$ and $\Delta\chi^2$ distributions through Monte Carlo simulations following \citet{Hirata04} for $30$ bootstrap areas in the sky, $10$ radial bins and $2$ free parameters (the amplitude and index of the power-law). 

To assess the significance of the detection, we compute $\Delta\chi^2 = \chi^2(A_{g+}=0)-\chi^2({\rm best\,fit})$, the difference between $\chi^2$ with no signal and the best fitting power-law. This is equivalent to performing a likelihood ratio test where $p(<\Delta\chi^2)$ gives the confidence level at which the results exclude the null hypothesis. In general, when we want to assess the significance of the detection, we compute $\chi^2$ for a null vector in $30$ regions and $10$ radial bins, which requires building a $\chi^2$ distribution that is different from the ones used in the previous two cases. 

Alternatively, we fit the NLA model described in Section \ref{sec:NLA} to the observed $gI^a$ and set constraints on the amplitude of the alignments of galaxies around clusters by constraining the combination of $b_LC_1\rho_{\rm crit}$ (Eq. \ref{eq:correl}). In this case, we need to simulate $\chi^2$ and $\Delta \chi^2$ distributions with one free parameter rather than two. 
 
\subsubsection{Error bar estimation}
\label{subsubsec:errorbars}

% Shape noise
There are three ways of estimating the error bars in the intrinsic alignment measurement. First, we consider shape noise only and derive the relevant expressions analytically, as described in Appendix A. The advantage of this method is the simplicity of the computation, but it does not take into account off-diagonal elements of the covariance matrix of the data or any spurious shear power due to systematics. 
% Jackknife		
Second, we divide the cluster sample in $30$ subareas and estimate the uncertainties in the lensing and intrinsic alignment signal using the jackknife. This method computes the variance of the correlation functions by removing one subarea at a time from the estimator. This takes into account variations across the Stripe that affect the quality of the shape measurements. However, it does not take into account fluctuations in the number of lenses across the Stripe caused by the change in depth.
% Bootstrap
Third, we compute the error bars by constructing $10^4$ bootstrap random realisations of Stripe 82, divided into $30$ fixed subareas with equal number of lenses in each. We have found that with $10^4$ bootstrap random realisations our results converge. The subareas are chosen such that there is an equal number of clusters in each subarea. 
%Comparison
The bootstrap and the jackknife procedures have a similar requirement for the size of the areas used to be large compared to the typical scales over which we expect to measure the alignment signal, to make the areas approximately independent. Similarly, for both we need to define a number of regions that is large compared to the number of data points \citep{Hirata04}.  For the bootstrap, there is an additional requirement over the the number of resampled datasets needed for convergence, which is not an issue for the jackknife.
% Overall
None of the methods account for Poisson noise due to the overall finite number of groups and clusters on the Stripe, which increases the error bars of the average shear behind the lenses by $\sim 2.2$ per cent. Both the bootstrap and the jackknife errors include correlated shape noise, the shape noise induced from summing over the same sources multiple times around different lenses, whereas the first method does not.

We measure all correlations up to a separation of $10$ $h^{-1}$Mpc from the centre of the lens. With our choice of binning in RA and DEC, the area subtended by a circle of $10$ $h^{-1}$Mpc radius at the median redshift of the cluster corresponds to $\sim 1/3$ of the area of a patch. Thus, the areas can approximately be considered independent. 

We compare our estimated errors as a function of scale in Figure \ref{fig:errors}. There are $10$ $r_p$ bins chosen to have lower boundaries starting at $0.1$ $h^{-1}$Mpc and reaching a maximum separation of $10$ $h^{-1}$Mpc. The bootstrap and jackknife estimates show good agreement. Moreover, there is significant covariance between the radial bins, as seen in the covariance matrix derived from the bootstrap technique in Figure \ref{fig:covarclusters}, with increasing importance on large scales. The shape noise analytic estimate of Appendix A gives a comparable result to the square root of the diagonal elements of the covariance matrix, but does not make a prediction for the off-diagonal terms.

%%%%%%%%%%%%%%%%%%%%%%%%%%%%%%%%%%%%%%%%%%%%%%%%%%%%%%%%%%%%%%%%%%
%covar_boot_clusters.pro
\bef
\centering
\subfigure[Diagonal elements of the covariance matrix of $gI^a(r_p)$ obtained by three different methods.]{
\includegraphics[width=0.45\textwidth]{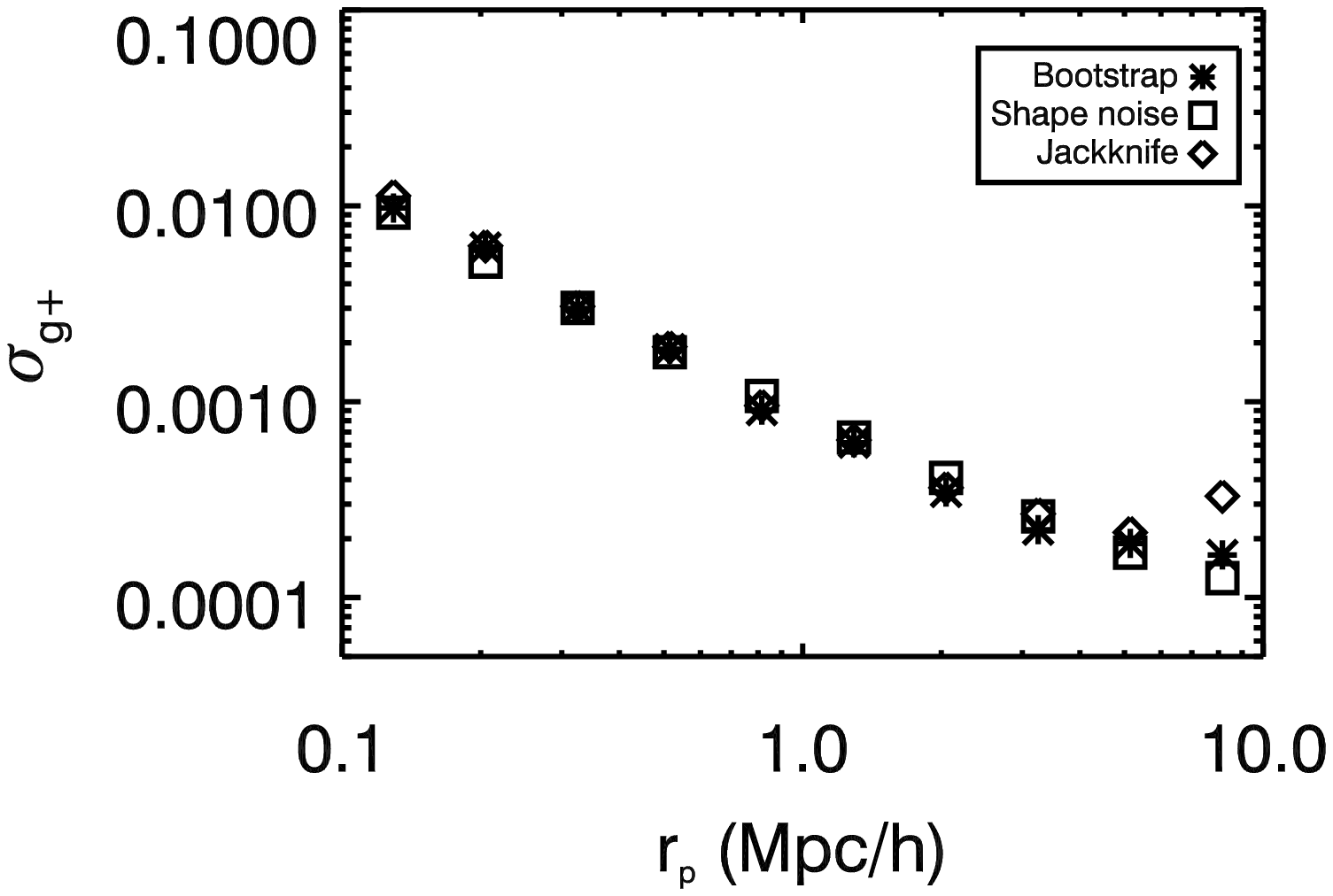}
}\hfill
\subfigure[Comparison of bootstrap and jackknife error bars to shape noise only calculation. In the last bin, the jackknife estimate is a factor of $\simeq 3$ larger than the shape noise and it is not shown in the scale of this figure.]{
\includegraphics[width=0.45\textwidth]{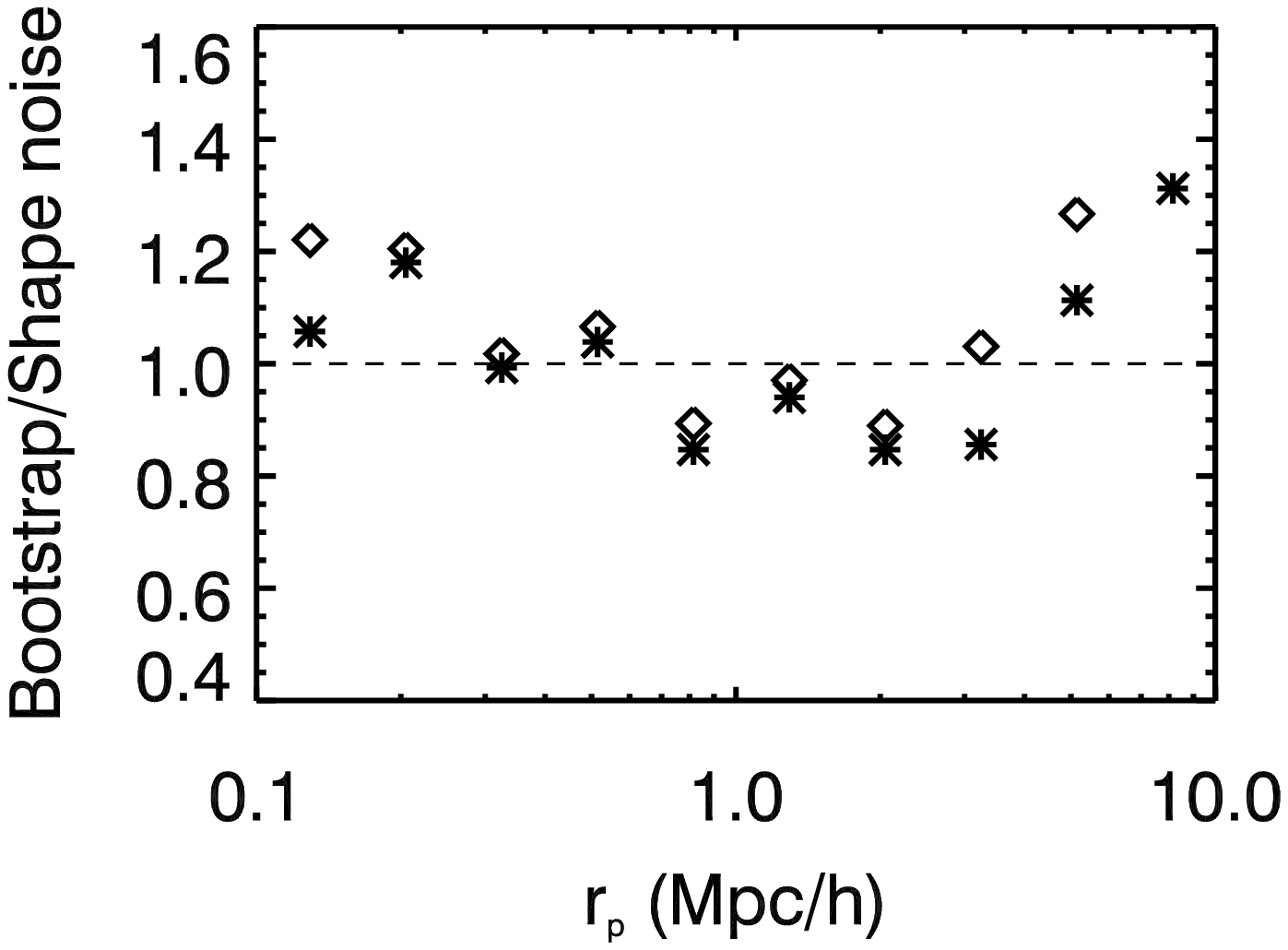}
}
\caption{Error bars ($1\sigma$) on $gI^a(r_p)$ obtained analytically (squares, Appendix A)
compared to the estimation by the bootstrap method (stars) and jackknife (diamonds) based on $i$-band shapes. Similar results hold for galaxies with $r$-band shapes (not shown).}
\label{fig:errors}
\enf
%%%%%%%%%%%%%%%%%%%%%%%%%%%%%%%
\bef
\subfigure[$i$-band]{
\includegraphics[width=0.45\textwidth]{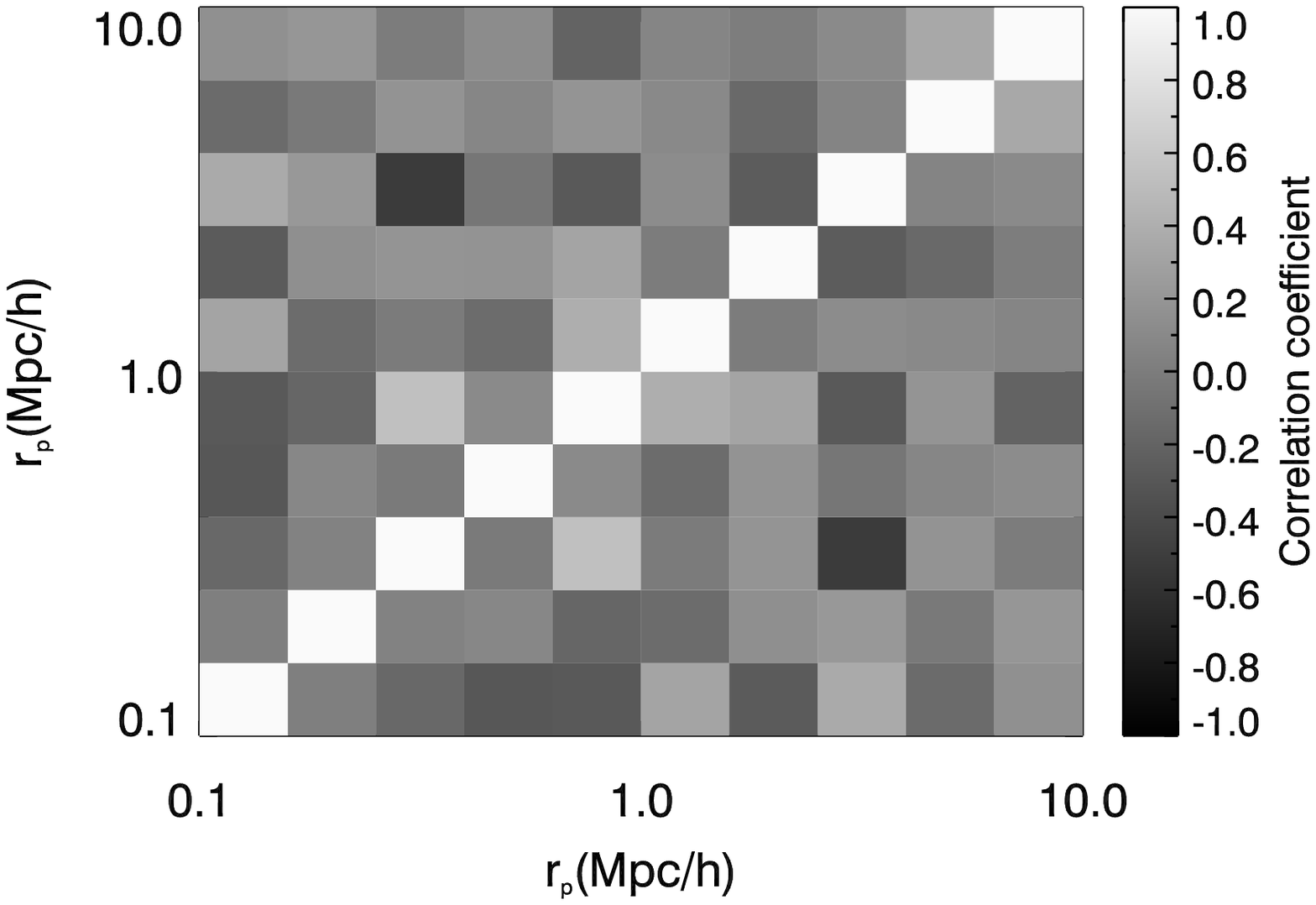}
}\hfill
\subfigure[$r$-band.]{
\includegraphics[width=0.45\textwidth]{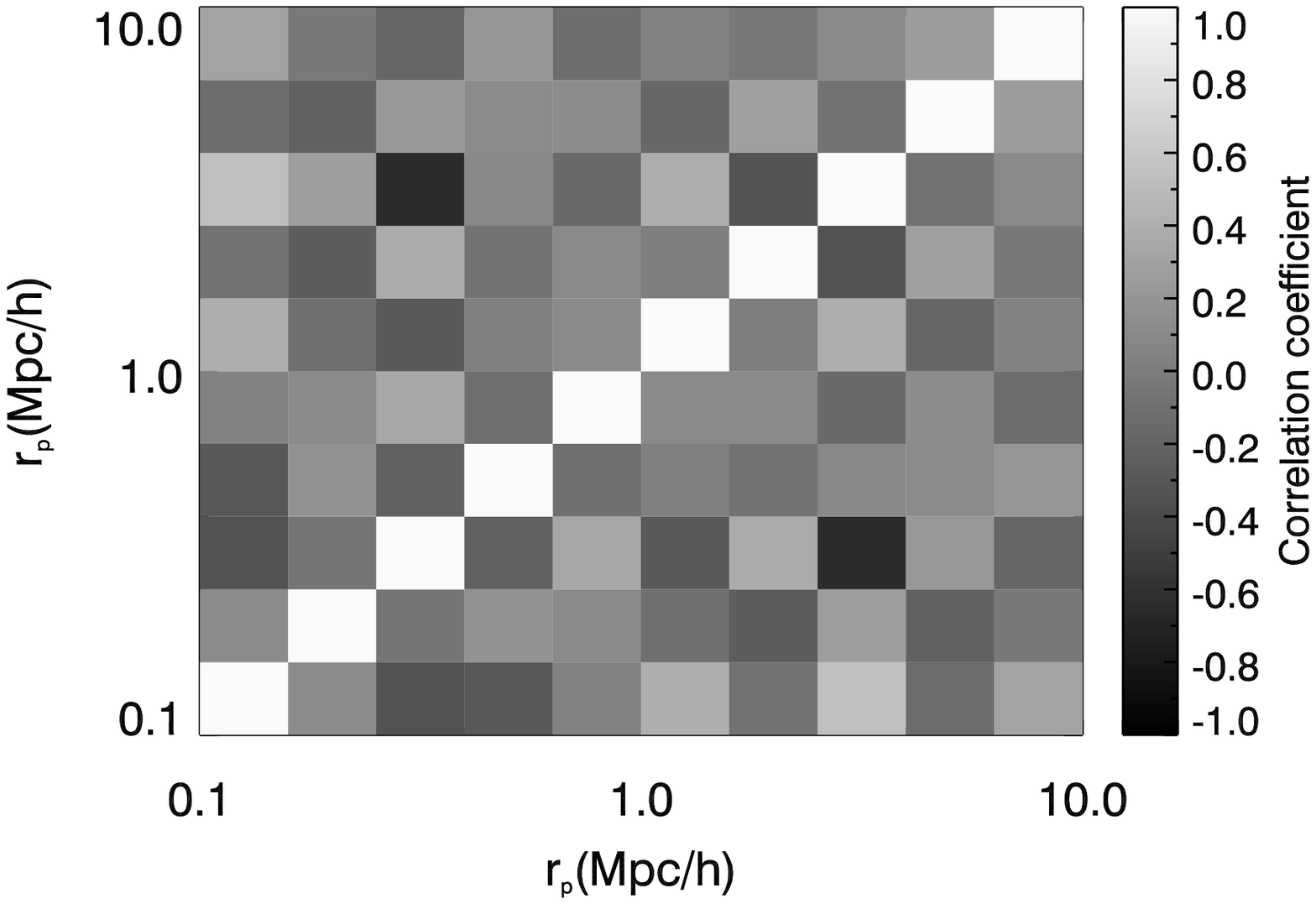}
}
\caption{Correlation coefficient, $r_{ij}=\frac{C_{ij}}{\sqrt{C_{ii}C_{jj}}}$, of the error of the covariance matrix, constructed from the bootstrap for $gI^a(r_p)$.}
\label{fig:covarclusters}
\enf
%%%%%%%%%%%%%%%%%%%%%%%%%%%%%%%%%%%%%%%%%%%%%%%%%%%%%%%%%%%%%%%%%%

%^^^^^^^^^^^^^^^^^^^^^^^^^^^^^^^^
%^^^^^^^^^^^^^^^^^^^^^^^^^^^^^^^^
\begin{table}
 \caption{Best-fitting and $A_{g+}=0$ parameters from $gI^a$ for different scenarios. The constraints on $A_{g+}$ in the second column correspond to leaving $\alpha_{g+}$ unconstrained, which is the reason for not quoting uncertainties in $\alpha_{g+}$ in the third column. Columns 4 and 5 indicate the quality of the best fit. Columns 6 and 7 indicate the parameters of the null fit and finally, column 9 shows the significance level at which the null hypothesis is rejected, with its corresponding $\Delta\chi^2$ shown in column 8.}
 \begin{tabular}{@{}lcccccccc}
  \hline
  Sample & $A_{g+} \times 10^3$ & $\alpha_{g+}$ & $\chi^2$ (best fit) & $p(<\chi^2)$ (best fit) & $\chi^2$ ($A_{g+}=0$) & $p(<\chi^2)$ ($A_{g+}=0$) & $\Delta\chi^2$ & $p(<\Delta\chi^2)$\\
  \hline
$i$-band & $0.06^{+0.88}_{-0.76}$ & $0.44$ & $10.1$ & $0.52$ & $11.4$ & $0.60$ & $1.37$ & $0.32$\\
$r$-band & $0.18^{+0.95}_{-0.86}$ & $0.26$ & $9.03$ & $0.45$ & $16.1$ & $0.80$ & $7.10$ & $0.81$\\
Geometrical centre & $0.2^{+1.0}_{-0.7}$ & $-0.22$ & $8.12$ & $0.38$ & $10.6$ & $0.55$ & $2.51$ & $0.48$\\
Removing edges & $0.12^{+0.83}_{-0.53}$ & $0.32$ & $8.50$ & $0.41$ & $14.3$ & $0.74$ & $5.80$ & $0.75$\\
$z_{\rm ML}$ & $0.3^{+1.2}_{-0.7}$ & $-0.04$ & $16.4$ & $0.81$ & $22.9$ & $0.93$ & $6.56$ & $0.79$\\
$\langle z \rangle$ & $0.0^{+1.1}_{-1.0}$ & $0.50$ & $8.39$ & $0.40$ & $9.53$ & $0.48$ & $1.15$ & $0.28$\\
\hline
\label{tab:powergplus}
\end{tabular}
\end{table}
%^^^^^^^^^^^^^^^^^^^^^^^^^^^^^^^^
%^^^^^^^^^^^^^^^^^^^^^^^^^^^^^^^^
%^^^^^^^^^^^^^^^^^^^^^^^^^^^^^^^^
%^^^^^^^^^^^^^^^^^^^^^^^^^^^^^^^^
\begin{table}
 \caption{Best-fitting NLA model constraints from $gI^a$ for different scenarios. The layout of the columns is the same as in Table \ref{tab:powergplus}.}
 \label{table:NLA}
 \begin{tabular}{@{}lccccccc}
  \hline
  Sample & $b_LC_1\rho_{\rm crit} \times 10^2$ & $\chi^2$ (best fit) & $p(<\chi^2)$ (best fit) & $\chi^2$ ($b_LC_1\rho_{\rm crit}=0$) & $p(<\chi^2)$ ($b_LC_1\rho_{\rm crit}=0$) & $\Delta\chi^2$ & $p(<\Delta\chi^2)$\\
  \hline
$i$-band & $-0.02^{+0.14}_{-0.14}$ &  $11.2$ & $0.47$ & $11.4$ & $0.49$ & $0.24$ & $0.34$\\
$r$-band & $-0.10^{+0.15}_{-0.15}$ &  $11.6$ & $0.49$ & $16.1$ & $0.71$ & $4.5$ & $0.85$\\
Geometrical centre & $-0.08^{+0.16}_{-0.15}$  & $8.5$ & $0.30$ & $10.8$ & $0.45$ & $2.3$ & $0.71$\\
Removing edges & $-0.06^{+0.14}_{-0.14}$ & $12.5$ & $0.55$ & $14.3$ & $0.64$ & $1.8$ & $0.66$\\
\hline
\end{tabular}
\end{table}
%^^^^^^^^^^^^^^^^^^^^^^^^^^^^^^^^
%^^^^^^^^^^^^^^^^^^^^^^^^^^^^^^^^

\subsubsection{Results from the full area in $i$ and $r$-band}
\label{sec:fiducres}

We construct the estimator for $gI^a$ and $\Delta\Sigma$ by summing over lens-source pairs on Stripe 82. Because the shape catalogue covers a smaller area than the group and cluster catalog ($\simeq 60$ per cent in $i$-band and $\simeq 55$ per cent in $r$-band), some clusters can lie on regions of the sky where there are no sources. We establish a criterion to include a cluster as part of the estimator only if the percentage of background sources ($\langle z \rangle > 0.6$) in a circular aperture of comoving radius $5$ Mpc$/h$ is at least $40$ per cent of the average surface density of galaxies with shapes across all of the Stripe 82. In $i$-band, there are $1839$ clusters after masking and there are $1791$ clusters in $r$-band. 

In Figure \ref{fig:iafiduc}, we present our estimate of $gI^a(r_p)$ for the sample of clusters and sources. The error bars correspond to the square root of the diagonal of the covariance matrix derived by the bootstrap procedure. Figure \ref{fig:covarclusters} shows there is significant covariance between projected radius bins.

We summarize the results of modelling the measured intrinsic alignment correlation as a power-law in Table \ref{tab:powergplus}. We show the confidence level contours on the grid of $(A_{g+},\alpha_{g+})$ in Figure \ref{fig:iafiduc} for $i$-band. $A_{g+}$ is consistent with zero at the $60$ per cent confidence level (C.L.). A positive value of $A_{g+}$ could indicate residual lensing (insufficient subtraction of the lensing contamination in bin $(b)$). However, we do not find consistent evidence for this hypothesis in this or other scenarios we analyze throughout this section. To test the impact of the uncertainty in the calibration factor, we increased the lensing term by $5$ per cent. This increase yields very similar results for the $gI$ power-law constraints. 

Overall, we do not detect a significant intrinsic alignment signal for the $i$-band case between $0.1$ Mpc$/h$ and $10$ Mpc$/h$. We repeat this analysis with shapes measured in $r$-band. We obtain a slightly negative trend in the $r$-band shape catalogue, with $A_{g+}$ consistent with zero at $80$ per cent C.L. There is, however, a second minimum of the $p$-value as a function of the fit parameters, as seen in Figure \ref{fig:iafiduc_r}, which corresponds to a positive value of $A_{g+}$.

We perform NLA model fits to the observed correlation function of Figure \ref{fig:iafiduc}. In this case, the presence of intrinsic alignments consistent with the NLA model would be indicated by a significantly positive value of $b_LC_1\rho_{\rm crit}$. The constraints on $b_LC_1\rho_{\rm crit}$ are presented in Table \ref{table:NLA}. In all cases, there is a slight negative trend in the fit parameter $b_LC_1\rho_{\rm crit}$, similarly to the case of power-law fits, rejecting the null hypothesis at the $85$ per cent C.L. at most. 

In Figure \ref{fig:wlfiduc}, we plot the lensing signal of the stacked clusters in the redshift range $z=[0.1,0.4]$. We overlay Navarro-Frenk-White (NFW) surface density profiles \citep[][]{NFW,Bartelmann96} for masses of $10^{13}$, $10^{13.5}$, $10^{14}$ and $10^{14.5}$ M$_\odot$ for reference; these are not intended to be fits to the stacked cluster profile. The mass corresponds to that within a sphere where the density is $200$ times the critical density of the Universe at approximately the median redshift of the clusters.

%%%%%%%%%%%%%%%%%%%%%%%%%%%%%%%%%%%%%%%%%%%%%%%%%%%%%%%%%%%%%%%%%%
\bef
\centering
\subfigure[Intrinsic alignment correlation function in $i$-band.]{
\includegraphics[width=0.45\textwidth]{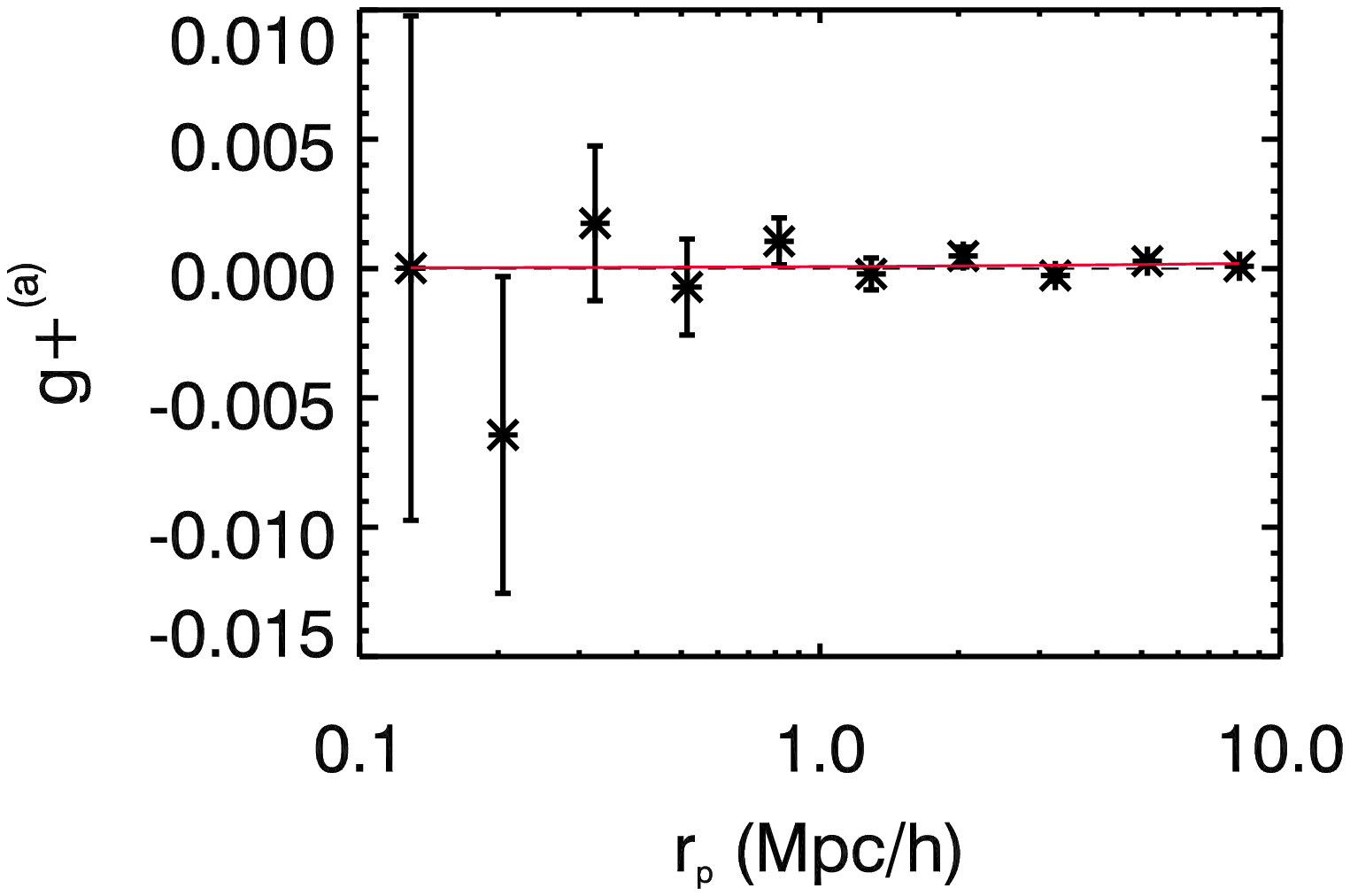}
}\hfill
\subfigure[Constraints on power-law parameters in $i$-band.]{
\includegraphics[width=0.45\textwidth]{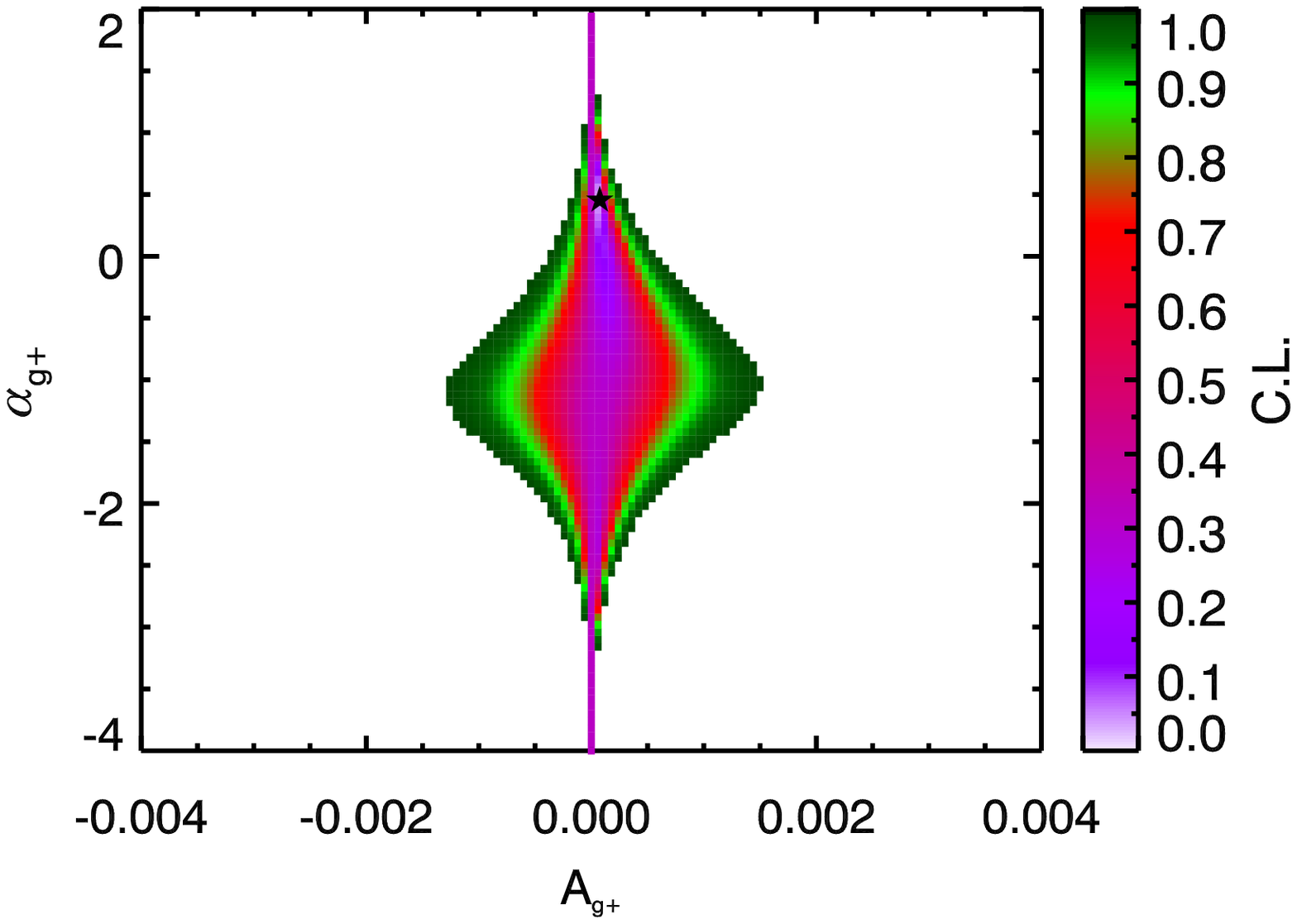}
\label{fig:iafiduc}
}
\subfigure[Intrinsic alignment correlation function in $r$-band.]{
\includegraphics[width=0.45\textwidth]{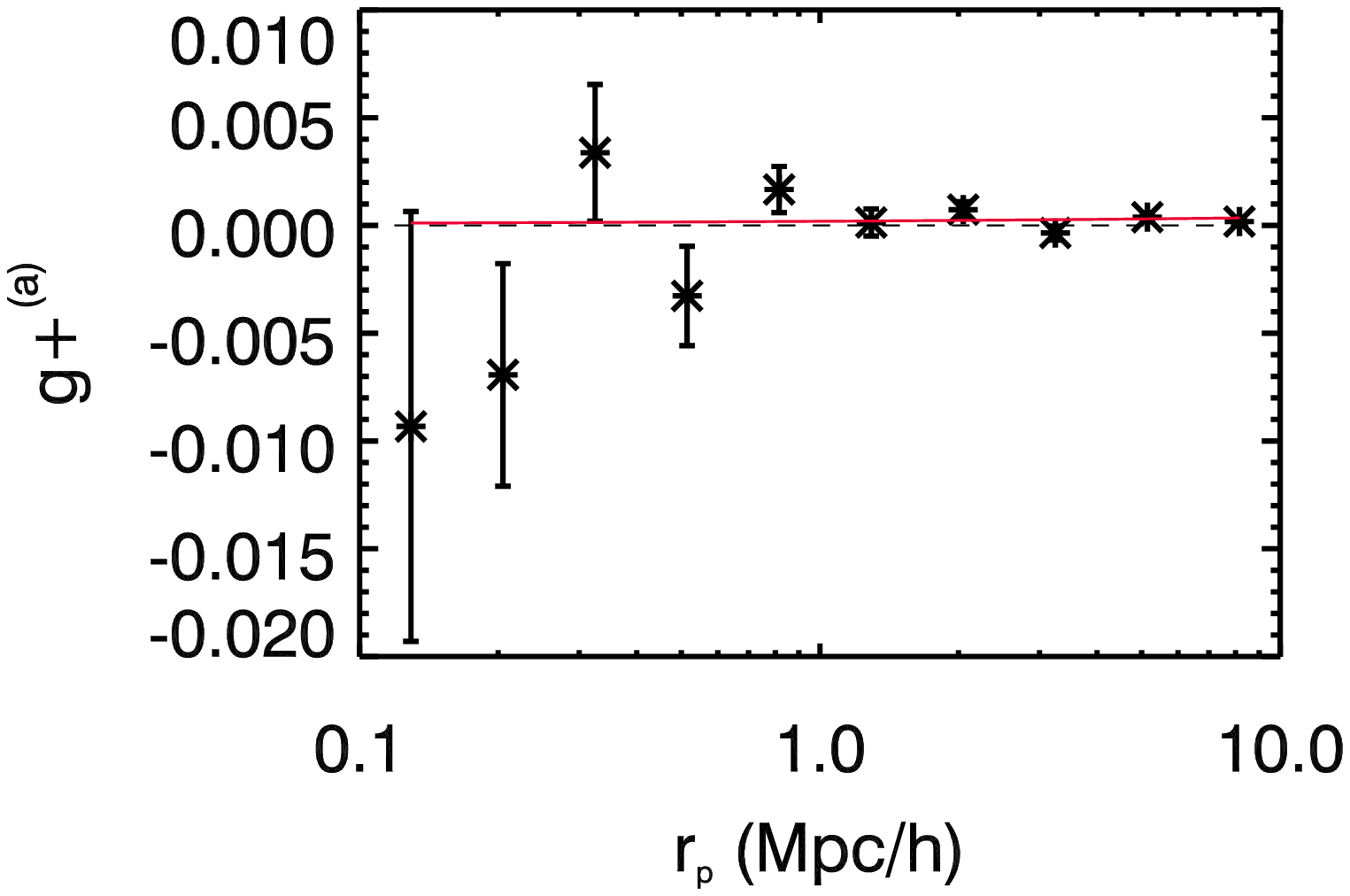}
}\hfill
\subfigure[Constraints on power-law parameters in $r$-band.]{
\includegraphics[width=0.45\textwidth]{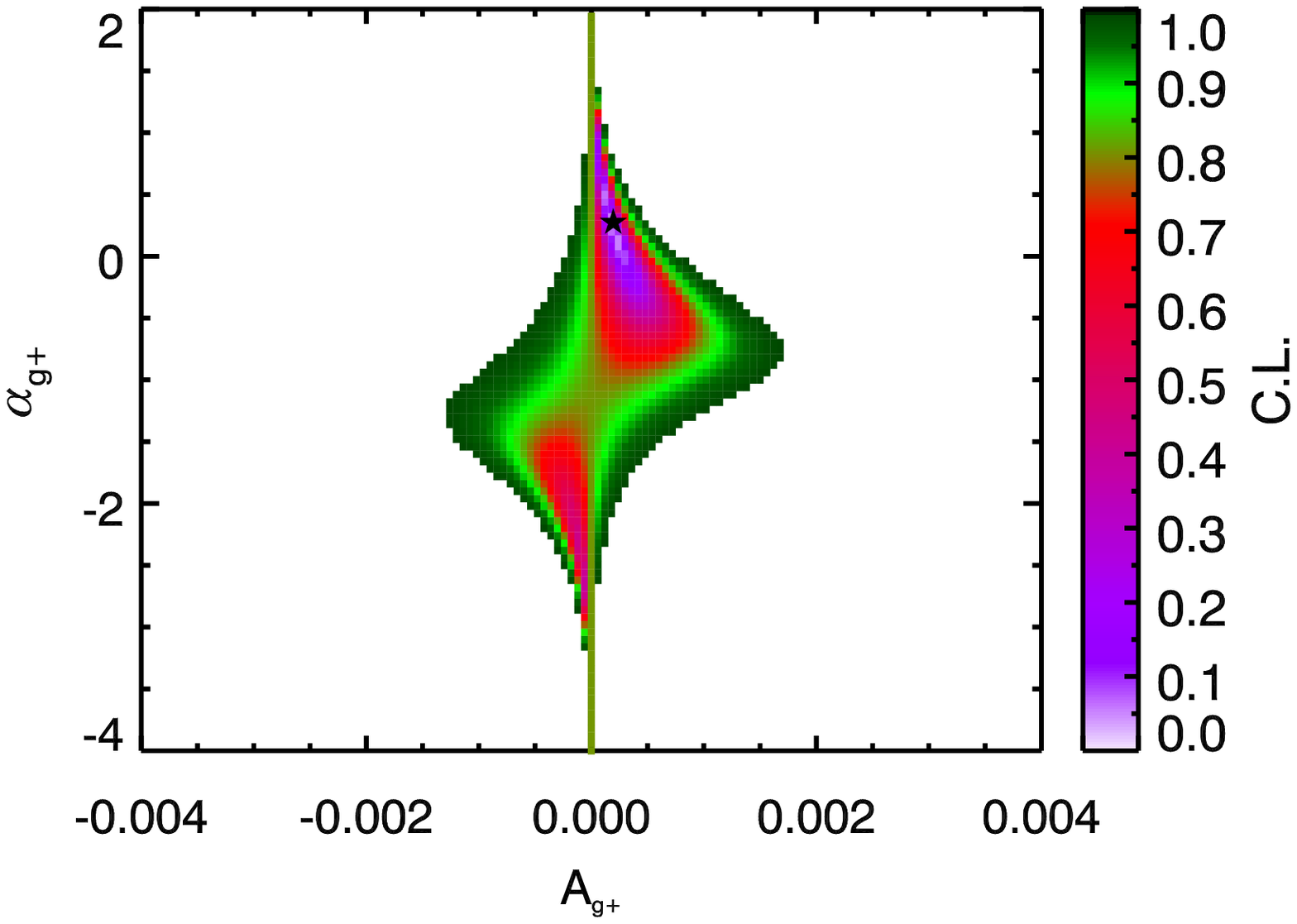}
\label{fig:iafiduc_r}
}
\caption{Intrinsic alignment correlation, $gI^{(a)}$ and constraints on power-law parameters for galaxies with $i$-band and $r$-band shapes around clusters in the range $0.1<z<0.4$.}
\enf
%%%%%%%%%%%%%%%%%%%%%%%%%%%%%%%%%%%%%%%%%%%%%%%%%%%%%%%%%%%%%%%%%%

%%%%%%%%%%%%%%%%%%%%%%%%%%%%%%%%%%%%%%%%%%%%%%%%%%%%%%%%%%%%%%%%%%
\bef
\centering
\subfigure[$i$-band.]{
\includegraphics[width=0.45\textwidth]{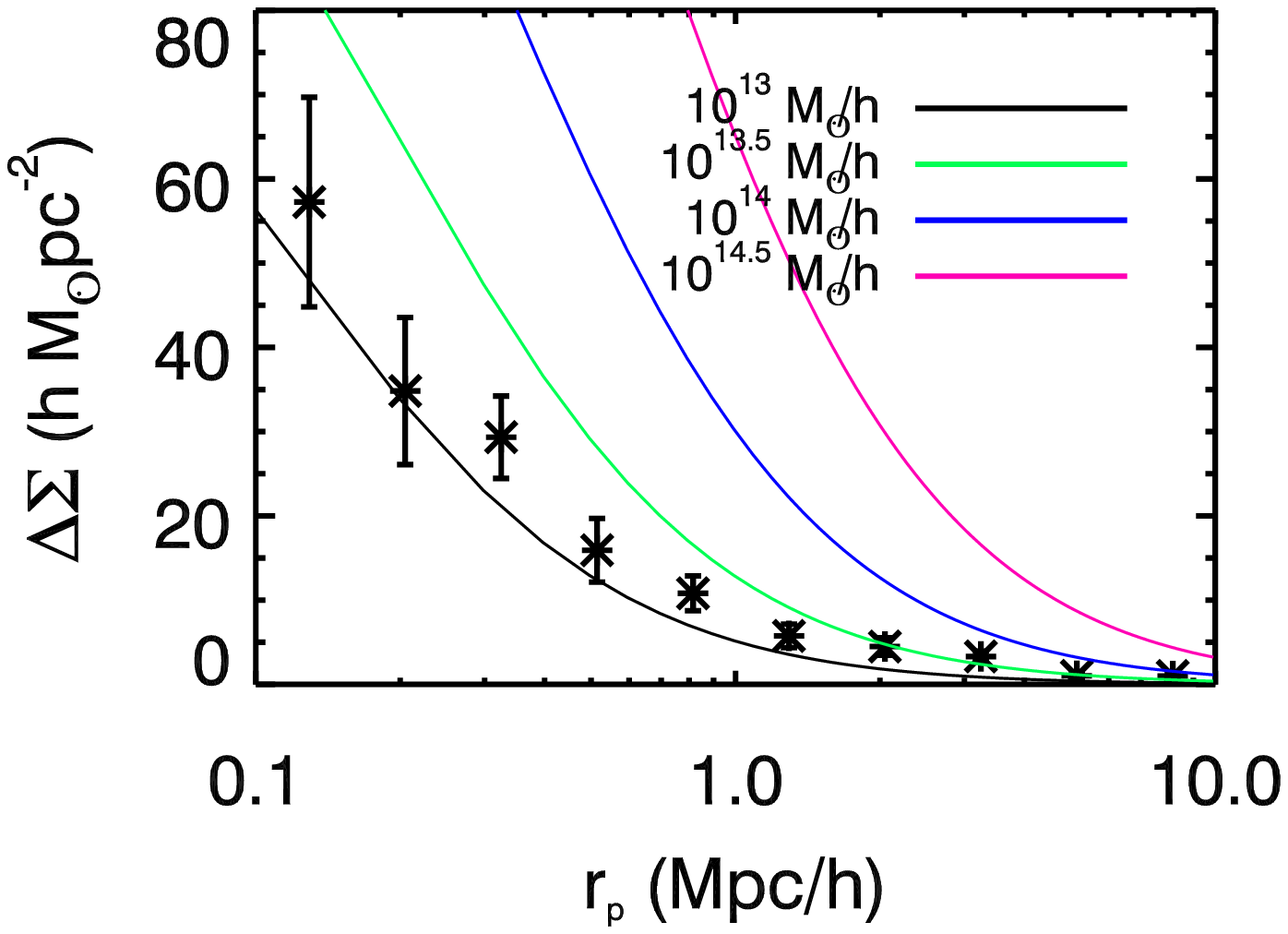}
}\hfill
\subfigure[$r$-band.]{
\includegraphics[width=0.45\textwidth]{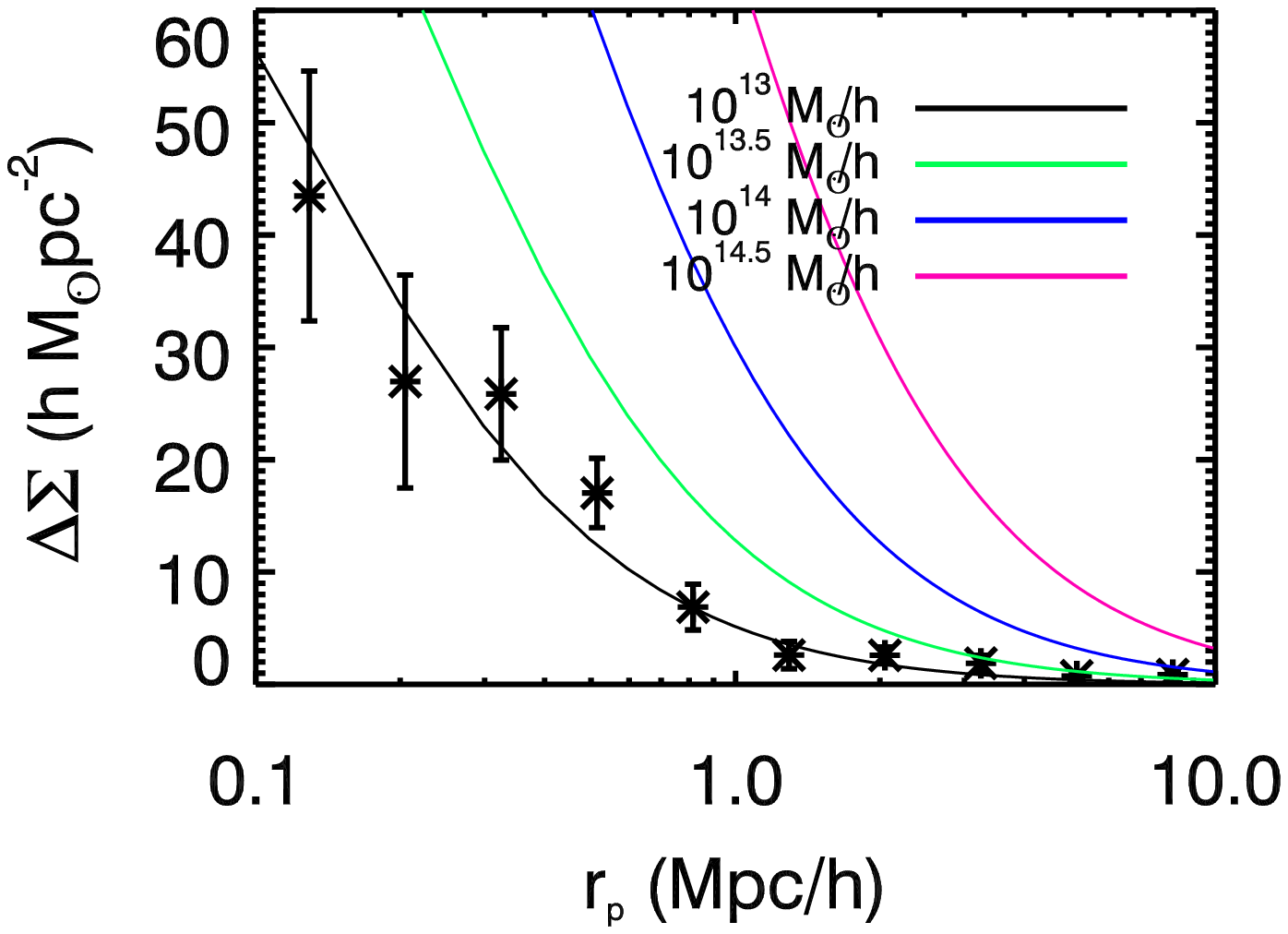}
}
\caption{Projected surface mass density of the stacked clusters, $\Delta\Sigma$, for clusters in $0.1<z<0.4$, as determined by weak lensing.}
\label{fig:wlfiduc}
\enf
%%%%%%%%%%%%%%%%%%%%%%%%%%%%%%%%%%%%%%%%%%%%%%%%%%%%%%%%%%%%%%%%%%

%=========================
%NEW SUBSECTION
%=========================
\subsubsection{Systematics tests}
\label{subsec:systemat}

We quantified the effect of systematics by several methods. The first test we perform is the computation of $g\times^a(r_p)$, which constitutes a null test. Similarly, we construct the surface mass density profile of the clusters in bin $(b)$ using the $\times$ component of the shape. In the absence of systematics, $\Delta\Sigma_{\times}$ should be consistent with zero. We compute the covariance matrices and error bars as in Section \ref{subsubsec:errorbars}. 

%%%%%%%%%%%%%%%%%%%%%%%%%%%%%%%%%%%%%%%%%%%%%%%%%%%%%%%%%%%%%%%%%%
\bef
\centering
\subfigure[$i$-band systematics in bin $(a)$.]{
\includegraphics[width=0.45\textwidth]{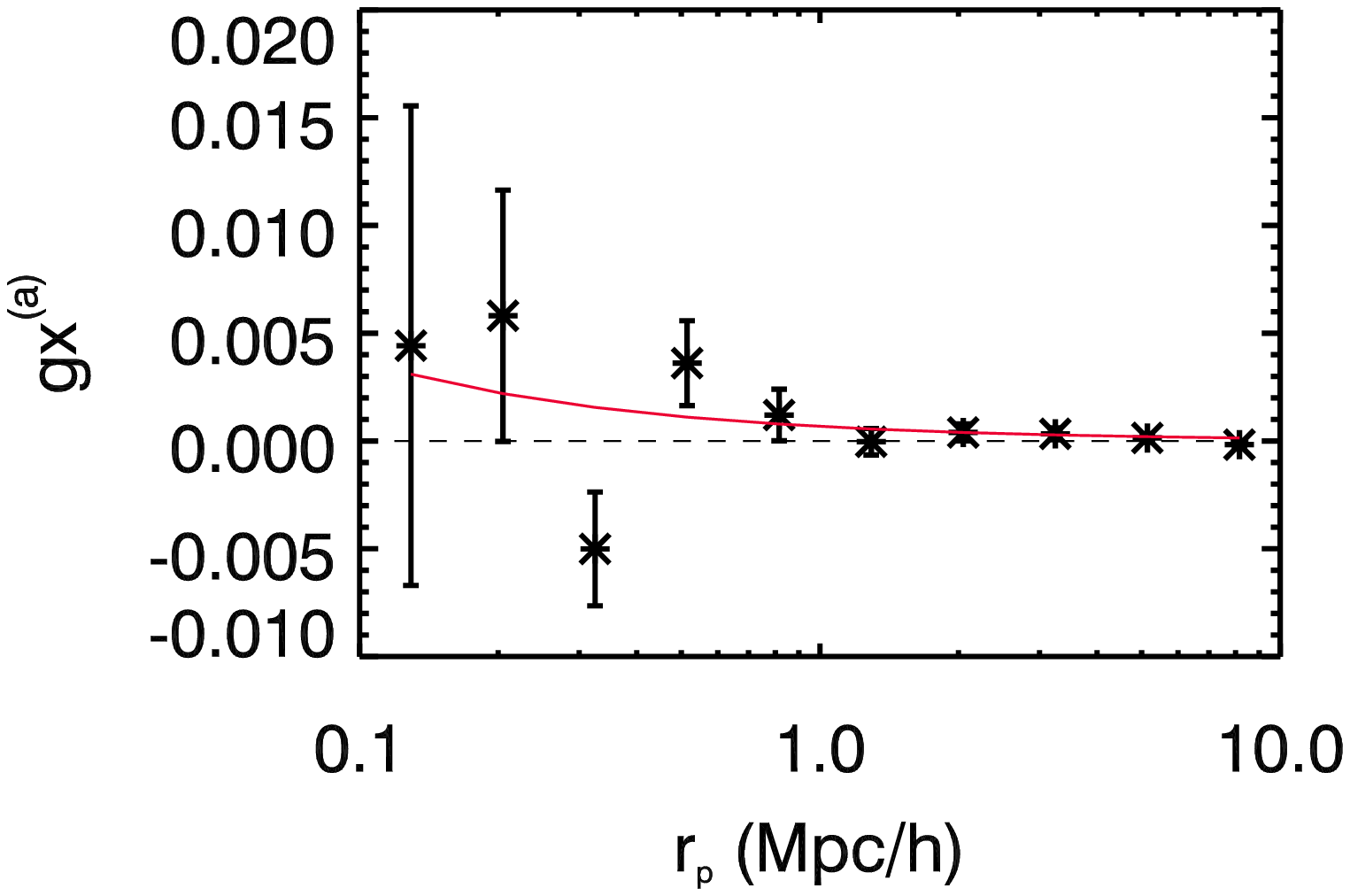}
}\hfill
\subfigure[$r$-band systematics in bin $(a)$.]{
\includegraphics[width=0.45\textwidth]{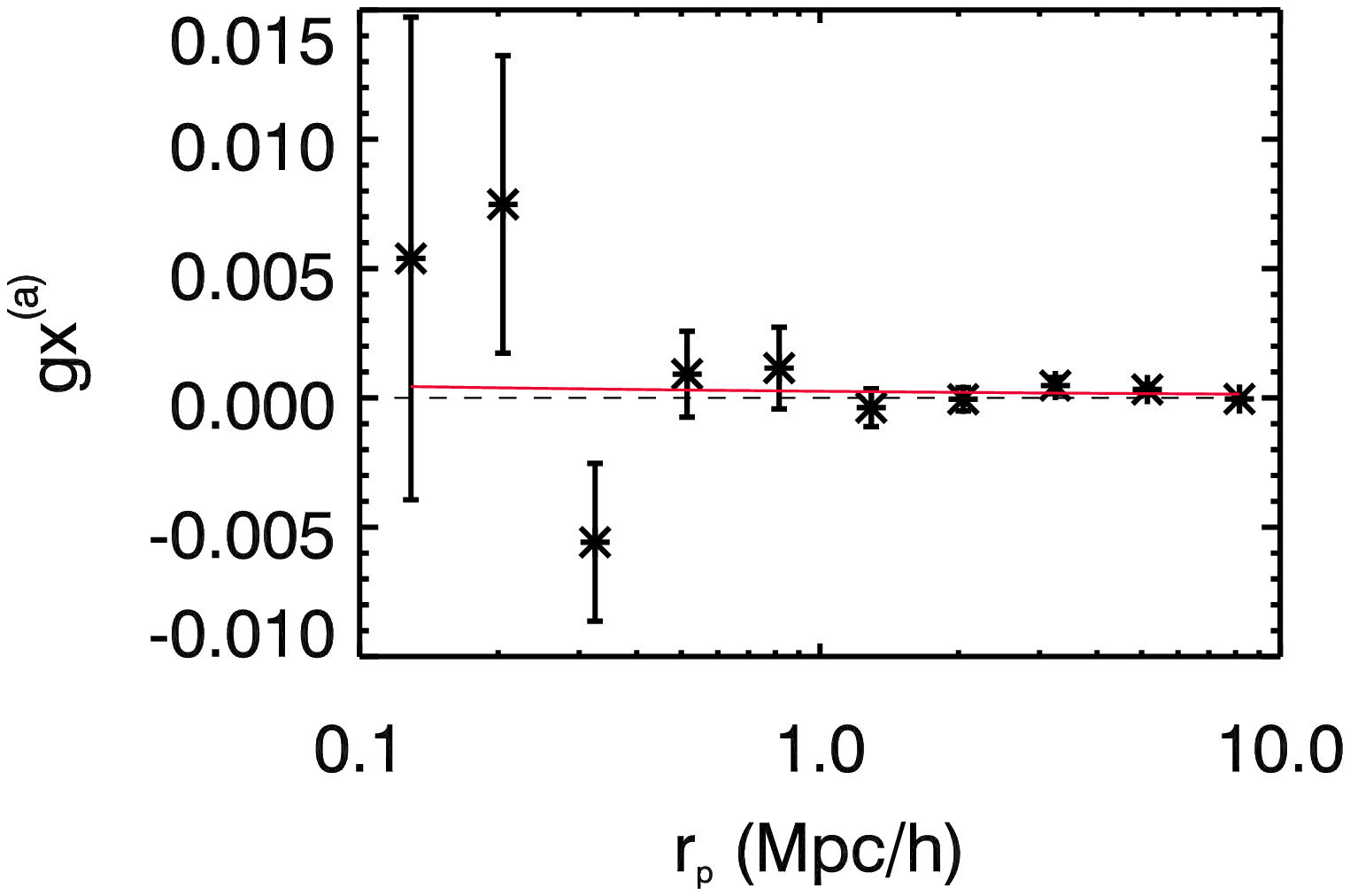}
}
\subfigure[$i$-band systematics in bin $(b)$.]{
\includegraphics[width=0.45\textwidth]{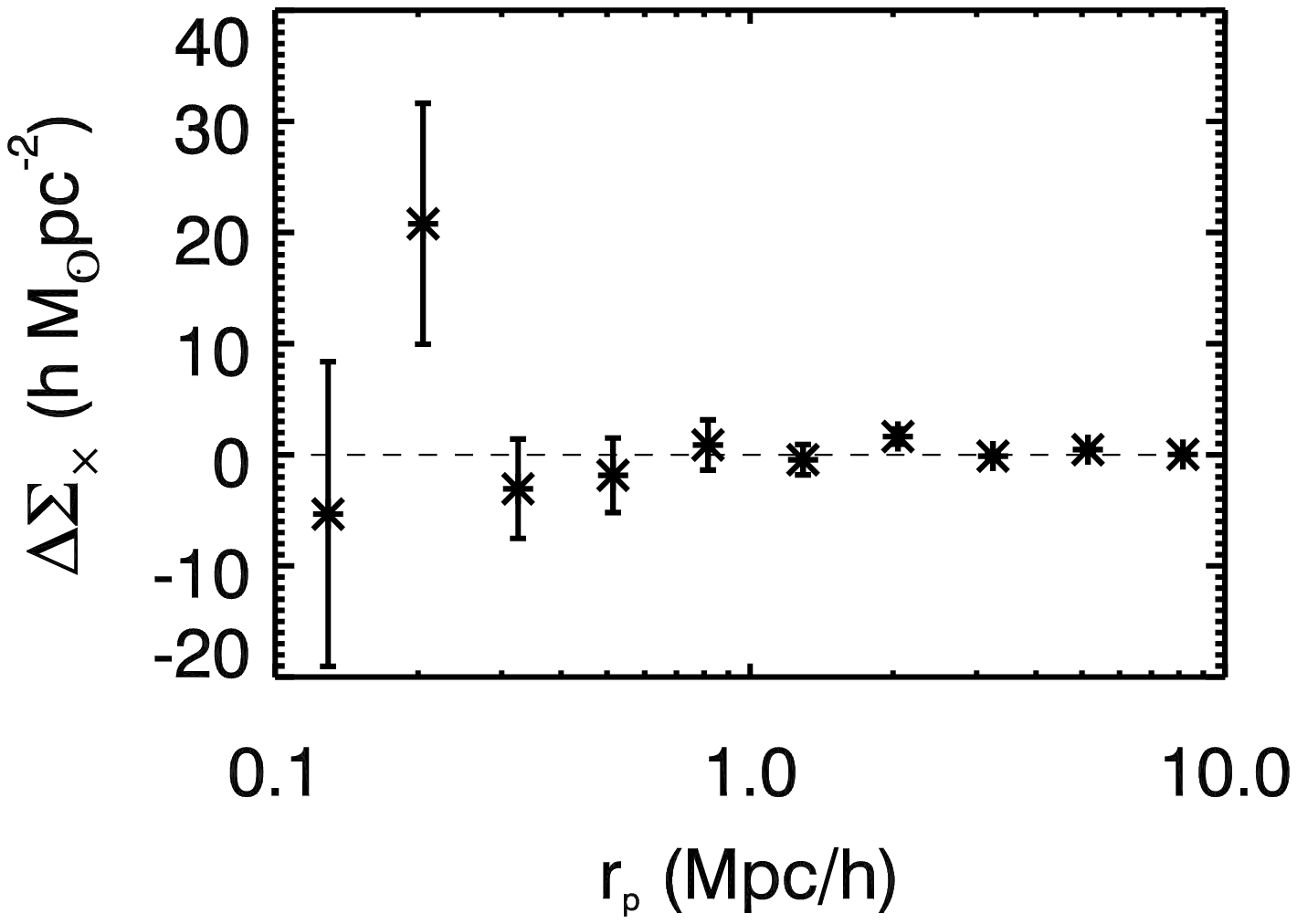}
}\hfill
\subfigure[$r$-band systematics in bin $(b)$.]{
\includegraphics[width=0.45\textwidth]{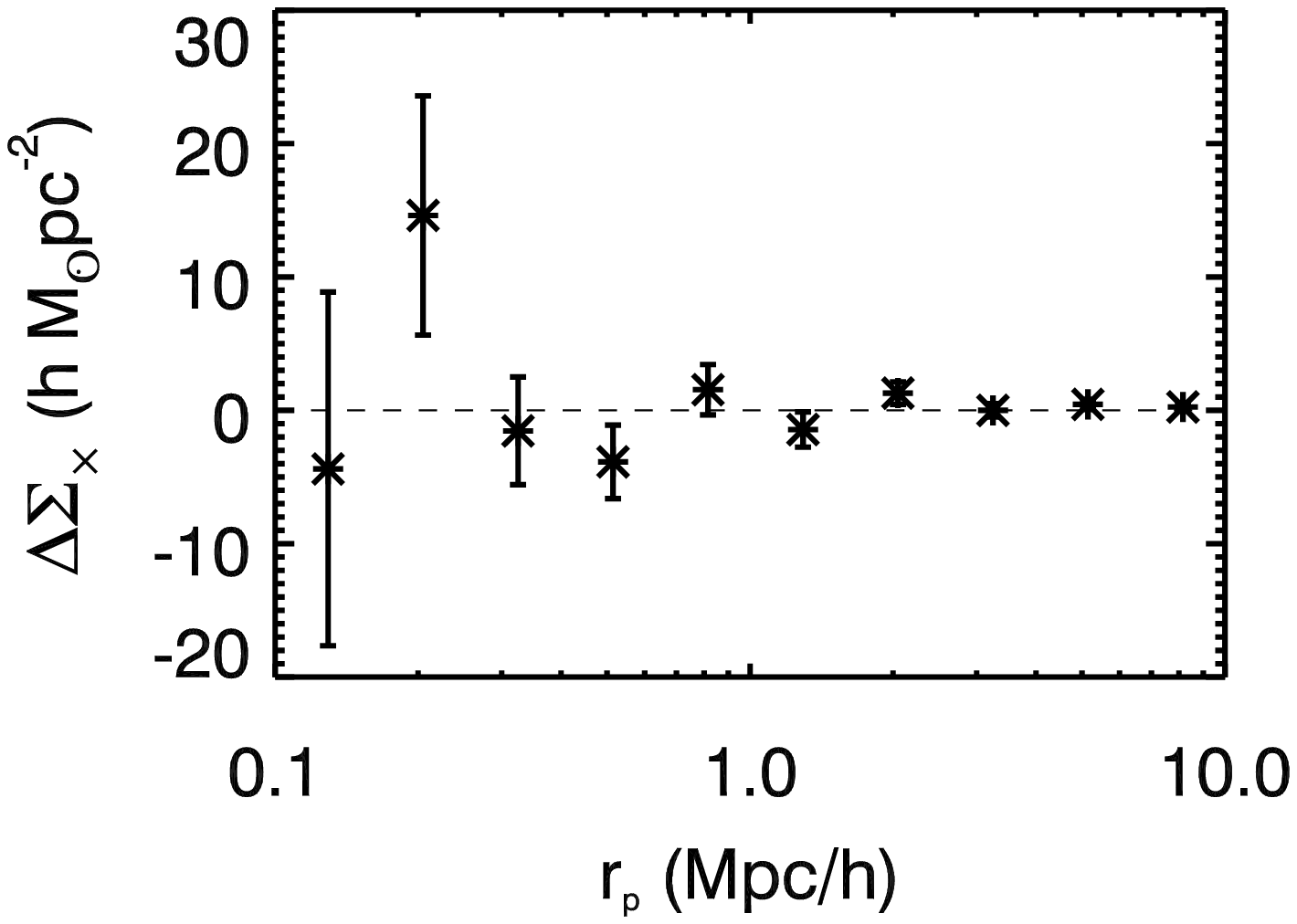}
}
\label{fig:systemat}
\caption{Systematics checks using the $\times$ component of the shapes for both $i$ and $r$-band. The upper panels show the analog of $gI^a$ where the ellipticity is replaced by the curl component (stars) and the best fitting power-law (solid red line). The bottom panels show the stacked cluster profile from the $\times$ component. The left column corresponds to $i$-band shapes and the right column, to $r$-band shapes.}
\enf
%%%%%%%%%%%%%%%%%%%%%%%%%%%%%%%%%%%%%%%%%%%%%%%%%%%%%%%%%%%%%%%%%%

We show the $g\times$ correlation in $(a)$ for galaxies with shapes in $i$ and $r$ in the top two panels of Figure \ref{fig:systemat} for both bands. The error bars correspond to the square root of the diagonal of the covariance matrices. We quote in Table \ref{tab:confgcross} the confidence level at which we can reject the null hypothesis when fitting $g\times(r_p)$ in bin $(a)$. 
For our $i$-band measurement, the null hypothesis is rejected at $51$ per cent for $\Delta\Sigma_{\times}$. For $r$-band, the rejection is at $37$ per cent. For $g\times$, shown in the same figure, we obtained rejection levels at $97$ per cent for $i$-band, and $92$ per cent for $r$-band. The significance is higher in $i$-band because masking bias can lead to spurious signals in the projected correlations \citep{Huff11}. We further checked the significance of the $\times$ component measurement by constructing aperture mass maps across the Stripe by integrating the total shear in a grid in circles of $10$ Mpc$/h$ centred in square bins of $\sim 0.2$ deg$^2$ along $({\rm RA},{\rm DEC})$. The aperture mass gives the radially filtered total shear within a circular aperture of angular radius $\theta_c$; it is defined as

\beeq
M_{ap}^{(+,\times)}(\theta_c) = \int_0^{\theta_c} d^2\theta Q_{Map}(\theta)\,\gamma_{(+,\times)}({\bf \theta}),
\label{eq:massap}
\eneq

\noindent where $Q_{Map}$ is a smooth filter function \citep{vanWaerbeke98} given by

\beeq
Q_{Map}(\theta) = \frac{4\theta^2}{\theta_c^4} \exp\left(-\frac{4\theta^2}{\theta_c^4} \right).
\label{eq:qmap}
\eneq

There is no evidence for biases in the aperture mass for the $+$ or $\times$ component of the shear, and the result is robust to doubling the number of bins along RA and DEC and to integrating the shear within circles of $1$ Mpc$/h$ radius. 

Finally, we considered removing the bins in $({\rm RA},{\rm DEC})$ that lie along the borders of the Stripe, where the optics of the SDSS camera give a somewhat worse image quality. We repeated our analysis of alignments, lensing and systematics including only clusters and sources that lie within the range $-1\deg<{\rm DEC}<1\deg$. In this case, we find the amplitude of the alignment signal to be slightly less consistent with zero than when using the full footprint; the confidence level at which we reject the null hypothesis for a power-law fit of $gI^a$ changes from $32$ per cent to $75$ per cent when we remove clusters and sources along the edges. The significance of the systematics in $\Delta\Sigma_\times$ and $g\times^a$ goes down from $51$ and $97$ per cent to $7$ and $76$ per cent, respectively.

%=========================
%NEW SUBSECTION
%=========================

\subsubsection{Centring}
\label{subsec:centre}

In this Section, we explore the effect of choosing the geometrical centre of the cluster to retrieve the intrinsic alignment and the lensing signal. \citet{George12} have performed an analysis of the impact of miscentring on weak lensing observables. They showed that the mean position of the member galaxies is a poor tracer of the halo centre, due to typical large statistical uncertainties ($50-150$ kpc) in the position of the centre in their sample. However, the group and cluster catalog we use in this work represents more massive overdensities than the $X$-ray group catalogue used in \citet{George12}, and we expect the uncertainties in the centring determined by the mean position of the members to be correspondingly smaller. There is a median offset of $\sim 100$ $h^{-1}$kpc between the BCG and the geometrical centre of the groups and clusters in the catalogue and $90$ per cent of the offsets are below $\simeq 200$ $h^{-1}$kpc. 

The analysis presented thus far assumes the BCG is at the cluster centre. We now consider the geometrical centre of the cluster and run our pipeline in this case as well. In Figure \ref{fig:gcenia} we show the constraints on the intrinsic alignment signal in this case. While there is a decrease in the measured average shear in the first radial bin, the lensing signal both in the $i$-band measurement centred on the BCG and on the geometrical centre (Figure \ref{fig:gcenlens}) are detected at $>99.999$ per cent C.L. In the geometrical centre case, the intrinsic alignment signal is inconsistent with zero at the $48$ per cent C.L. 

%%%%%%%%%%%%%%%%%%%%%%%%%%%%%%%%%%%%%%%%%%%%%%%%%%%%%%%%%%%%%%%%%%
%covar_boot_clusters.pro
\bef
\includegraphics[width=0.45\textwidth]{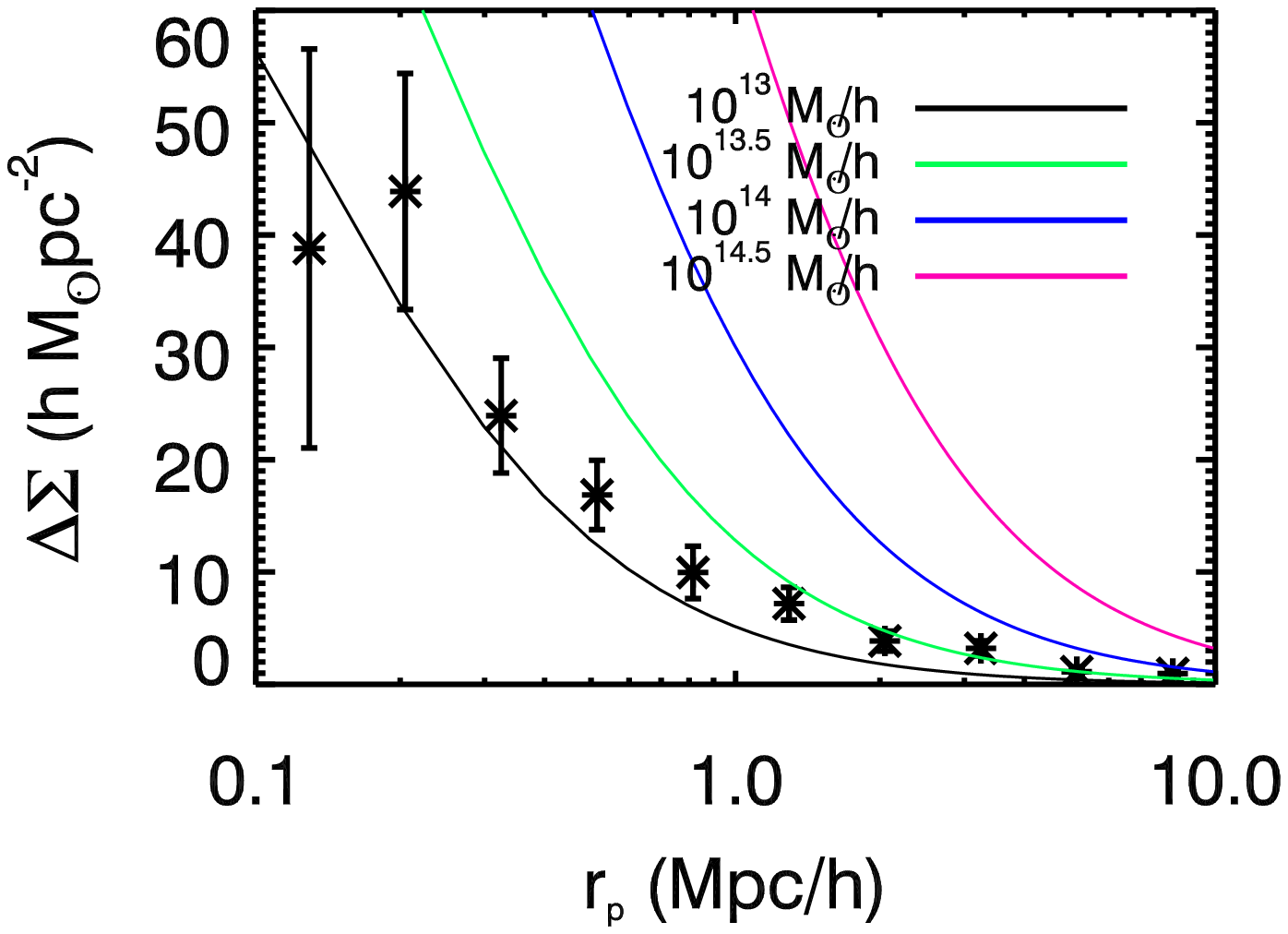}
\label{fig:gcenlens}
\caption{Projected surface mass density of the stacked clusters, $\Delta\Sigma$, for clusters in $0.1<z<0.4$ and $i$-band shapes when the geometrical centre is chosen as a proxy for the true location of the cluster centre.}
\enf
\bef
\centering
\subfigure[Intrinsic alignment correlation function, $gI^a$, for clusters in $0.1<z<0.4$.]{
\includegraphics[width=0.45\textwidth]{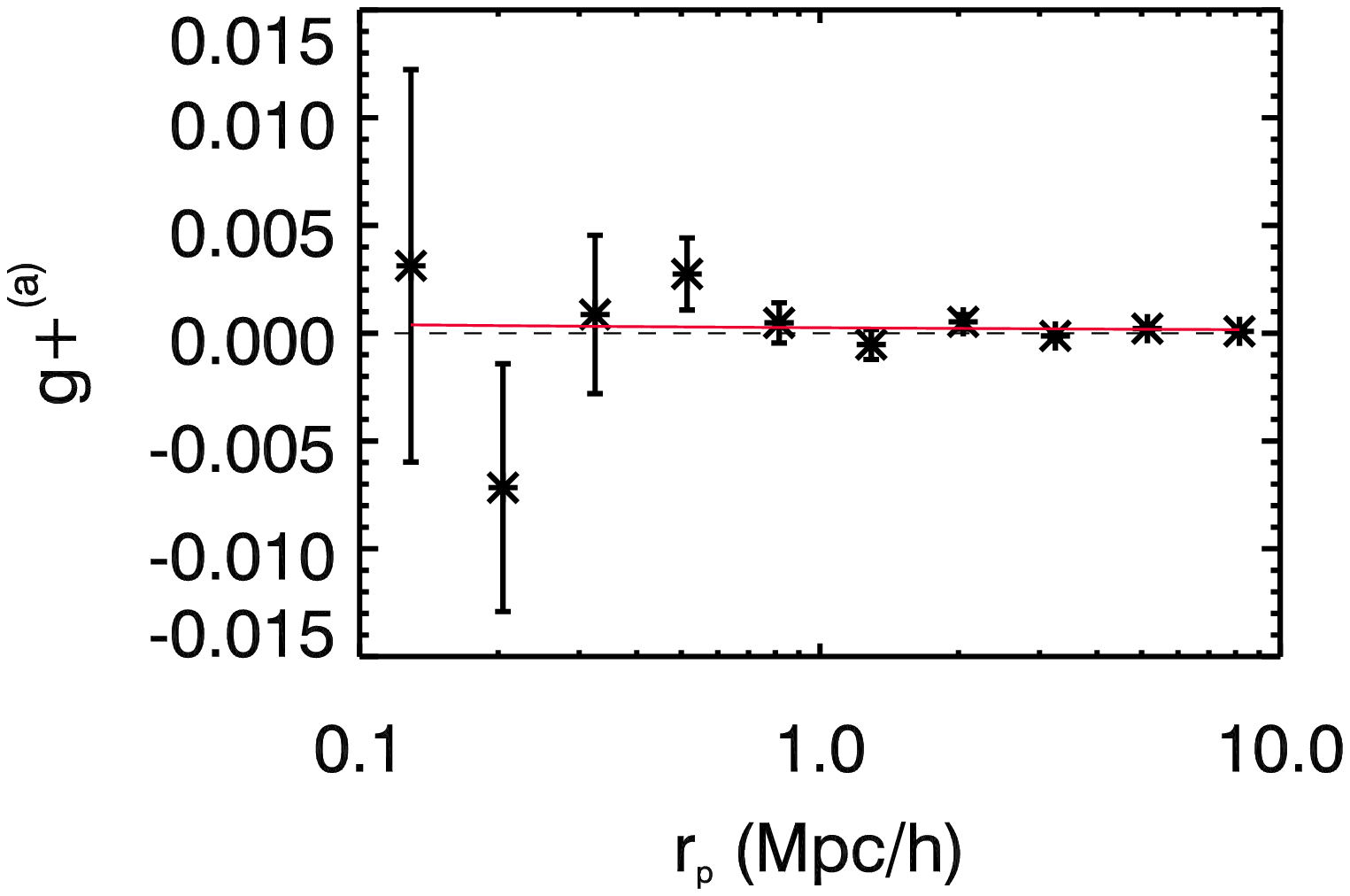}
\label{fig:gcenia}
}\hfill
\subfigure[Constraints on parameters of the power-law fits.]{
\includegraphics[width=0.45\textwidth]{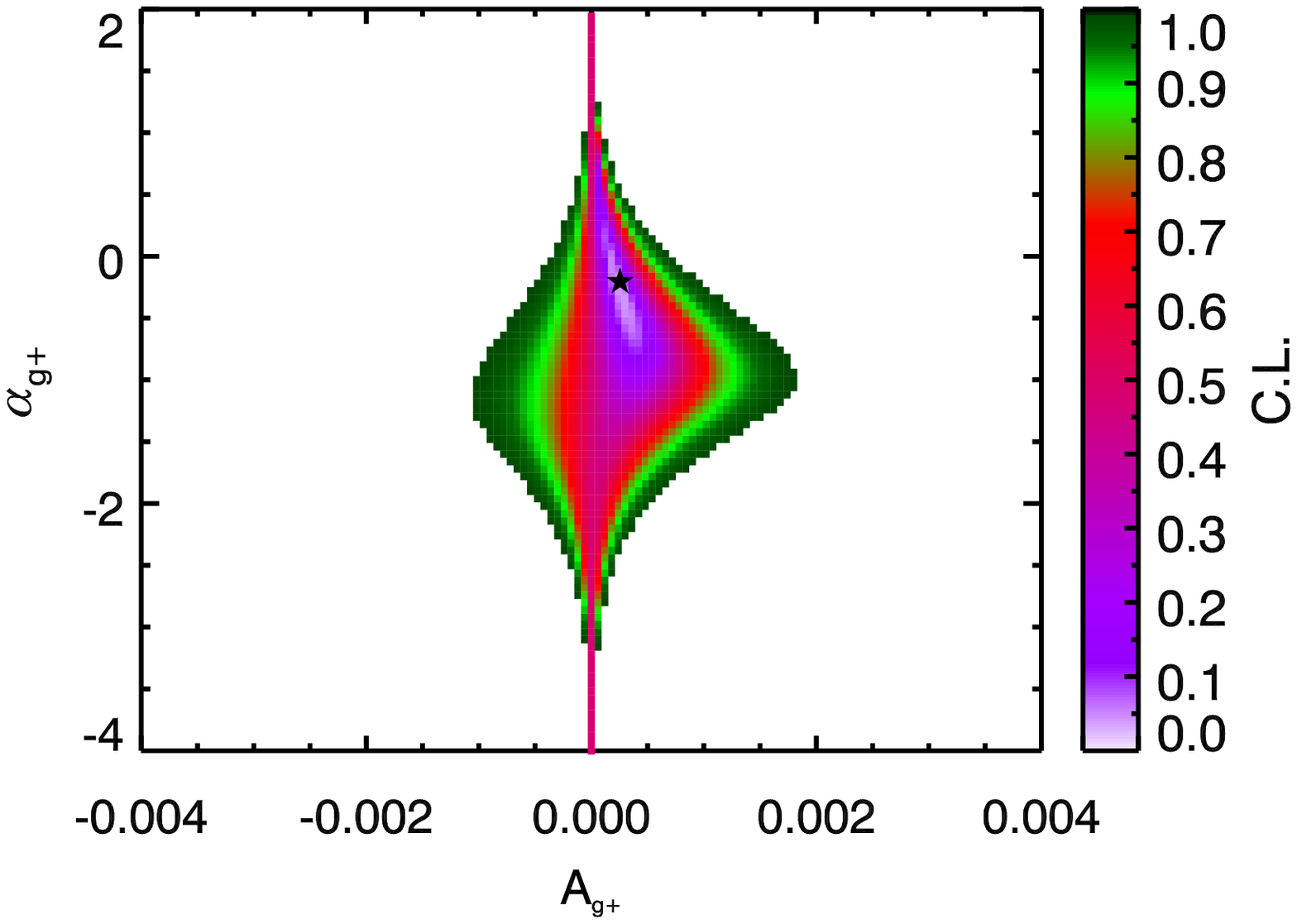}
}
\caption{Intrinsic alignment results when the geometrical centre is chosen to centre the cluster
and using galaxies with shapes in the $i$-band. These results are not signficantly different from those obtained assuming the BCG is at the centre of the cluster.}
\enf
%%%%%%%%%%%%%%%%%%%%%%%%%%%%%%%%%%%%%%%%%%%%%%%%%%%%%%%%%%%%%%%%%%

%=========================
%NEW SUBSECTION
%=========================
\subsubsection{Intrinsic alignments using single photo-$z$ estimates}
\label{ss:point}

In this section, we present the alignments results when the photo-$z$ are modeled as $\langle z \rangle$ or $z_p^{\rm ML}$ instead of using the $P(z)$, as we described in Section \ref{subsec:separate}. The distinction between this and the fiducial case lies in the weights assigned to each galaxy in bins $(a)$ and $(b)$. A galaxy will contribute a non-negligible weight to bin $(a)$ if $z_L-\Delta z <\langle z\rangle$($z_p^{\rm ML}$)$<z_L+\Delta z$ and it will contribute to bin $(b)$ only if $\langle z\rangle$($z_p^{\rm ML}$)$>z_L+\Delta z$. The lensing signal and intrinsic alignment correlation function measured using the $\langle z \rangle$ redshift estimate are shown in Figure \ref{fig:avgz}. Figure \ref{fig:avgz2} shows the analog results for $z^{\rm ML}_{\rm p}$.

The level of contamination of the lensing signal in $(b)$ to the intrinsic alignment signal in $(a)$ differs in these cases from that quoted in Section \ref{subsec:boostcalib}. For the $z_{\rm ML}$ estimator, $\mathcal{C}^{(a)}/\mathcal{C}^{(b)}=0.3052\pm0.0096$, and for the $\langle z \rangle$ estimator, we find $\mathcal{C}^{(a)}/\mathcal{C}^{(b)}=0.2921\pm0.0085$. While the level of contamination is comparable to that of our fiducial estimator, this does not take into account photometric redshift biases and can yield a biased $gI^a$ as a final result. This is particularly worrisome when estimating the intrinsic alignment signal, because it relies on galaxies that are clustered with the source. Leakage of those galaxies to redshifts outside of bin $(a)$ significantly reduces the amplitude of the measured signal. In the case of lensing, photo-$z$ bias can lead to a mis-estimate of the lensing signal due to a biased value of the effective $\Sigma_{c}$ of the sample. 
The lensing efficieny ratios are very similar for $z_{\rm ML}$ and $\langle z \rangle$ to that obtained using the full $P(z)$ and despite the fact that $z_{\rm ML}$ gives a worse redshift estimate than $\langle z \rangle$. The differences mainly arise at very low and high redshift (Figure \ref{fig:zMLvsz}), but most sources are at intermediate redshifts (Figure \ref{fig:below233}).

The case with $\langle z \rangle$ yields an $A_{g+}$ inconsistent with zero at the $28$ per cent C.L. Using $z_p^{\rm ML}$, we obtain $A_{g+}$ inconsistent with zero at the $77$ per cent C.L. (Table \ref{tab:powergplus}). 

%%%%%%%%%%%%%%%%%%%%%%%%%%%%%%%%%%%%%%%%%%%%%%%%%%%%%%%%%%%%%%%%%%
\bef
\subfigure[Lensing signal.]{
\includegraphics[width=0.45\textwidth]{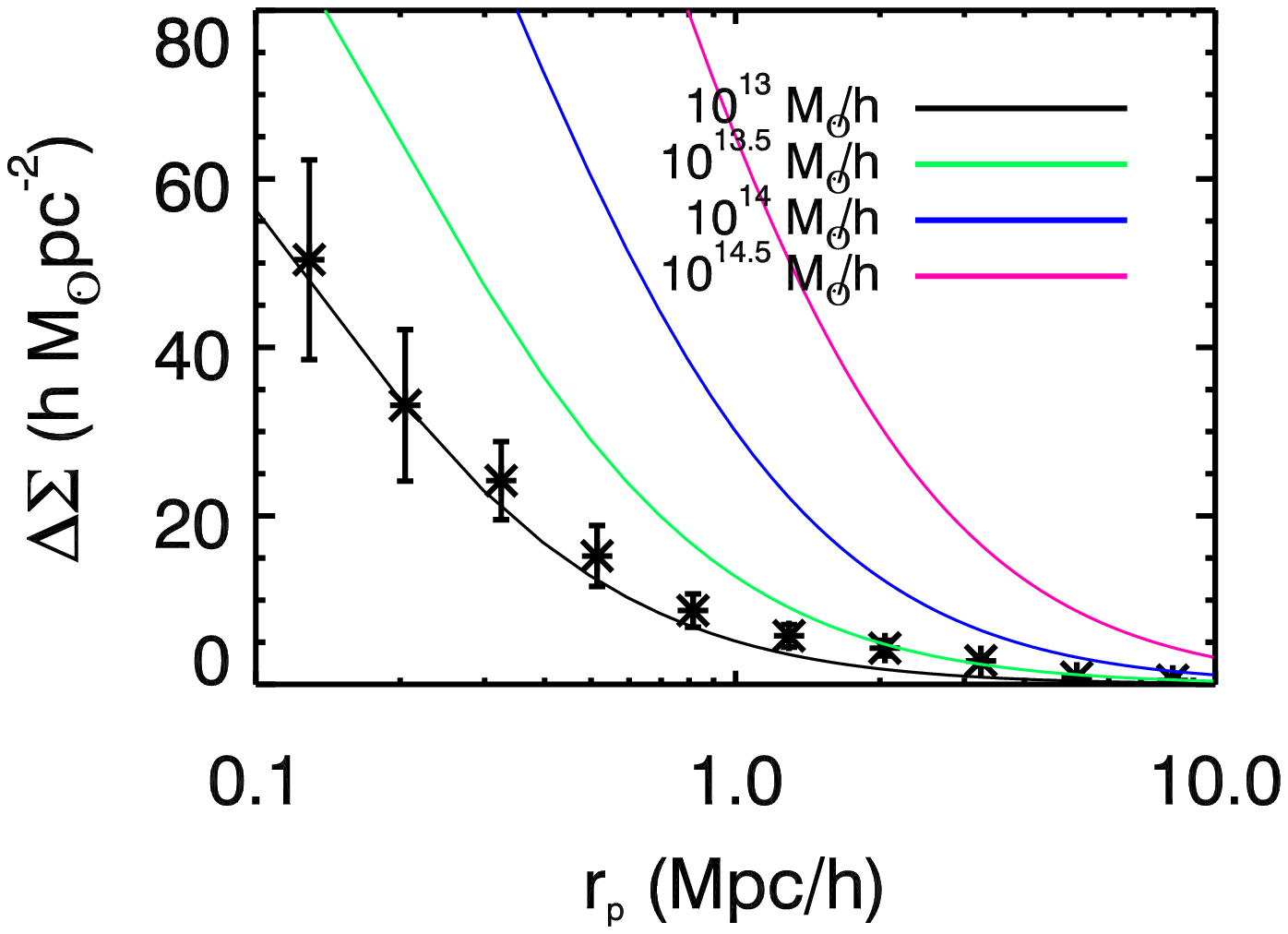}
}\hfill
\subfigure[$gI^a$.]{
\includegraphics[width=0.45\textwidth]{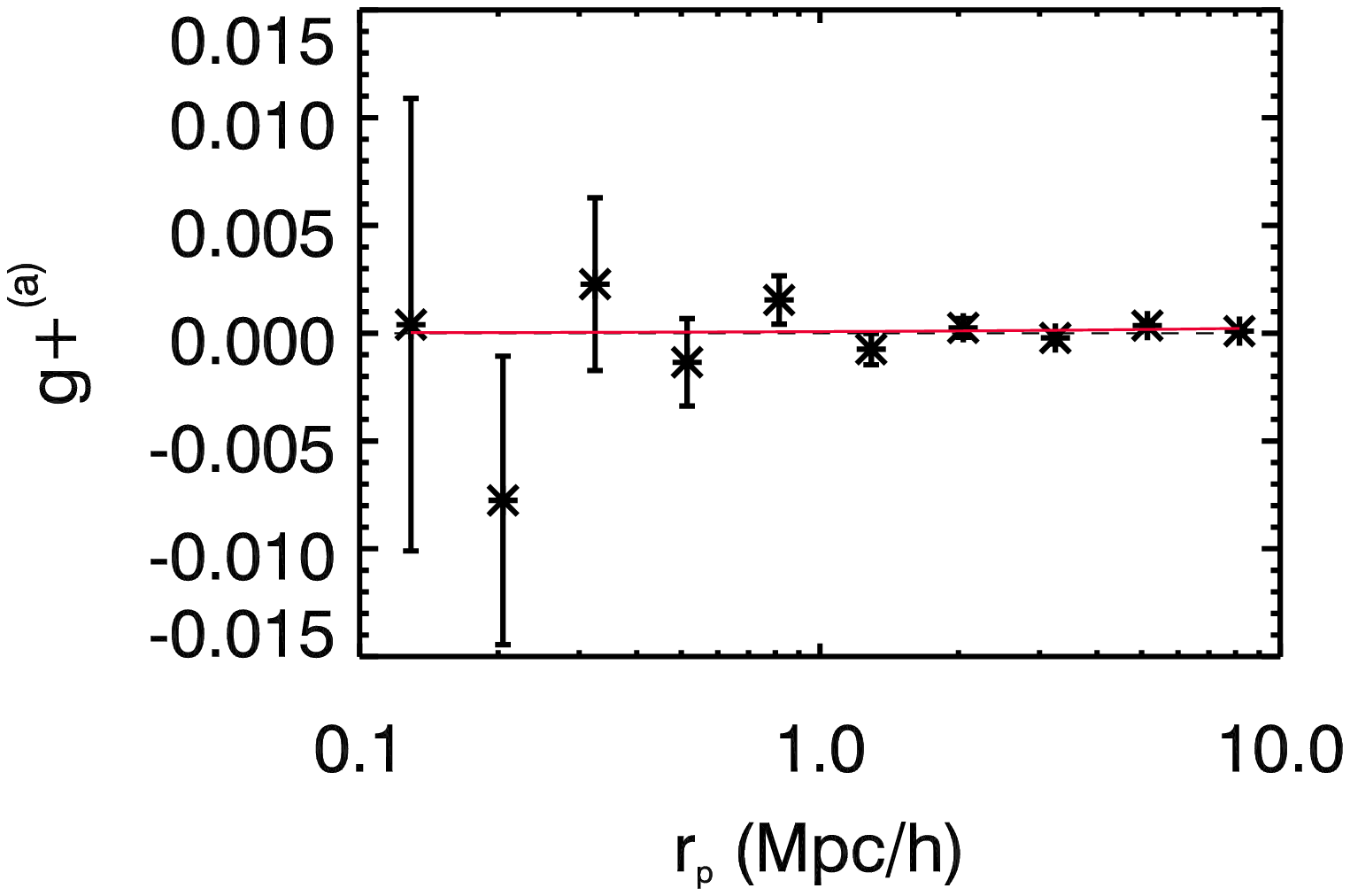}
}
\caption{Results for lensing and intrinsic alignments based on the expectation value estimator for photometric redshifts, $\langle z \rangle$, instead of $P(z)$.}
\label{fig:avgz}
\enf
%%%%%%%%%%%%%%%%%%%%%%%%%%%%%%%%%%%%%%%%%%%%%%%%%%%%%%%%%%%%%%%%%%
%%%%%%%%%%%%%%%%%%%%%%%%%%%%%%%%%%%%%%%%%%%%%%%%%%%%%%%%%%%%%%%%%%
\bef
\subfigure[Lensing signal.]{
\includegraphics[width=0.45\textwidth]{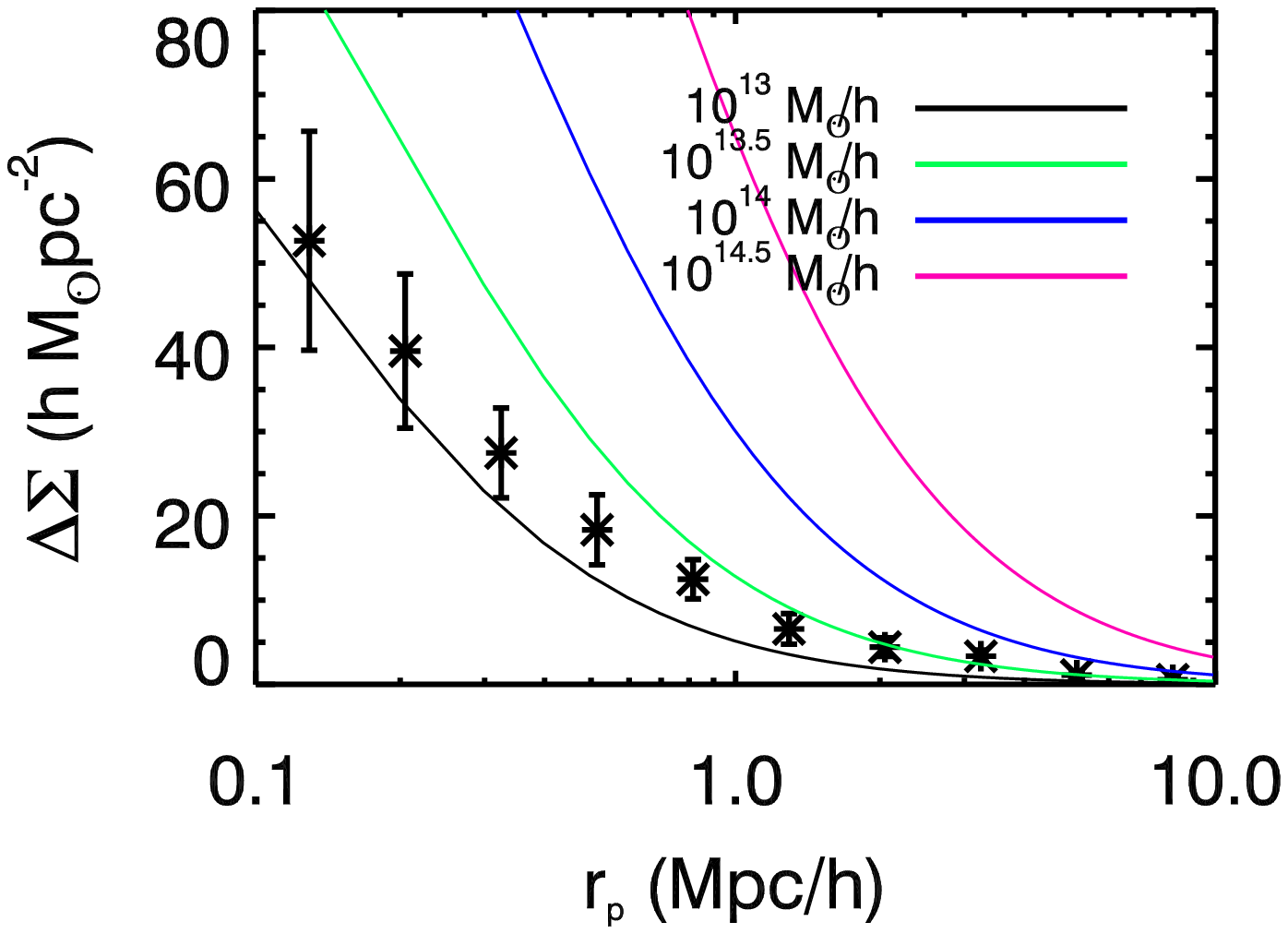}
}\hfill
\subfigure[$gI^a$.]{
\includegraphics[width=0.45\textwidth]{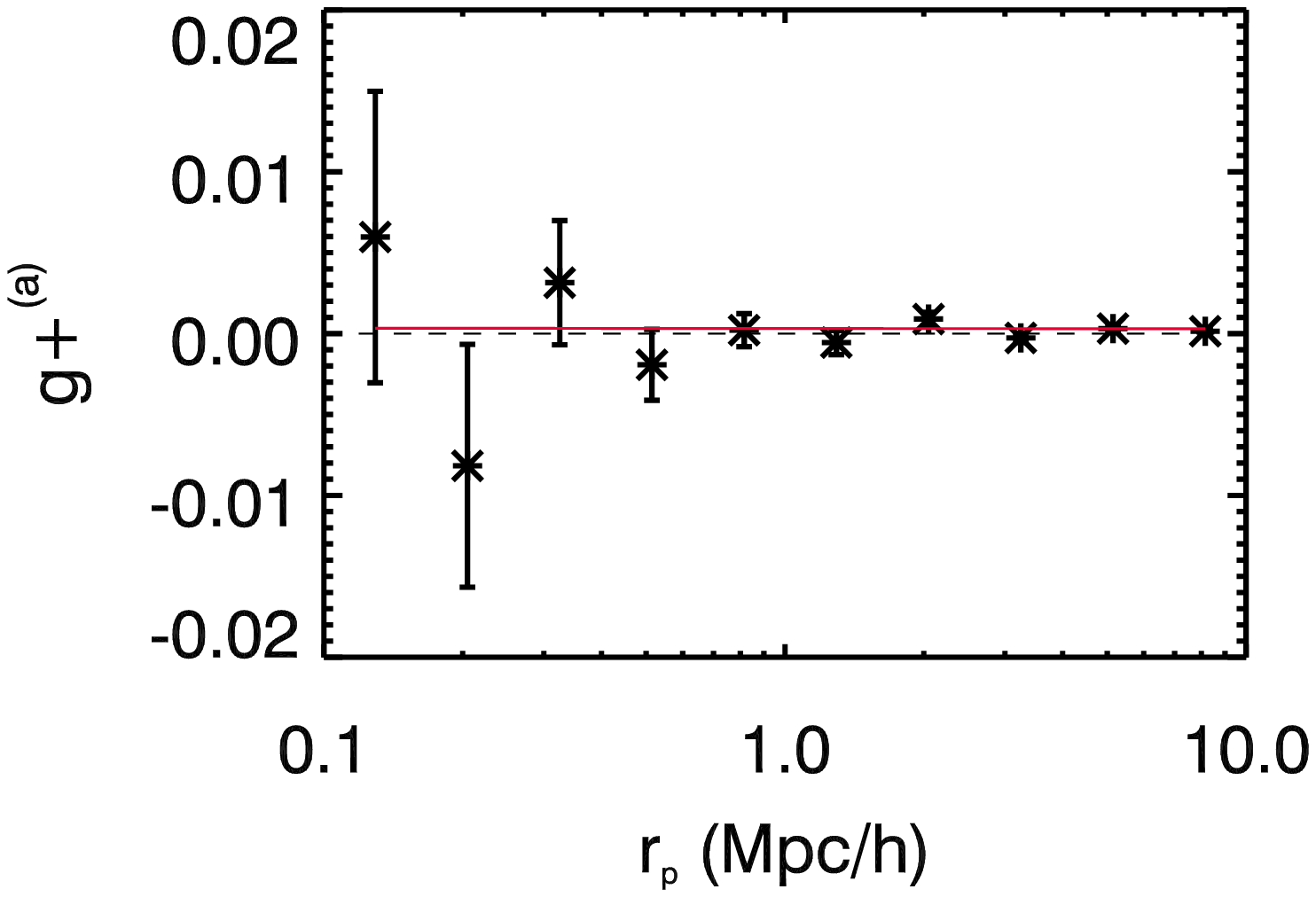}
}
\caption{Results for lensing and intrinsic alignments based on $z_p^{\rm ML}$.}
\label{fig:avgz2}
\enf
%%%%%%%%%%%%%%%%%%%%%%%%%%%%%%%%%%%%%%%%%%%%%%%%%%%%%%%%%%%%%%%%%%

%^^^^^^^^^^^^^^^^^^^^^^^^^^^^^^^^
%^^^^^^^^^^^^^^^^^^^^^^^^^^^^^^^^
\begin{table}
 \caption{Confidence levels from $\chi^2$ testing of $g\times$. $p(<\chi^2)$ represents the confidence level at which the null hypothesis is rejected.}
 \label{tab:confgcross}
 \begin{tabular}{@{}lcccccccc}
  \hline
  $p(<\chi^2)$ & $i$-band & $r$-band & Geometrical centre & Removing edges \\
  \hline
  $g\times^a$ & $0.97$ & $0.92$ & $0.72$ & $0.76$ \\
  $\Delta\Sigma_\times$ & $0.51$ & $0.37$ & $0.04$ & $0.07$ \\
\hline
\end{tabular}
\end{table}
%^^^^^^^^^^^^^^^^^^^^^^^^^^^^^^^^
%^^^^^^^^^^^^^^^^^^^^^^^^^^^^^^^^

\subsection{Intrinsic alignment contamination in future weak lensing surveys}
\label{sec:contamination}

Current models of intrinsic alignments in the non-linear regime are based on an extrapolation of the alignment model from large scales to small scales \citep{Hirata07,Bridle07}. Assuming this extrapolation is valid, our constraints shown in this section allow us to predict the typical contamination of the intrinsic alignment signal compared to the lensing of background galaxies. Our constraints on $b_LC_1$ allow us to set limits on the contamination of the alignment signal to the lensing in $(b)$. We quantify this contamination fraction as the ratio $gI^{a\rightarrow b}/gG^b$, which is also equal to the fractional contamination to the surface mass density profile of the stacked lenses. We obtain $gG^b$ in two possible ways. First, we consider the results of the fiducial case. Second, we assume the stacked clusters have an NFW profile consistent with $10^{13}$ M$_\odot$ (Figure \ref{fig:wlfiduc}) and the lensing efficiency is obtained from our fiducial analysis carried out in the $i$-band. We use the derived $95$ per cent C.L. on $b_LC_1\rho{\rm crit}$ to estimate the
contamination fraction in Figure \ref{fig:contam}. 

Our results indicate a contamination fraction between approximately $-18$ and $23$ per cent within 1 $h^{-1}$Mpc at $95$ per cent C.L. A negative contamination is due to radial alignments. A positive contamination is due to tangential alignments, which are allowed at this level. At increased radii, our upper bounds on this contamination increase. Our NFW estimate does not include the effect of a two halo term, which should be relevant in the range $1 $-$ 10$ $h^{-1}$Mpc, decreasing the contamination fraction. This fraction is specific to our sample and the quality of our photometric redshifts. A lower scatter in photometric redshifts would lead to lower contamination.

%%%%%%%%%%%%%%%%%%%%%%%%%%%%%%%%%%%%%%%%%%%%%%%%%%%%%%%%%%%%%%%%%
\bef
\centering
\includegraphics[width=0.45\textwidth]{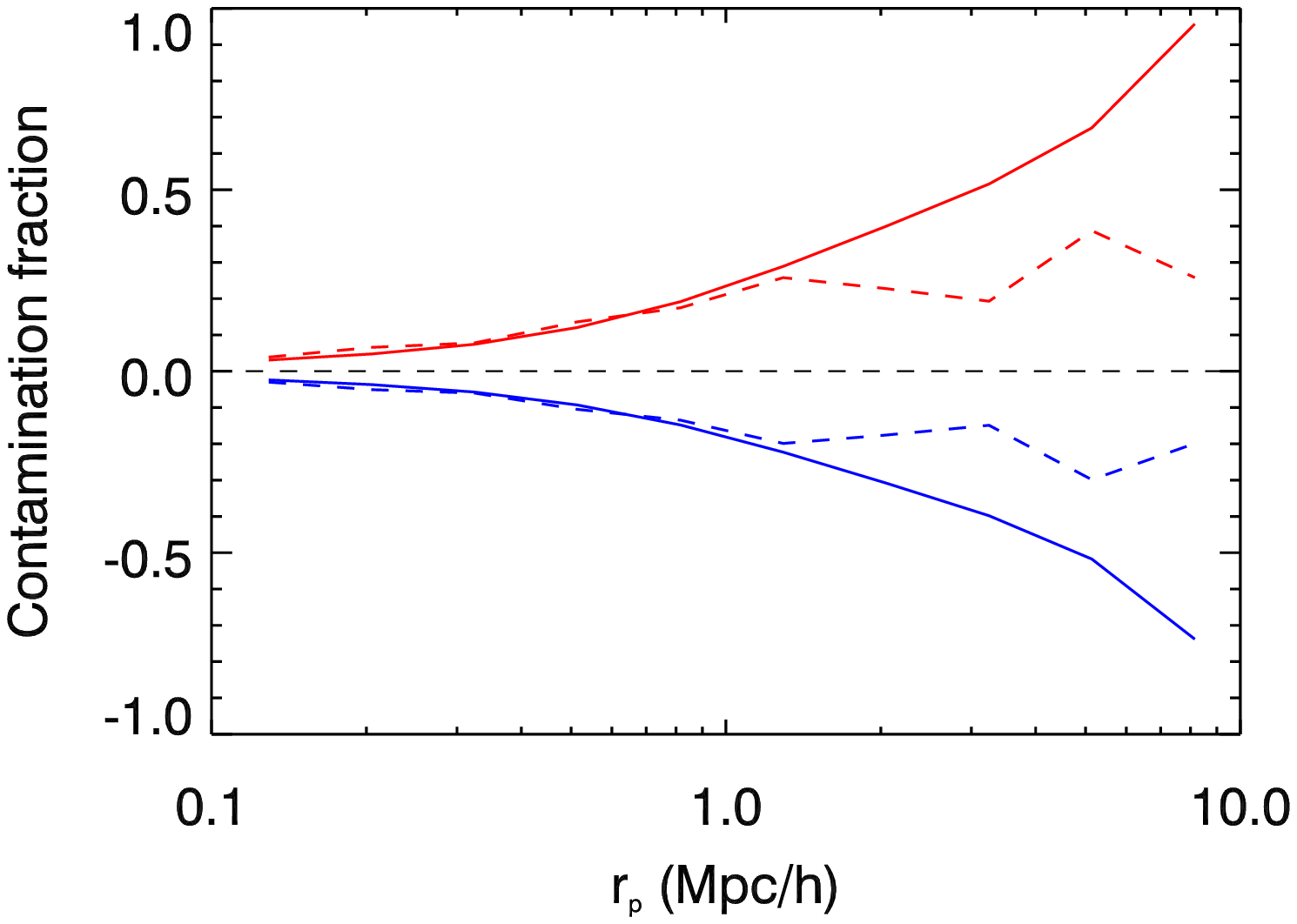}
\caption{Contamination fraction, $gI^{a\rightarrow b}/gG^b$. The solid lines indicate the contamination fraction when $gG^b$ is obtained assuming an NFW profile consistent with $10^{13}$ M$_\odot$ and the amplitude of $gI^{a\rightarrow b}$ given by the $95$ per cent C.L. constraints. The dashed lines give the contamination fraction when $gG^b$ is calculated from the $i$-band measurements. Red indicates upper limits and blue, lower limits. Positive contamination fraction is due to tangential alignments, while negative contamination is due to radial alignments.}
\label{fig:contam}
\enf
%%%%%%%%%%%%%%%%%%%%%%%%%%%%%%%%%%%%%%%%%%%%%%%%%%%%%%%%%%%%%%%%%%

%---------------------------------------------------------------------------------------------------------------------------
%NEW SECTION------------------------------------------------------------------------------------
%---------------------------------------------------------------------------------------------------------------------------
\section{Discussion}
\label{sec:discuss}

Our findings suggest that the alignments of satellites in and around groups and clusters are consistent with zero. Our results are consistent with those from previous work by \citet{Schneider13}. Those authors found that alignments of satellites around groups in the GAMA survey were consistent with null at the $2\sigma$ level. They also found that their constraints are sensitive to the shape measurement method. When galaxy ellipticities were measured from two-dimensional profile fits, a marginal detection of $3\sigma$ was observed. \citet{Hao11} measured the alignments of satellites around a highly pure sample of clusters in SDSS. Satellites were selected by red sequence color cuts around each cluster. The redshift range spanned by that study is similar to that in our work. They observed that alignments were detected only when using isophotal position angles to define the orientation of the satellite, in contrast to using model fit position angles. Different shape measurement methods can probe different scales of each galaxy, being more or less sensitive to the alignment mechanism, but they might also be subject to different systematics. We have not explored other shape measurement methods in this work. While this is interesting from the point of view of testing alignment models, weak lensing shapes should be used when attempting to constrain the alignment contamination to lensing surveys. We should note as well that weak lensing shapes can probe different physical scales of a galaxy depending on its resolution and that this introduces difficulties in the interpretation of intrinsic alignment signals measured with weak lensing shapes. Moreover, different instruments are likely to obtain different constraints on alignments for the same population of galaxies due to this dependence on their typical PSFs.

\citet{Hirata04b} measured the average shear due to intrinsic alignments of satellites around SDSS main galaxies using photometric redshifts. Our method provides an extension of theirs by using $P(z)$ for modelling the source redshifts and by using the spectroscopic calibration set to compute the lensing contamination. Their results (integrated in projected separation up to $1$ $h^{-1}$Mpc proper distance at most) were consistent with no alignment within their uncertainties. Phrased in terms of the average shear obtained by \citet{Hirata04b} (their Eqs. 37 and 38), we obtain an average shear of $\Delta\gamma = 0.0035\pm 0.0076$ within 10 $h^{-1}$ Mpc of the lens centre from the full area in $i$-band at the 95 per cent C.L. We find $\Delta\gamma = -0.0013\pm 0.0060$ and $\Delta\gamma =  0.0012 \pm 0.0054$ within $500\,h^{-1}$ kpc and $1\,h^{-1}$ Mpc, respectively. In comparison, \citet{Hirata04b} find $-0.0062 < \Delta\gamma < 0.0066$ at 99.9 per cent C.L. for the lenses in the SDSS main sample within a proper distance of $450\,h^{-1}$ kpc of the lens. For the brightest galaxies in their sample, with $M_r<-22$, they find $-0.008 < \Delta\gamma < 0.018$ within proper distance $1\,h^{-1}$ Mpc. While the samples we are comparing are different in terms of redshift distribution, luminosity and environment, we obtain similar order of magnitude constraints of the alignment strength. The redshift ranges and area coverage are also different, as \citet{Hirata04b} measures alignments at $z<0.15$ using a much larger area, $2,500$ deg$^2$.

\citet{Joachimi11} measured the alignment of Luminous Red Galaxies at $z<0.7$ using data from the SDSS. They found that the alignment signal has a scaling with luminosity of $L^\beta$, where $\beta \simeq 1.1$ from their combined sample. While the galaxies in our source sample are not selected by color, we can extrapolate their findings to estimate the alignment strength of red galaxies at the typical luminosity of our sources. The resulting alignment amplitude is a factor of $\sim 13$ larger than the maximum $b_LC_1\rho_{\rm crit}$ allowed by our constraints at the 95 per cent C.L. If we consider the steepest luminosity scaling allowed at $2\sigma$ level in the \citet{Joachimi11} results, we obtain a factor $\sim 5$ larger alignment amplitude. In practice, the fact that our sources are not all red ellipticals implies that this is an upper limit to the strength we could have expected to measure. A small fraction of red ellipticals, together with a low purity of the cluster sample, could bring our 95 per cent C.L. constraints into agreement with the results of \citet{Joachimi11}.

\citet{Blazek12} obtained constraints on the alignments of satellites around Luminous Red Galaxies, finding them consistent with null for both typical $\simeq L*$ red and blue galaxies. The typical haloes inhabited by these galaxies are more massive (by a factor of $\simeq 3$) than those probed in this work, as seen from the constraints obtained on the surface mass density profile of the lenses. The sources used in this work are considerably fainter than the sources used by \citet[][$m_r<21.8$]{Blazek12}, but the redshift range probed is similar ($0.16<z<0.36$). In comparison, we find a larger lensing contamination: we obtain an lower and upper limits of approximately $-18$ and $23$ per cent contamination to galaxy-galaxy lensing at $1$ $h^{-1}$Mpc, respectively, compared to their result of $10$ per cent. We emphasize that this result depends on the lens-source sample and the quality of the photometric redshifts, which are different for the two works. With more stringent cuts on photometric redshifts and modeling the alignment signal, \citet{Blazek12} find constraints at the $1-2$ per cent level.

In our model-dependent constraints, we have assumed that the non-linear alignment model is valid. This model is based on expressions derived for linear theory. Corrections in the non-linear regime are expected to have a significant impact on our constraints. Moreover, the non-linear alignment model assumes that the dominant mechanism for alignment is the stretching by the tidal field, although in cluster environments, other mechanisms might be at play. Satellites might be torqued by the tidal field or accreted along preferential directions. The scale dependence of the non-linear alignment model predictions need not represent those mechanisms. The strength of alignment is a free parameter of the model and cannot be predicted from first principles given our current understanding of galaxy formation. Hence, while our constraints on $b_LC_1\rho_{\rm crit}$ cannot serve to test this model, they can aid in the estimation of the contamination of intrinsic alignments to weak lensing in photometric surveys. 

As we were completing this work, we became aware of the work of \citet{Sifon14}, who studied the intrinsic alignments of galaxies around $91$ massive clusters in the range $0.05<z<0.55$, with a median redshift ($z_{\rm med}\simeq 0.145$) lower than the one in this work. Cluster members, identified from spectroscopic archival data from different surveys, were selected by their peculiar velocities and red sequence cuts.  The typical mass of the clusters in that work ($\simeq 10^{15}\,h^{-1}$ M$_{\odot}$) is significantly larger than the average stacked mass in our work. \citet{Sifon14} find no detection of alignments around clusters and no dependence on scale, galaxy color and luminosity. Their overall constraint on the average radial component of the shape of member galaxies within the virial radius of the clusters is of the same order of magnitude as our constraint on $\Delta\gamma$ within $1\,h^{-1}$Mpc. Our work differs from theirs in a number of ways. Mainly, our method, which includes the treatment of photometric redshift posterior, is suitable to be applied to ongoing and upcoming photometric surveys. By avoiding color selection, we do not restrict to red galaxies (which can result in a bias when selecting cluster galaxy populations at high redshift). We also probe scales beyond the virial radius of the clusters, as galaxies in the surrounding large-scale structure could also present significant alignment, and we consider the effect of miscentring. Finally, \citet{Sifon14} focus on constraining the contamination of the cross-correlation of intrinsic shapes of cluster galaxies and weak lensing shear to the cosmic shear power spectrum, while the focus of this work has been to determine the contamination of intrinsic alignments to galaxy-galaxy lensing measurements.

Group and cluster haloes are not spherical \citep{Oguri10}. In our analysis, we have stacked all lenses regardless of their shape and orientation. The alignment signal, however, can depend on the ellipticity of haloes, and thus, stacking regardless of shape and orientation can contribute to dilute the alignment. Tighter constraints on alignments might be obtained in the future by stacking clusters following the orientation of the major axis of the BCG \citep{Binggeli82,Niederste10}.

%---------------------------------------------------------------------------------------------------------------------------
%NEW SECTION------------------------------------------------------------------------------------
%---------------------------------------------------------------------------------------------------------------------------
\section{Conclusions}
\label{sec:conclude}

We have presented a new method to simultaneously constrain the lensing and intrinsic alignment signals around clusters of galaxies, using the photometric redshift posterior distribution of each source galaxy. We have shown that doing this results in a better photometric redshift calibration than when single point estimates of the photo-$z$ are used. In particular, there is significant improvement on redshift bias and scatter of the galaxies in the calibration set of spectroscopic redshifts at low redshift ($z<0.25$). While previous studies had been forced to remove starburst galaxies from their samples due to poor photo-$z$ calibration, we have found that using the full $P(z)$ results in improved redshift estimation for those galaxies, and they can thus be included in the analysis. This is of particular relevance to weak lensing studies in future surveys, which will observe a larger fraction of starburst galaxies at high redshift. 

The formalism developed in this work allows for the simultaneous constraint of the lensing and the alignment signal. For this to be possible, a representative subset of galaxies with redshifts is necessary to assess the biases induced by photometric redshift scatter in the lensing efficiency of galaxies in the background of the clusters. For our separation between the cluster population and the background galaxies of $\Delta z=0.2$, we find that there is a $30$ per cent contamination of lensing from the background bin on the mean shear measured using the sources in the redshift bin near the lenses. The choice to have such a large contamination fraction is driven by the assumptions of the formalism, which requires that we neglect the intrinsic alignment contamination to the lensing in the background. Improvements in the formalism or stronger priors on the intrinsic alignment strength (from spectroscopic studies) could allow us to reduce this contamination and thus obtain tighter constraints on alignments from photometric data. 

Our results are consistent with a non-detection of alignments around groups and clusters. We find no evidence of systematics affecting our results. We have performed model-independent (power-law) and model-dependent fits to the data. Our quoted results will be of use to future studies of lensing and alignments around clusters, but it should be pointed out that our constraints depend on the redshift distribution, purity and completeness of the group and cluster catalog used in this study. In particular, the typical low mass of the clusters in the \citet{Geach11} catalog ($\sim 10^{13}$ M$_\odot$, Figure \ref{fig:wlfiduc}) could be indicative of a low purity sample and this encourages us to explore the use of other cluster catalogs in the future.
The quoted contamination level also depends on the photo-$z$ quality. By relating our formalism to the tidal alignment model of \citet{Catelan01}, we have set constraints on the product of the effective amplitude of the bias of a sample of clusters and the response of cluster members and galaxies in the cluster outskirts to the tidal field of large-scale structure. 

Our formalism has been developed assuming the use of $P(z)$ to represent the redshift of each galaxy. An extension to include templates as a $P(z,t)$, as described in Section \ref{ss:photoz}, would naturally allow for separation of galaxies by type, which would be interesting from the point of view of determining the strength of alignment and the physical mechanism behind it for each type of galaxy. However, the incorporation of $P(z)$ is already computationally costly (see Appendix B for possible improvements), and expanding the parameter space to include galaxy templates will also require additional storage capacity and computational time in the processing of the data. 

State-of-the-art lensing measurements of stacked clusters are subject to $\sim10$ per cent systematic uncertainties \citep{Sheldon09}, where the dominant contribution is the uncertainty in the redshift distribution of the sources.  In this work, we have shown that the uncertainty in the stacked surface mass density profile from intrinsic alignment systematics is of order $\simeq 20$ per cent at projected radii below 1 $h^{-1}$ Mpc. However, percent level constraints can be placed already by applying similar methods to larger area datasets (i.e., \citealt{Blazek12}). Stage IV dark energy surveys will require mass calibration uncertainties at the per cent level to contrain the equation state of dark energy \citep{Weinberg13}. We conclude that intrinsic alignments continue to be potential systematics to weak lensing mass calibration of groups and clusters, and that constraining them should be a priority in the preparation for Stage IV dark energy surveys. We are optimistic that the combination of data from ongoing surveys and the advent of new methods for studying alignments will enable better constraints in the near future.

Ongoing and upcoming large-scale structure surveys (such as the the Kilo-Degree Survey\footnote{\url{http://kids.strw.leidenuniv.nl/}}, Dark Energy Survey\footnote{\url{http://www.darkenergysurvey.org/}}, Pan-STARRS\footnote{\url{http://pan-starrs.ifa.hawaii.edu/public/}}, Hyper Suprime-Cam\footnote{\url{http://www.naoj.org/Projects/HSC/}}, {\it Euclid}\footnote{\url{http://sci.esa.int/euclid/}}, the Large Synoptic Survey Telescope\footnote{\url{http://www.lsst.org/lsst/}} and {\it WFIRST-AFTA}\footnote{\url{http://wfirst.gsfc.nasa.gov/}}), will improve on the following: increased number density of sources, better shape measurements, larger area coverage, better identification of clusters of galaxies and better separation of cluster members from foreground and background sources. A combination of spectroscopic and photometric redshifts would ease the separation of background and foreground galaxies. Color selections that isolate red sequence galaxies \citep{Medezinski07} can also be applied to separate lensing from alignments. \citet{Hao11} applied this procedure to measure the alignments of satellites around clusters in SDSS in the same redshift range as probed in this work. However, due to the unconstrained evolution of the Butcher-Oemler \citep{Butcher78} effect, whereby cluster galaxies are increasingly blue at high redshift, it is important not to bias the cluster member selection when applying color cuts. With several deeper wide-field lensing surveys underway, we are optimistic that use of the methods described in this paper with the new data will result in tighter constraints on intrinsic alignments of galaxies around clusters in the next few years.

\section*{Acknowledgements}

We are grateful to Jonathan Blazek, Renyue Cen, James Gunn and Kevin Bundy for fruitful discussions. We thank Atsushi Nishizawa for his comments on the modeling of the redshift posteriors. We thank the anonymous referee's for comments that helped improve the quality of this manuscript. We also thank Cora Dvorkin for sharing of algorithms developed for \citet{Chisari13}, which were applied towards the NLA model results presented in this work. Finally, we thank the PRIMUS and DEEP2 survey teams and the authors of \citet{Geach11} for making their data publicly available.

RM was supported for the duration of this work by the Department of Energy Early Career Award program.

Funding for the SDSS and SDSS-II has been provided by the Alfred P. Sloan Foundation, the Participating Institutions, the National Science Foundation, the U.S. Department of Energy, the National Aeronautics and Space Administration, the Japanese Monbukagakusho, the Max Planck Society, and the Higher Education Funding Council for England. The SDSS Web Site is \url{http://www.sdss.org/}.

The SDSS is managed by the Astrophysical Research Consortium for the Participating Institutions. The Participating Institutions are the American Museum of Natural History, Astrophysical Institute Potsdam, University of Basel, University of Cambridge, Case Western Reserve University, University of Chicago, Drexel University, Fermilab, the Institute for Advanced Study, the Japan Participation Group, Johns Hopkins University, the Joint Institute for Nuclear Astrophysics, the Kavli Institute for Particle Astrophysics and Cosmology, the Korean Scientist Group, the Chinese Academy of Sciences (LAMOST), Los Alamos National Laboratory, the Max-Planck-Institute for Astronomy (MPIA), the Max-Planck-Institute for Astrophysics (MPA), New Mexico State University, Ohio State University, University of Pittsburgh, University of Portsmouth, Princeton University, the United States Naval Observatory, and the University of Washington.

Funding for SDSS-III has been provided by the Alfred P. Sloan Foundation, the Participating Institutions, the National Science Foundation, and the U.S. Department of Energy Office of Science. The SDSS-III web site is \url{http://www.sdss3.org/}.

SDSS-III is managed by the Astrophysical Research Consortium for the Participating Institutions of the SDSS-III Collaboration including the University of Arizona, the Brazilian Participation Group, Brookhaven National Laboratory, Carnegie Mellon University, University of Florida, the French Participation Group, the German Participation Group, Harvard University, the Instituto de Astrofisica de Canarias, the Michigan State/Notre Dame/JINA Participation Group, Johns Hopkins University, Lawrence Berkeley National Laboratory, Max Planck Institute for Astrophysics, Max Planck Institute for Extraterrestrial Physics, New Mexico State University, New York University, Ohio State University, Pennsylvania State University, University of Portsmouth, Princeton University, the Spanish Participation Group, University of Tokyo, University of Utah, Vanderbilt University, University of Virginia, University of Washington, and Yale University. 

Funding for PRIMUS is provided by NSF (AST-0607701, AST-0908246, AST-0908442, AST-0908354) and NASA (Spitzer-1356708, 08-ADP08-0019, NNX09AC95G). Funding for the DEEP2 Galaxy Redshift Survey has been provided by NSF grants AST-95-09298, AST-0071048, AST-0507428, and AST-0507483 as well as NASA LTSA grant NNG04GC89G.

%*******************************
%Bibliography
%\bibliographystyle{mn2e}
%\bibliography{chisari_IA_2014.bib}

%********************************

%---------------------------------------------------------------------------------------------------------------------------
%APPENDIX----------------------------------------------------------------------------------------------
%---------------------------------------------------------------------------------------------------------------------------
\onecolumn
\nopagebreak

\appendix

\subsection*{APPENDIX A: Analytic error estimates due to shape noise}

In this appendix, we present a set of equations to describe the uncorrelated shape noise of the intrinsic alignment and lensing correlations discussed in Section \ref{sec:formalism}. The correlation function of lens positions and galaxy shapes in Eq. \ref{eq:corra} can be expressed as a weighted sum,

\beeq
gI^a+gG^{a\rightarrow b} = \sum_j^{\rm lens} \omega_j^{(a)} \tilde{\gamma}_j
\label{eq:appA_1}
\eneq

\noindent where the weights are given by

\beeq
\omega_j^{(a)} = \frac{N_R}{N_L}\frac{\twia\wipzLa}{\sum_j^{\rm random} \twia\wipzLa}
\eneq

\noindent and analogously for bin $(b)$. Assuming that galaxy ellipticities are uncorrelated, we obtain the variance of Eq. \ref{eq:appA_1},

\bear
{\rm Var}\left[ \frac{N_R}{N_L}\sum_j^{\rm lens} \omega_j^{(a)} \tilde{\gamma}_j \right]
&=& \left( \frac{N_R}{N_L}\right)^2 \frac{\sum_j^{{\rm lens}} \left[\twia\wipzLa\,\tilde\gamma_j\right]^2}{\left[\sum_j^{{\rm  random}} \twia\wipzLa\right]^2},
\enar

\noindent and analogously for $(b)$.

We can propagate the uncertainties to obtain expressions for the shape noise in $\Delta\Sigma^{\rm stack}$ (Eq. \ref{eq:DS}) and $gI^a$ (Eq. \ref{eq:wfinal2}). We propage the uncertainty in the average shear behind the lenses and the lensing efficiency and we obtain

\bear
{\rm Var}\left[\Delta\Sigma^{\rm stack}(r_p)\right] &=& {\rm Var}\left[\frac{\sum_j^{\rm lens} \twia\wipzLb \tilde\gamma_j}{\sum_j^{{\rm lens}} \twia\wipzLb} \frac{\sum_j^{{\rm lens}} \twia\wipzLb}{\sum_j^{\rm lens} \twia\wipzLb\Sigma_{j,c}^{-1}}\right], \nonumber\\
 &=&  {\rm Var}\left[\frac{\sum_j^{\rm lens} \twia\wipzLb \tilde\gamma_j}{\sum_j^{{\rm lens}} \twia\wipzLb}\right] {\rm Var}\left[\frac{\sum_j^{{\rm lens}} \twia\wipzLb}{\sum_j^{\rm lens} \twia\wipzLb\Sigma_{j,c}^{-1}}\right] +
 \left[\frac{\sum_j^{\rm lens} \twia\wipzLb \tilde\gamma_j}{\sum_j^{{\rm lens}} \twia\wipzLb}\right]^2 {\rm Var}\left[\frac{\sum_j^{{\rm lens}} \twia\wipzLb}{\sum_j^{\rm lens} \twia\wipzLb\Sigma_{j,c}^{-1}}\right] +  \nonumber\\
&+&{\rm Var}\left[\frac{\sum_j^{\rm lens} \twia\wipzLb \tilde\gamma_j}{\sum_j^{{\rm lens}} \twia\wipzLb}\right] \left[\frac{\sum_j^{{\rm lens}} \twia\wipzLb}{\sum_j^{\rm lens} \twia\wipzLb\Sigma_{j,c}^{-1}}\right]^2 ,
\enar

\noindent which we approximate by Taylor expanding and neglecting the uncertainty in the calibration factor,

\bear
{\rm Var}\left[\Delta\Sigma^{\rm stack}(r_p)\right]  &\simeq&  
{\rm Var}\left[\frac{\sum_j^{\rm lens} \twia\wipzLb \tilde\gamma_j}{\sum_j^{{\rm lens}} \twia\wipzLb}\right] \left[\frac{\sum_j^{{\rm lens}} \twia\wipzLb}{\sum_j^{\rm lens} \twia\wipzLb\Sigma_{j,c}^{-1}}\right]^2 
\label{eq:appA_2}
\enar

We have assumed that because the calibration set is a small representative subset and covers a small area of the galaxies in the shape catalogue, it is independent of the lens-source pairs. 

We also propagate the uncertainties in $gI^a$, obtaining

\bear
{\rm Var}\left[gI^a \right] &=& \left(\frac{N_R}{N_L}\right)^2{\rm Var}\left[ \frac{\sum_j^{\rm lens} \twia\wipzLa \tilde\gamma_j}{\sum_j^{\rm random} \twia\wipzLa}\right] 
+ \left(\frac{N_R}{N_L}\right)^2{\rm Var} \left[ \frac{\sum_j^{\rm lens} \twia\wipzLb \tilde\gamma_j}{\sum_j^{\rm random} \twia\wipzLa}\right]\left[ \frac{\sum_j^{\rm random} \twia\wipzLa\Sigma_{j,c}^{-1}}{\sum_j^{\rm random} \twia\wipzLb\Sigma_{j,c}^{-1}}\right]^2  \nonumber\\
&-&2 \left(\frac{N_R}{N_L}\right)^2\left[  \frac{\sum_j^{\rm random} \twia\wipzLa\Sigma_{j,c}^{-1}}{\sum_j^{\rm random} \twia\wipzLb\Sigma_{j,c}^{-1}}\right]{\rm Cov}\left[\frac{\sum_j^{\rm lens} \twia\wipzLa \tilde\gamma_j}{\sum_j^{\rm random} \twia\wipzLa},\frac{\sum_j^{\rm lens} \twia\wipzLb \tilde\gamma_j}{\sum_j^{\rm random} \twia\wipzLa}\right],
\label{eq:fullvarIA}
\enar

\noindent where the covariance is given by 

\bear
{\rm Cov}\left[\frac{\sum_j^{{\rm lens}} \twia\wipzLa\,\tilde\gamma_j}{\sum_j^{{\rm random}} \twia\wipzLa} ,\frac{\sum_j^{{\rm lens}} \twia\wipzLb\,\tilde\gamma_j}{\sum_j^{{\rm  random}} \twia\wipzLb}\right]
&=& \frac{\sum_j^{{\rm lens}} \wipzLa\wipzLb\,\left[\twia\tilde\gamma_j\right]^2}{\left[\sum_j^{{\rm  random}} \twia\wipzLa\right]^2}.
\enar

The variance of the ratio of the calibration factors that appears in Eq. \ref{eq:fullvarIA} can be computed by bootstrapping over the galaxies in the calibration set. We have not considered the calibration factor uncertainties in the propagation because they are of the order of $<3$ per cent. Instead, in Section \ref{sec:results}, we evaluate the impact on our results of increasing the calibration factor ratio by $5$ per cent. In this appendix, we have not propagated the shape noise due to the subtraction of systematics. This propagation is straightforward and the variance of the signal and the systematics are added in quadrature because the two quantities are not correlated.

We have assumed a constant shear calibration factor $\mathcal{R}$ and no uncertainty. Moreover, in the presence of a systematic average shear in Eq. \ref{eq:shape}, the error bars have to take into account the covariance between lenses and random points. While the positions of these two samples are not correlated, the catalogue of source galaxies used in both cases is the same and this results in a covariance between the average shear around the lenses and the random points. In Sec. \ref{subsec:intrinsic}, we compare the shape noise error bars added in quadrature for the signal and the systematics to the result obtained by the bootstrap method, which takes into account that covariance.

\subsection*{APPENDIX B: Parametrizing photometric redshift posteriors}
\renewcommand\thefigure{B.\arabic{figure}}   
\setcounter{figure}{0} 
 
We have shown in Section \ref{subsec:speccalib} that the use of the $P(z)$ offers many advantages with respect to the single photo-$z$ estimates. The $P(z)$ guarantees a better redshift calibration and its use is of critical importance in the case of star-forming galaxies, where single point estimates of the redshifts have in the past meant that these galaxies had to be excluded from the measurements. While we use the full $P(z)$ sampled with $101$ points in a redshift range between $0<z<1.5$, this would become computationally prohibitively in ongoing and upcoming photometric surveys, such as DES, HSC or LSST among others, which will measure and store redshift information for billions of galaxies. These surveys would benefit from an efficient parametrization of the $P(z)$ that can allow for faster input/output operations and lower the cost of information storage.

With this in mind, we fit the $P(z)$ distributions for each galaxy in the calibration set with a sum of multiple Gaussian functions. We choose to fit $1$ Gaussian, $2$ Gaussian or $4$ Gaussian components to each $P(z)$. We show the results of approximating the $P(z)$ by a sum of Gaussian functions in Figure \ref{fig:gaussianfits}. For each case, we compute the median and the $1\sigma$ dispersion of $\langle z \rangle - z_s$, as a function of $\langle z \rangle$, which is obtained from the multiple Gaussian model. In the top panels, we compare the effect that using $1$, $2$ or $4$ Gaussian components has in the median and the dispersion of $\langle z \rangle - z_s$. All cases yield very similar results. In the bottom panels, we compare the width of the $P(z)$, given by the 68 per cent probability intervalf, $2\sigma_{68}$, to the same quantity obtained from the multiple Gaussian model. The $2$ component case reproduces the true width of the $P(z)$ better than the $1$ component case. The $4$ component case does not visibly improve the calibration nor does it significantly improve the comparison of the second moment of the $P(z)$. We conclude that, for our sample, $2$ component Gaussian fits seem sufficient to reproduce the properties of the $P(z)$. 

While our result suggests that studies similar to the one presented in this paper could be undertaken with bi-Gaussian fits to the $P(z)$, we emphasize that the details of how to best perform this parametrization depend on the population of galaxies, their apparent magnitude and redshift range. The use of spectroscopic calibration samples representative of the galaxies in the photometric surveys is needed to determine the trade-off between the number of parameters used to describe the $P(z)$ and the amount of information to be stored.

\citet{Carrasco14} applied the publicly available {\sc tpz} machine-learning algorithm \citep{Carrasco13} to obtain photometric redshifts in 154 deg$^2$ of photometric data from the CFHTLens survey\footnote{\url{http://www.cfhtlens.org/}} \citep{Heymans12,Erben13}. They studied the optimal description of photometric redshift posterior distribution functions (PDF). Among the options considered were single Gaussian fits, multi-Gaussian and sparse basis decomposition (a combination of Gaussian and Voigt profile functions). They found that the former provides a more accurate description than a multiple Gaussian fit. The typical number of parameters needed for describing each PDF is between $10$ and $20$. 

\citet{Carrasco14} multi-Gaussian decompositions are performed by determining the number of Gaussian components from counting the number of peaks in the PDF. In comparison, we have considered decompositions with the same number of components for all galaxies. This fixes the number of parameters to be stored for each galaxy. We seldom find more than 4 peaks in a given posterior ($\leq 8$\%, with an average of $\simeq 2$ peaks and a median of $1$). The number of peaks could be conditioned by the smoothing over the prior.

%%%%%%%%%%%%%%%%%%%%%%%%%%%%%%%%%%%%%%%%%%%%%%%%%%%%%%%%%%%%%%%%%%
\bef
\centering
\subfigure[Median difference between $\langle z \rangle$ from the $P(z)$ and from the fit measured as the deviation between the median and the identity.]{
\includegraphics[width=0.45\textwidth]{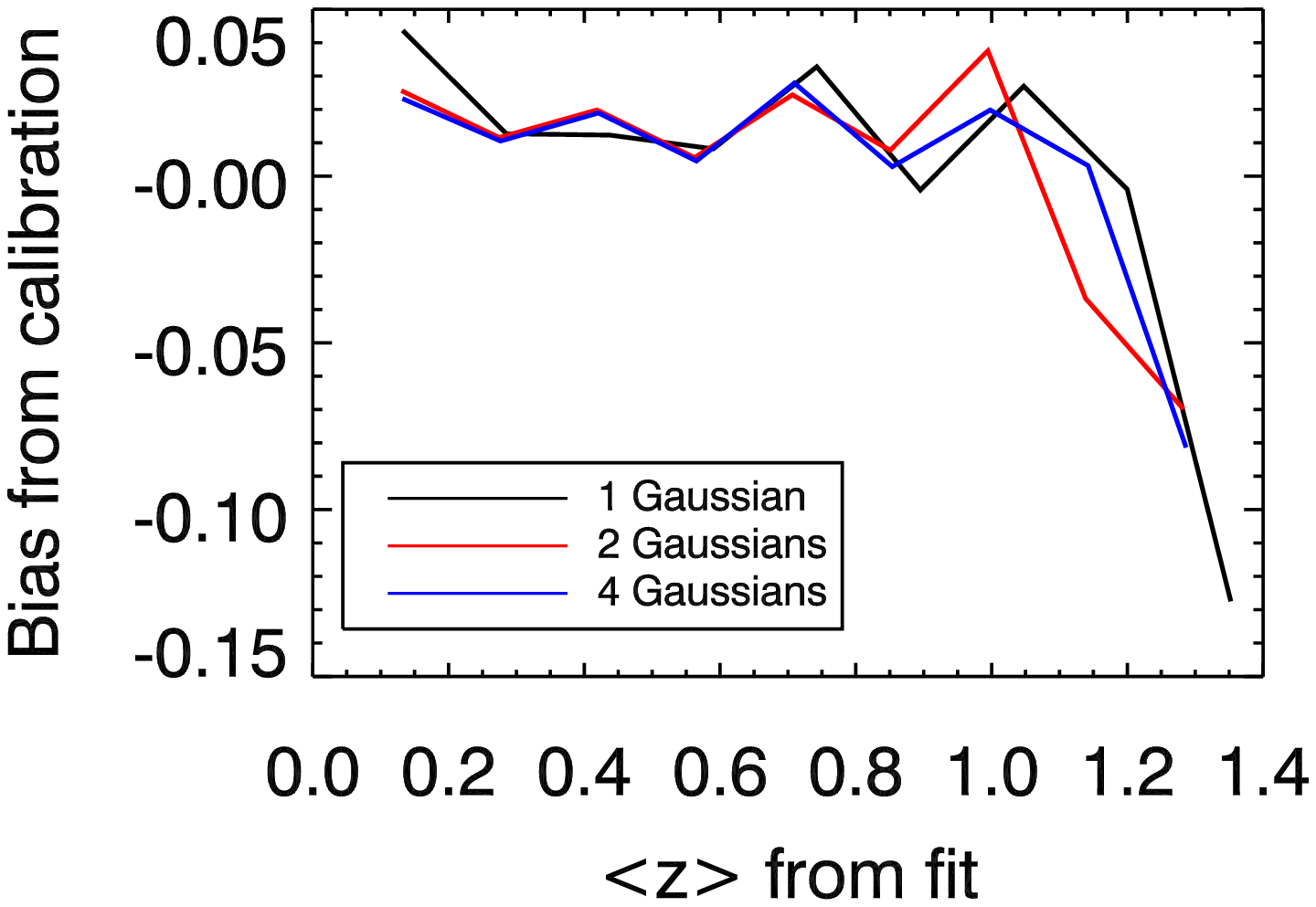}
}\hfill
\subfigure[RMS scatter in the comparison between $\langle z \rangle$ from the $P(z)$ and from the multi-Gaussian fit measured as the $2\sigma_{68}$.]{
\includegraphics[width=0.45\textwidth]{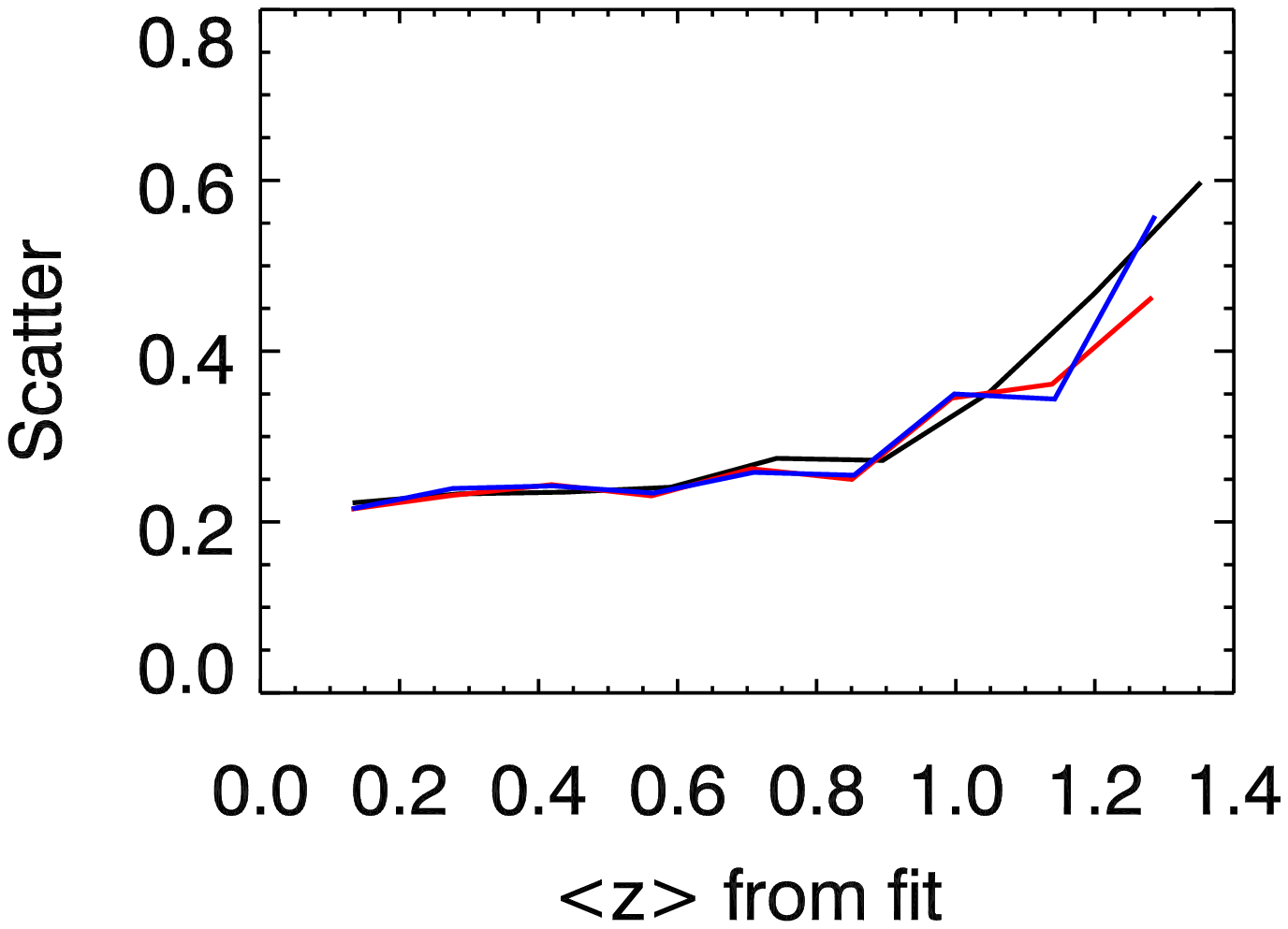}
}
\subfigure[Median difference between the dispersion $\sigma$ from the $P(z)$ and from the fit measured as the deviation between the median and the identity.]{
\includegraphics[width=0.45\textwidth]{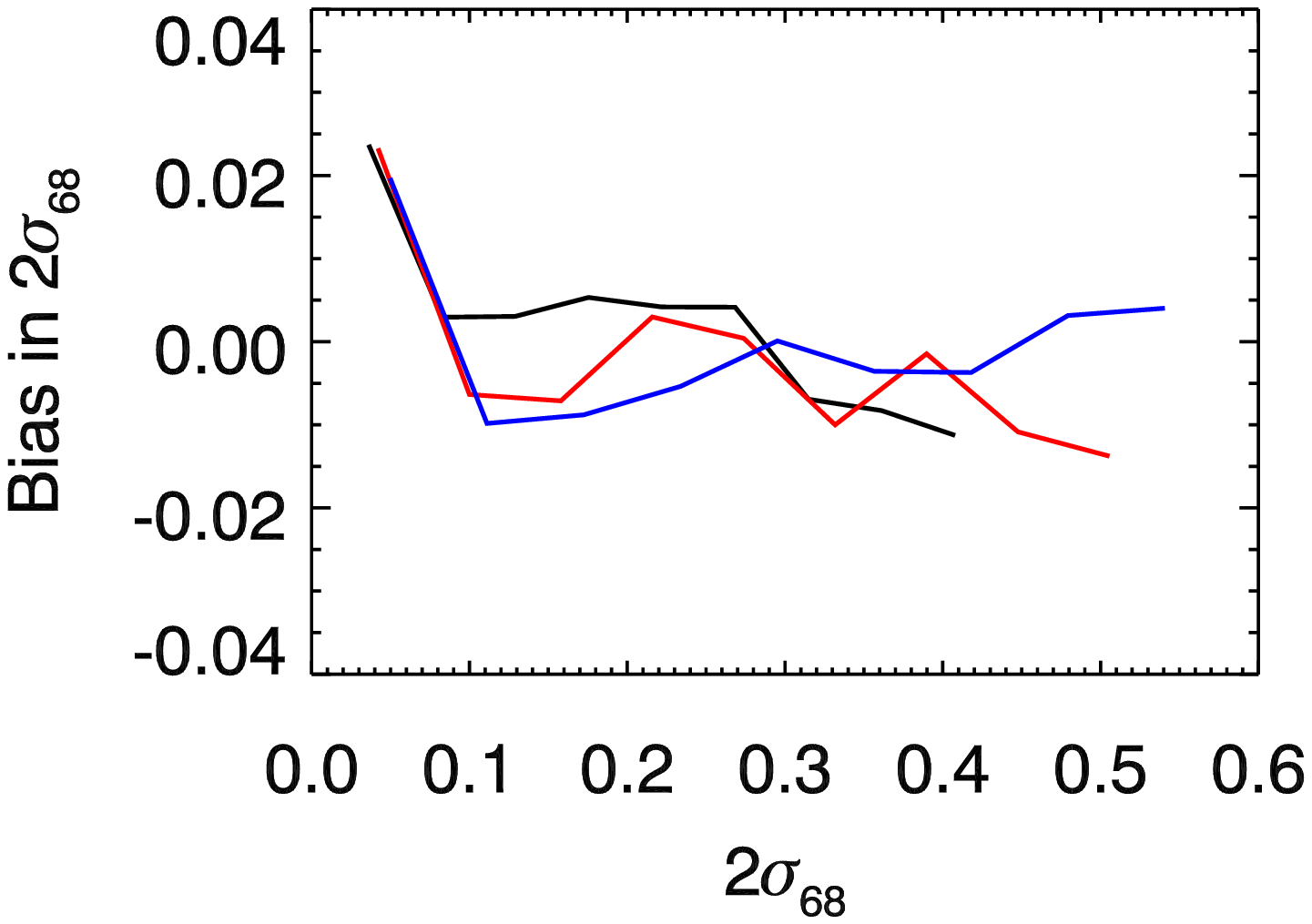}
}\hfill
\subfigure[RMS scatter in the comparison between the dispersion $\sigma$ from the $P(z)$ and from the multi-Gaussian fit measured as the $2\sigma_{68}$.]{
\includegraphics[width=0.45\textwidth]{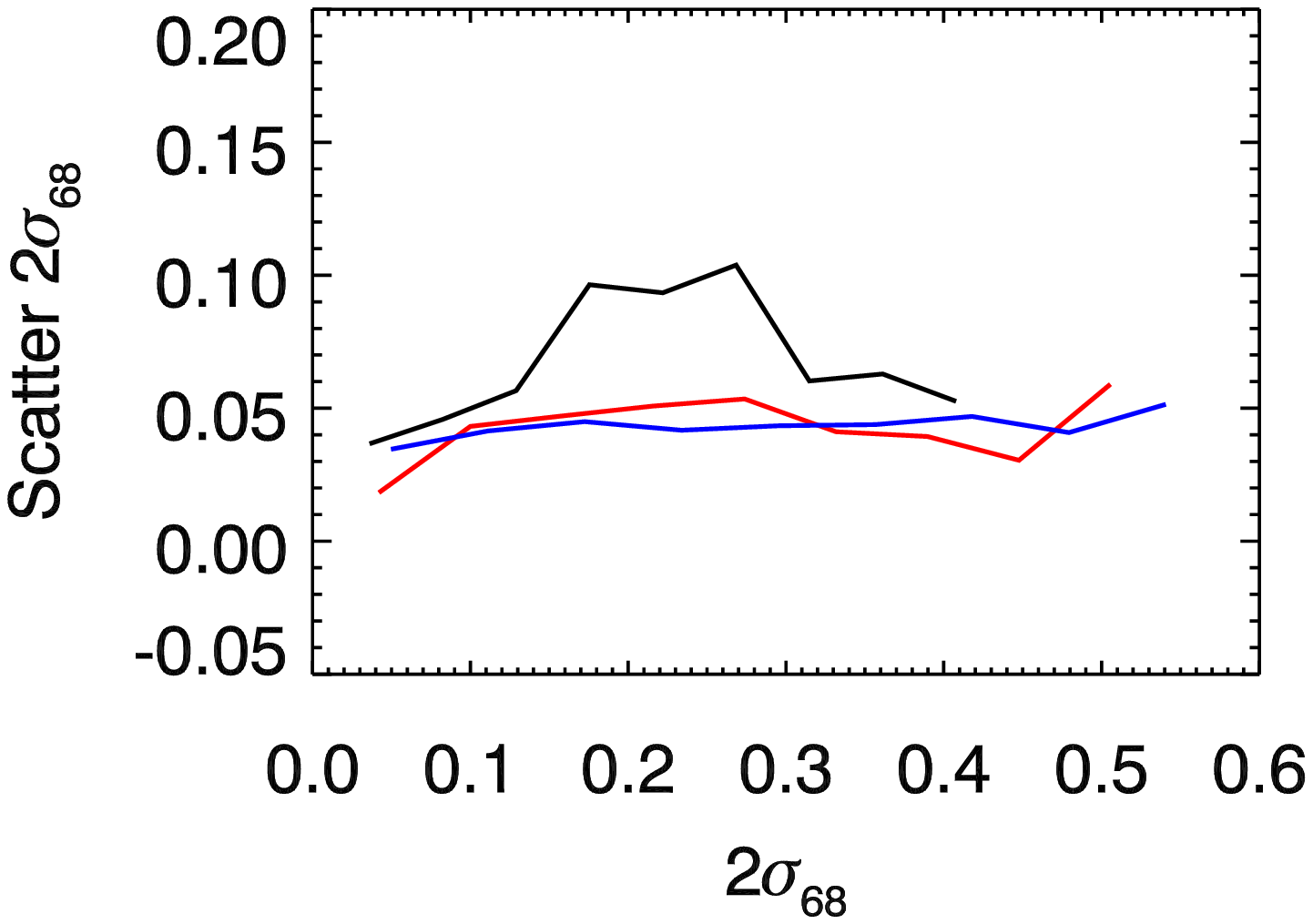}
}
\label{fig:gaussianfits}
\caption{Bias and scatter in the comparison between $\langle z \rangle$ from the $P(z)$ and from the fit (top) and between the dispersion from the $P(z)$ and from the fit (bottom).}
\enf
%%%%%%%%%%%%%%%%%%%%%%%%%%%%%%%%%%%%%%%%%%%%%%%%%%%%%%%%%%%%%%%%%%

\end{document}